\documentclass[fleqn,usenatbib]{mnras}

\usepackage{mathptmx}

\usepackage[T1]{fontenc}

\DeclareRobustCommand{\VAN}[3]{#2}
\let\VANthebibliography\thebibliography
\def\thebibliography{\DeclareRobustCommand{\VAN}[3]{##3}\VANthebibliography}

\usepackage{graphicx}	
\usepackage{amsmath}	
\usepackage{amssymb}	
\usepackage{pdflscape}

\usepackage{hyperref}

\defcitealias{luhman17}{L17}
\defcitealias{povich13}{P13}
\defcitealias{getman17}{G17}
\defcitealias{dunham15}{D15}
 
\title[W-band III]{A Novel Survey for Young Substellar Objects with the W-band filter III: Searching for very low-mass brown dwarfs in Serpens South and Serpens Core}

\author[S.C. Dubber et al.]{
Sophie Dubber$^{1,2}$\thanks{E-mail: dubber@roe.ac.uk}, 
Beth Biller$^{1,2}$\thanks{Visiting Astronomer at the Infrared Telescope Facility, which is operated by the University of Hawaii under Cooperative Agreement No. NCC 5-538 with the National Aeronautics and Space Administration, Office of Space Science, Planetary Astronomy Program.},
Katelyn Allers$^{3}$, 
Jessy Jose$^{4}$, 
Lo\"{\i}c Albert$^{5}$, 
\newauthor
Blake Pantoja$^{3}$, 
Clémence Fontanive$^{6}$,
Michael Liu$^{7}$\footnotemark[2], 
Zhoujian Zhang$^{7}$, 
\newauthor
Wen-Ping Chen$^{8}$,
Bhavana Lalchand$^{8}$, 
Belinda Damian$^{9}$,
Tanvi Sharma$^{8}$
\\
$^{1}$SUPA, Institute for Astronomy, Royal Observatory, University of Edinburgh, Blackford Hill, Edinburgh EH93HJ, UK\\
$^{2}$Centre for Exoplanet Science, University of Edinburgh, Edinburgh, UK\\
$^{3}$Department of Physics and Astronomy, Bucknell University, Lewisburg, PA 17837, USA\\
$^{4}$Indian Institute of Science Education and Research (IISER) Tirupati, Rami Reddy Nagar, Karakambadi Road, Mangalam (P.O.), Tirupati 517 507, India\\
$^{5}$Institut de recherche sur les exoplan\`etes, Universit\'e de Montr\'eal, Qu\'ebec, Canada\\
$^{6}$Center for Space and Habitability, University of Bern, Gesellschaftsstrasse 6, 3012 Bern, Switzerland
\\
$^{7}$Institute for Astronomy, University of Hawaii at Manoa, Honolulu, HI 96822, USA\\
$^{8}$Graduate Institute of Astronomy, National Central University, Zhongli, Taiwan\\
$^{9}$CHRIST (Deemed to be University), Hosur Road, Bengaluru 560029, India\\
}

\date{Accepted XXX. Received YYY; in original form ZZZ}

\pubyear{2019}

\begin{document}
\label{firstpage}
\pagerange{\pageref{firstpage}--\pageref{lastpage}}
\maketitle

\begin{abstract}

We present CFHT photometry and IRTF spectroscopy of low-mass candidate members of Serpens South and Serpens Core ($\sim$430 pc, $\sim$0.5 Myr), identified using a novel combination of photometric filters, known as the W-band method. We report SC182952+011618, SS182959-020335 and SS183032-021028 as young, low-mass Serpens candidate members, with spectral types in the range M7-M8, M5-L0 and M5-M6.5 respectively. Best-fit effective temperatures and luminosities imply masses of $<$ 0.12M$_{\odot}$ for all three candidate cluster members. We also present Hubble Space Telescope imaging data (F127M, F139M and F850LP) for six targets in Serpens South. We report the discovery of the binary system SS183044-020918AB. The binary components are separated by $\approx$45 AU, with spectral types of M7-M8 and M8-M9, and masses of 0.08--0.1 and 0.05--0.07~M$_{\odot}$. We discuss the effects of high dust attenuation on the reliability of our analysis, as well as the presence of reddened background stars in our photometric sample.  
\end{abstract} 




\section{Introduction} \label{sec:intro}

Nearby young star-forming regions, such as Perseus \citep{bally08}, Taurus-Auriga \citep{kenyon08}, Serpens South \citep{gutermuth08} \& Chamaeleon \citep{luhman08}, have been extensively surveyed with the goal of discovering low-mass members. A key motivating factor for such a survey is to improve the understanding of the statistical properties of stellar populations. 

The initial mass function (IMF) is an empirical function that describes the distribution of stellar masses at formation. Discovering low-mass members of star-forming regions constrains the substellar IMF in such clusters, which can then be directly compared to the substellar IMF in the local solar neighbourhood \citep[][]{kirkpatrick18}. The IMF may be environmentally dependent \citep[e.g][]{dokkum10,lu13,gennaro18,hosek19}, meaning investigations of its form in specific regions are crucial. Furthermore, different theories predict different values of the minimum mass \citep[e.g][]{larson92,whitworth07}, highlighting the importance of finding the lowest-mass members in every region.

There are other motivations to push detection thresholds into the planetary-mass regime in young star-forming regions. Planetary-mass brown dwarfs are often targets of atmospheric investigations \citep[e.g][]{knapp04,saumon06,cushing08}. They can be studied as analogues of directly-imaged planets, and increase our understanding of a different part of the temperature-surface gravity grid. Investigating the full range of possible atmospheric parameters is key in understanding how planets and brown dwarfs form and evolve. Doing so again requires a large sample of such objects.

Surveys of star-forming regions have predominantly used optical and infrared (IR) photometry to identify young, very low-mass members. Wide-field surveys such as 2MASS \citep{skrutskie06}, PanSTARRS \citep{chambers16}, and WISE \citep{wright10} play a crucial role in initial photometric identification. Typically, survey objects are placed on colour-magnitude diagrams, with the parameter space chosen such that late- and early-type populations lie in (somewhat) distinct regions \citep[e.g.~][]{briceno02,luhman99,rebull10}. Spectroscopic follow-up is then used to confirm the spectral type of suspected late objects, and to constrain their physical properties. This method is challenging as interstellar reddening by dust changes the observed properties of objects. This can cause the spectra of older, background stars in the field to be reddened to the extent that they have extremely similar colours to young brown dwarfs that are bonafide members of the star-forming region. As a result, spectroscopic follow-up can often reveal that objects chosen photometrically are in fact contaminants - and surveys identifying candidates in this way can suffer from low confirmation rates. In cases where photometry can be combined with proper motion information or additional photometric bands, confirmation rates can be much higher \citep[e.g.~][$\approx 80\%,$]{zapatero17,lodieu18}

The W-band technique uses standard $J$- and $H$-bands in combination with a custom medium-band (6\%) IR filter centred at 1.45$\mu$m \citep{allers20}, the W-band.  The filter is located at a wavelength that is sensitive to the depth of the H$_{2}$O absorption feature present in objects with spectral types of M6 or later, the approximate stellar/substellar boundary in typical star-forming regions \citep{alves10}. Photometry from these three bands can be combined to calculate a reddening-insensitive index, $Q$, which can be used to distinguish between early- and late-type objects. The calculated $Q$-values are used to identify candidates for spectroscopic follow-up, greatly improving the success rate of detecting brown dwarfs \citep{allers20,jose20}. \citet{najita00} were the first to use a tailored combination of filters to look for young low-mass stars and brown dwarfs. They used three narrow-band filters in the Hubble Space Telescope/NICOMOS imager, with one centred on the water absorption feature at 1.9$\mu$m, and the first and third measuring continuum flux either side - directly comparable to the W-band technique.

This is the third paper in the W-band series. In the first, \citet{allers20} present the initial proof-of-concept results, and report a confirmation rate of $84\%$ for the initial W-band survey - comparable to other surveys that use many more photometric bands. In the second, \citet{jose20} present the first results from the W-band survey of Serpens South. In this work, we present full results from the W-band survey of Serpens South and Serpens Core.
The Serpens star-forming region is more distant than other frequently-observed nearby regions ($d = 436.0 \pm 9.2$ pc, \citet{ortiz17,ortiz18}) and highly attenuated by dust (generally Av = 10--30~mag, see Section \ref{sec:serpens}). However, it is also young ($\sim$0.5Myr), compact and active in star-formation. The first candidate L-type dwarf member was reported in 2002 \citep{lodieu02}, and further deep IR photometric surveys \citep[e.g.~][]{klotz04,spezzi12} have each reported candidate low-mass objects in Serpens Core. \citet{winston18} detail a recent survey dedicated to finding the lowest mass members of Serpens South, reaching a lower mass limit of $\approx$ 0.1M$_{\odot}$. In the second W-band paper, \citet{jose20} present the first survey dedicated to finding ultracool dwarfs (the very lowest mass stars, brown dwarfs and planetary mass objects) in Serpens South. As such, the four discoveries presented in \citet{jose20} are the coolest and lowest mass candidate members of Serpens South identified to date. 
In this work, we use the same photometric catalogue of Serpens South as \citet{jose20}, and present follow-up of additional targets that were not discussed in this first paper.

We describe the properties of Serpens South and Serpens Core in Section \ref{sec:serpens}, and explain the $Q$-index in Section \ref{sec:q}. We present the observations undertaken in Section \ref{sec:obs}, which include: photometry obtained using the W-band filter on the Canada-France-Hawaii Telescope (CFHT), follow-up spectroscopy using SpeX on the NASA Infrared Telescope Facility (IRTF), and imaging of a subsample of objects obtaining using the Hubble Space Telescope (HST). In Section \ref{sec:extinc}, we consider the effects of dust attenuation on our survey results. In Section \ref{sec:results}, we discuss the late-type candidate members of Serpens South and Serpens Core discovered in this work, as well as their physical properties. Finally, we present the results for a newly discovered binary system, identified using our HST data. 

\section{The Serpens star-forming Region} \label{sec:serpens}

\begin{figure*}
    \centering
    \includegraphics[width=\textwidth]{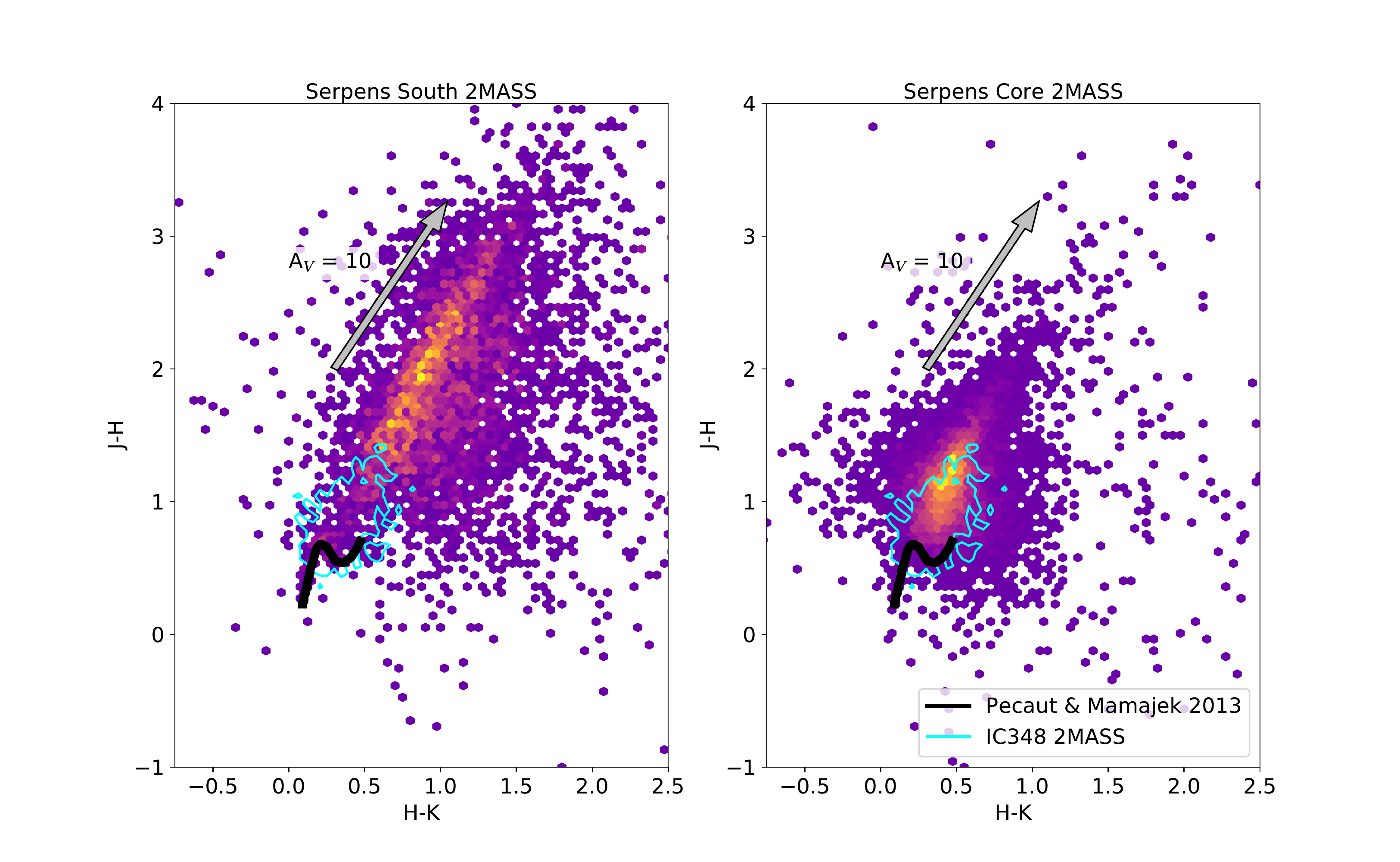}
    \caption{2MASS $J-H$ vs $H-K$ colour-colour diagrams for Serpens South (left) and Serpens Core (right). Binned colour-colour number density distributions of objects in each region detected by 2MASS are shown, with yellow indicating the highest density of sources. Also shown in each panel is the $A_{\text{V}} = 10$ extinction vector (grey arrow). The black (thick) line is the intrinsic colour sequence of G-M stars from \citet{pecaut13}, and plotted in cyan (thin) is a 1$\sigma$ contour of objects in the IC348 star forming region, again obtained using 2MASS.}
    \label{fig:jhhk}
\end{figure*}

Serpens is a highly compact, highly dust attenuated region containing at least 2000 stars \citep{herczeg19}. The physical properties of the cluster vary significantly across its extent. Furthermore, the region has low proper motion, and is located near the galactic plane. Both factors present significant challenges when searching for young, low-mass objects in this region. Proper motion selection of candidates is nearly impossible, and there are more background contaminant objects compared to other areas of sky. 

The Serpens Core subcluster is part of Serpens Main, historically the best studied part of the complex \citep[see][for a detailed review of Serpens Core]{eiroa08}. Serpens Main is actively star-forming, with a few hundred known candidate members. \citet{herczeg19} report optical counterparts for many of these members. The Serpens South subcluster, discovered by \citet{gutermuth08} in a {\it Spitzer} survey, lies approximately $3\deg$ south of Serpens Main in a separate cloud \citep{herczeg19}. It contains a large fraction of protostars \citep[77$\%$,][]{gutermuth08}, and \citet{herczeg19} find that few of these have optical counterparts, suggesting this region may be younger than the other Serpens subclusters.  The distance to Serpens remains a topic of debate, with accepted values ranging from $\approx 260-460$ pc \citep[see][for a detailed review of recent distance estimates to Serpens South and Serpens Core]{winston18}. 
Whilst the two regions are known to be spatially distinct, various works have shown that the radial extent of the Serpens cloud is small, and as such we adopt the same distance for both subclusters, $d = 436.0 \pm 9.2$ pc \citep{ortiz17,ortiz18}.

Stars in the Serpens star-forming region are affected by high levels of dust extinction \citep{herczeg19}. To quantify this, and to investigate any differences between the two subclusters, we compared their 2MASS \citep{skrutskie06} colour-colour diagrams. Using the same survey areas as our CFHT photometric observations (see Section \ref{sec:obs}), we found 2MASS $J-H$ and $H-K$ colours for all objects detected in these fields. Figure \ref{fig:jhhk} shows colour-colour number density plots for Serpens South (left) and Serpens Core (right) (where the brightest areas indicate the highest density of objects). Also shown is the intrinsic colour sequence for G-M stars given in \citet{pecaut13}. $A_{\text{V}}$ = 10 extinction vectors are plotted, along with the 1$\sigma$ contour of objects from a 2MASS query of the IC348 star-forming region, which has comparatively low extinction \citep[$A_{\text{V}} \approx 1-7$,][]{herbst08}. \\
When examining the colour-colour diagrams in conjunction with this additional information, the high extinction of both subclusters is clear. Both 2MASS distributions extend well above the low extinction contour of IC348 and the intrinsic G-M colour sequence, suggesting high values of extinction for many objects in both regions. We can quantify the approximate peak extinction values by comparing to the average intrinsic $J-H$ colour of the \citet{pecaut13} sequence, $J-H \approx 0.5$. We can see from Figure \ref{fig:jhhk} that the maximum value of $J-H$ for sources in Serpens South is $\approx 3.25$ (not including a scattering of bins populated by just one object that are above this value), and is similarly $\approx 2.5$ for Serpens Core. This is equivalent to a colour excess $E(J-H)$ = 2.75 mag and $E(J-H)$ = 2.0 mag for Serpens South and Serpens Core respectively, when compared to the intrinsic colours from \citet{pecaut13}. Using the extinction law of \citet{fitzpatrick99} we can convert these colour excesses to approximate upper limits on the extinction in each region: $A_{\text{V,max}} \approx 33$ for Serpens South, and $A_{\text{V,max}} \approx 24$ for Serpens Core. This is comparable to the results obtained from analysis of our own photometric catalogue, discussed in Section \ref{sec:extinc} (extinction maps shown in Figure \ref{fig:av_map}).

\section{The Reddening Insensitive Index ($Q$)} \label{sec:q}

\subsection{CFHT Photometry} \label{sec:q_cfht}

\begin{figure}
    \centering
    \includegraphics[width=\columnwidth]{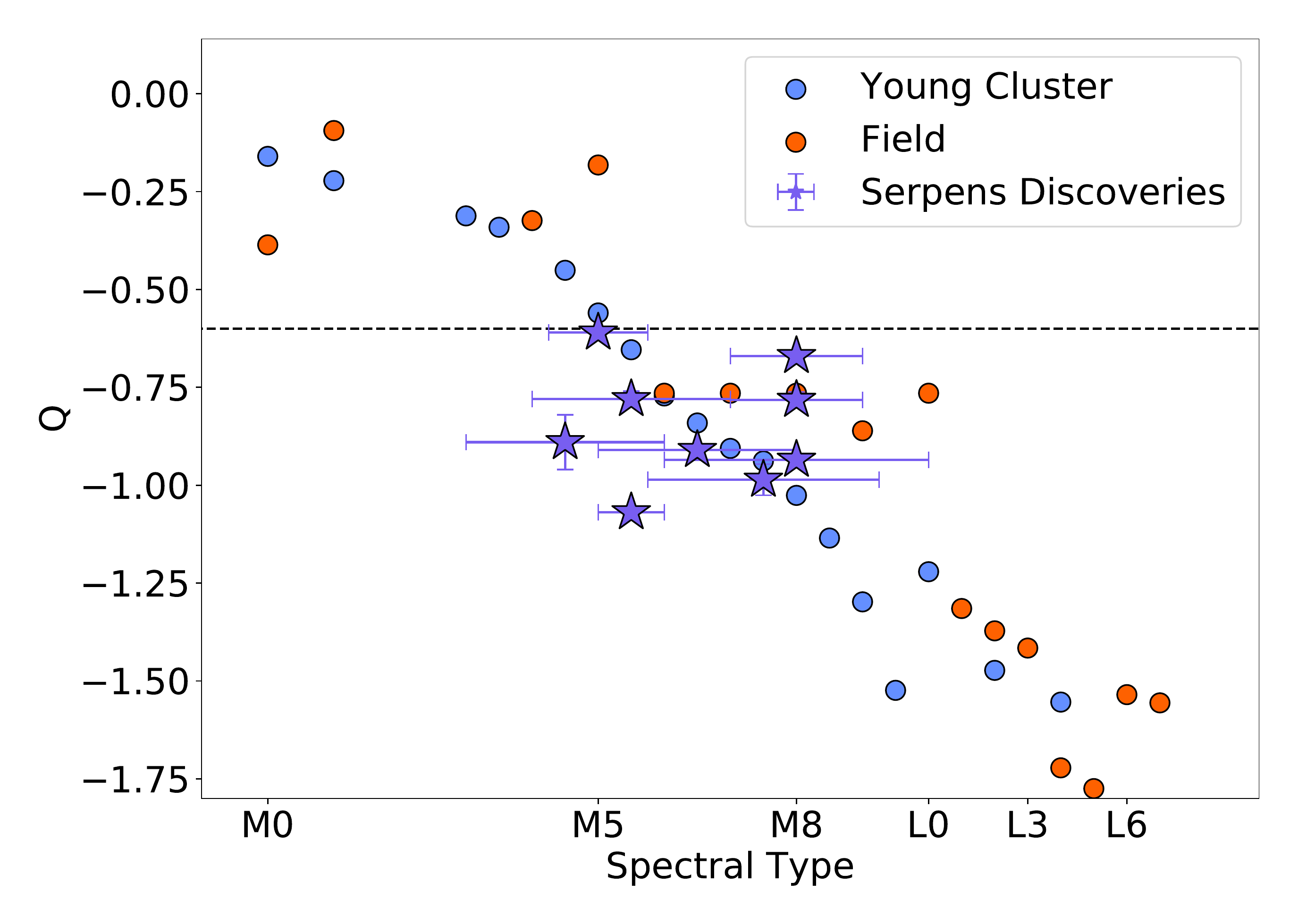}
    \caption{$Q$ vs SpT for field objects and young cluster members. Orange shows the synthetic $Q$ values of field-age standards of varying spectral types, taken from the SpeX spectral library \citep[]{cushing05,rayner09}. Pink shows the synthetic $Q$ values of young, low surface gravity standards \citep{luhman17}, modelled at the distance of Serpens. Purple stars show our young, late type candidate Serpens members.}
    \label{fig:q_spt}
\end{figure}

For the W-band photometric survey of Serpens, we used the WIRCam instrument on CFHT \citep{puget04}. We imaged the Serpens South and Serpens Core fields across 5 nights in 2016-2017. The W-band method uses a custom filter at 1.45$\mu$m combined with standard CFHT $J$ and $H$ photometry to build a reddening insensitive index, $Q$. Our custom filter is centred on the 1.45$\mu$m H$_{2}$O absorption feature, the depth of which is an indicator of spectral type (SpT). The construction of the $Q$-index is explained in detail in \citet{allers20}. For each object, we calculate $Q$:

\begin{equation}
    \centering
    Q = (J-W) + e(H-W)
    \label{eq:q}
\end{equation}

where $J$,$H$, and $W$ are the magnitudes of the object observed in these photometric bands, and $e$ is a ratio of extinction in said bands:

\begin{equation}
    \centering
    e = (A_{\text{J}} - A_{\text{W}}) / (A_{\text{H}} - A_{\text{W}})
    \label{eq:e}
\end{equation}

A value of $e$ must be adopted based on the type of contaminant that is most common. This value was determined for the entire W-band survey, not specifically for observations of Serpens. The most common background contaminants in the survey are M0 stars, so the $Q$ scale was fixed such that a star with SpT = M0 corresponds to $Q=0$. Consequently, increasingly negative values of $Q$ correspond to objects with increasingly later spectral types. This is demonstrated in Figure \ref{fig:q_spt}. Here we show $Q$ vs SpT for field and young cluster objects (calculated from synthetic photometry of standard targets), as well as for the new Serpens candidate members discovered in \citet{jose20} and in this work.

To fix the value of $e$, we used synthetic photometry of an M0 standard spectrum from \citet{kirkpatrick10}, reddened by $A_{\text{V}} = 10$ mag (an average value along all lines of sight observed in the W-band survey) using the $R_{\text{V}} = 3.1$ reddening law of \citet{fitzpatrick99}. We compared reddened synthetic photometry for this object to unreddened values, and used this to determine the selective extinctions. These were then used to calculate $e$, resulting in $e$ = 1.85. This value was used in all W-band $Q$ calculations.

Based on this $Q$-scale, we can estimate which values should correspond to the spectral types of interest for this survey. Using synthetic photometry in $J$,$H$ and $W$ of young objects \citep{allers13,muench07} and field standards \citep{cushing05}, we calculated that objects with spectral types later than M6 \citep[the approximate stellar/substellar boundary,][]{alves10} should have values of $Q < -0.6$. 

\subsection{HST Photometry} \label{sec:q_hst}

We obtained HST WFC3 photometry for a subsample of our Serpens South objects. Due to the different filter set available with HST WFC3, we required a second version of the reddening insensitive index, $Q_{\text{HST}}$. We use a combination of filters: F139M (1.34-1.43$\mu$m) gives information on the H$_{2}$O absorption feature, and is used in conjunction with F127M (1.22-1.32$\mu$m) and F850LP (0.82-1.09$\mu$m) fluxes to build $Q_{\text{HST}}$:

\begin{equation}
    Q_{\text{HST}} = -2.5 \text{log} \left(\frac{F_{\text{F850LP}}}{F_{\text{F127M}}} \right) + e \times 2.5 \text{log} \left( \frac{F_{\text{F127M}}}{F_{\text{F139M}}} \right)
    \label{eq:q_hst}
\end{equation}

where $F_{\text{F850LP}}$, $F_{\text{F127M}}$ and $F_{\text{F139M}}$ are the filter fluxes, and $e$ is calculated as in Eq. \ref{eq:e}, but using this HST filter set. Using $Q_{\text{HST}}$, objects with late-M, L \& T spectral types will have $Q_{\text{HST}} > 1$, while background stars will have $Q_{\text{HST}}$  $\approx$ 0.

\section{Observations} \label{sec:obs}

\subsection{Photometric survey observations} \label{sec:phot}

\begin{figure*}
    \centering
    \includegraphics[width=\textwidth]{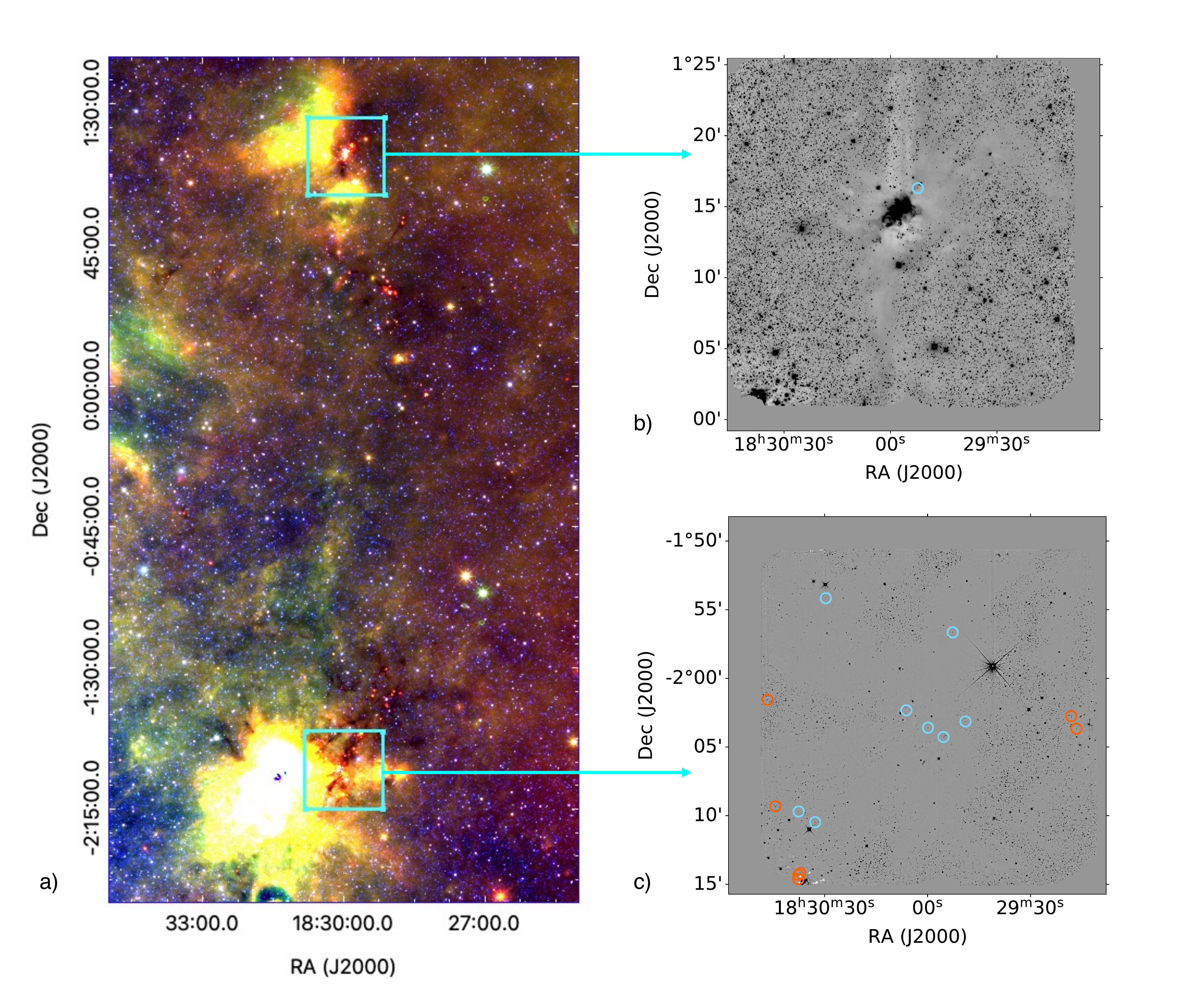}
    \caption{a) Colour-composite image of Serpens Core, Serpens South and the surrounding regions, obtained using WISE 22$\mu$m (red), 12$\mu$m (green) and 3.4$\mu$m (blue) images. The cyan boxes show the areas covered by the WIRCam observations of Serpens Core (upper) and Serpens South (lower). Both contain the dark filament structures that run through each subcluster. The nebulosity that contributes to the high visual extinctions of both regions is also visible. b) $J$-Band WIRCam image of Serpens Core. The blue circle indicates the Serpens Core object, SC182952+011618, followed up spectroscopically in this paper. c) $J$-Band WIRCam image of Serpens South. The orange circles show objects followed up spectroscopically in \citet{jose20}. The blue circles show the 8 Serpens South objects with spectroscopic follow-up reported in this work.}
    \label{fig:serpens}
\end{figure*}

Photometry for both Serpens South and Serpens Core was obtained using WIRCam on CFHT \citep{puget04}. The WIRCam field of view is $\sim 20' \times 20'$, with a sampling of 0.3 arcseconds per pixel. The Serpens South photometric catalogue was first discussed in \citet{jose20}. The Serpens Core catalogue is presented for the first time in this paper. Both were processed and calibrated according to the methods described in \citet{jose20}.

A single pointing was used to image all of Serpens South, centred on RA = 277.5125$^{\circ}$, Dec = -2.0327$^{\circ}$. We used a 21-point dithering pattern to fill the gaps between the four detectors of WIRCam and to accurately subtract the sky background. This photometry was obtained during 14-15th July 2016. In addition to $J$- and $H$-bands, we also obtained photometry using our custom 1.45$\mu$m ($W$-band) filter. The integration times used were 1890, 1920, and 12285s for $J$, $H$ and $W$ respectively. 

A single pointing was also used to image all of Serpens Core, centred on RA = 277.4729$^{\circ}$, Dec = 1.2055$^{\circ}$. We again used a 21-point dithering pattern and obtained photometry using the $J$,$H$ and $W$ filters. The integration times used were 1350, 1650, and 14625s for $J$,$H$ and $W$ respectively, with the images taken on 12th, 13th and 15th April 2017. 

The stacked $J$-band exposures of Serpens Core and Serpens south are shown in panels b) and c) of Figure \ref{fig:serpens}. Figure \ref{fig:serpens} also shows a WISE colour-composite image (22$\mu$m (red), 12$\mu$m (green) and 3.4$\mu$m (blue)) covering $\sim2.5^{\circ} \times 4.5^{\circ}$, along with the WIRCam imaging areas for each subcluster (cyan boxes). Comparing these imaging areas to Figure 4 of \citet{herczeg19} - which shows the spatial extent of both Serpens South and Serpens Core, based on Gaia DR2 data \citep{gaia18} - it is clear that they cover the dense central regions of each subcluster, but are likely not large enough in size to encompass all known members.

\subsection{Photometric criteria for candidate member selection} \label{sec:selection}

\subsubsection{Serpens South} \label{sec:phot_ss}

\begin{figure}
    \centering
    \includegraphics[width=\columnwidth]{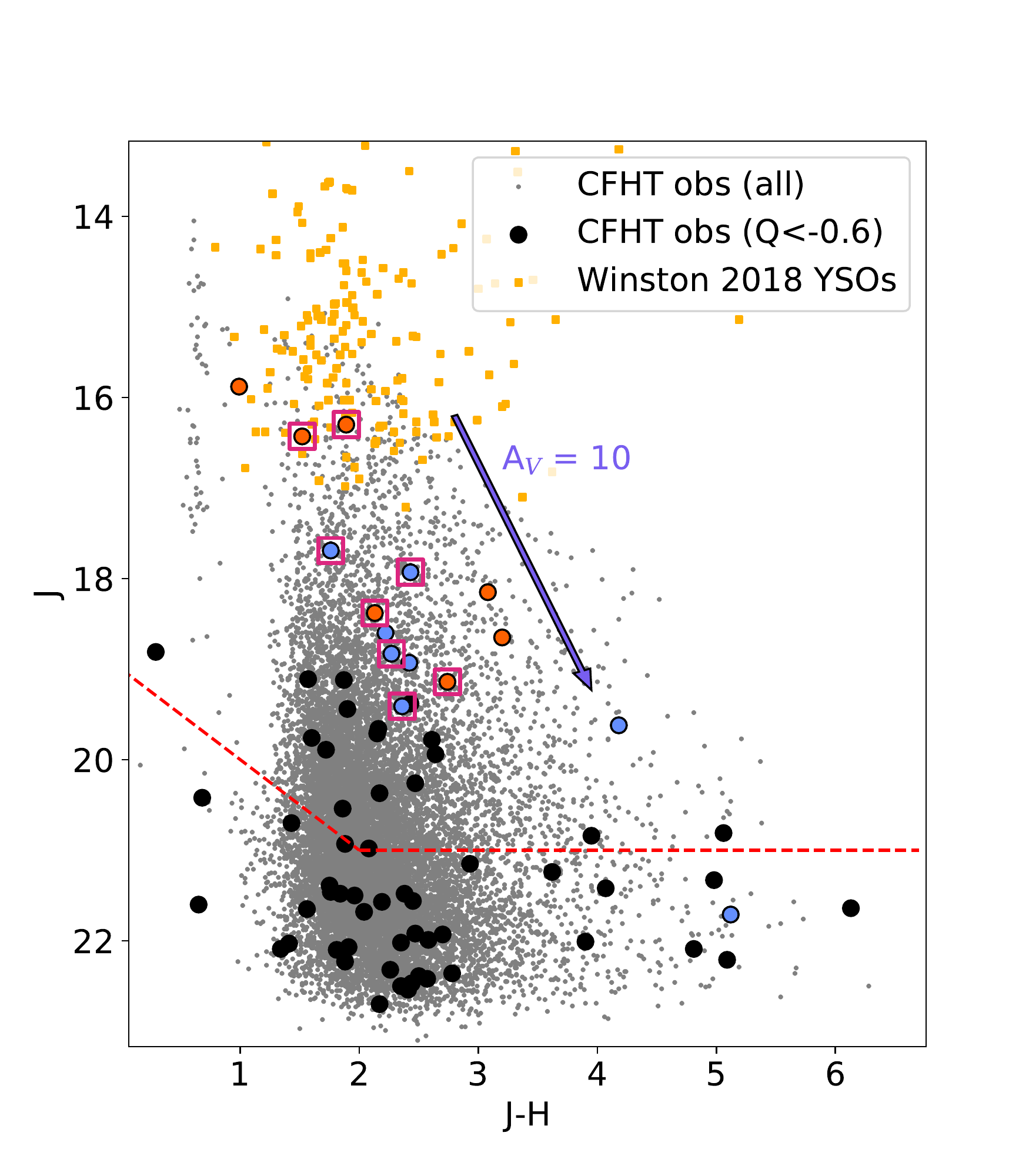}
    \caption{$J$-$H$ vs $J$ colour-magnitude diagram for Serpens South objects with magnitude errors $< 0.1$ mag (grey). The black points show all of the objects that satisfy the photometric criteria $Q \le -(0.6 +Q_{\text{err}})$ if $H < 18$ mag or $Q \le -(0.6 +3 \times Q_{\text{err}})$ if $H > 18$ mag (69 targets). The large orange circles show objects followed up spectroscopically in \citet{jose20}; large blue circles are objects followed up spectroscopically in this work. Pink boxes indicate confirmed young, low-mass Serpens South candidates. Likely YSO members of Serpens South from \citet{winston18} are shown as yellow squares. The dashed red line indicates the completeness limit for the CFHT photometric survey, and the purple arrow shows the $A_{\text{V}} = 10$ extinction vector.}
    \label{fig:all_phot_ss}
\end{figure}

\begin{figure}
    \centering
    \includegraphics[width=\columnwidth]{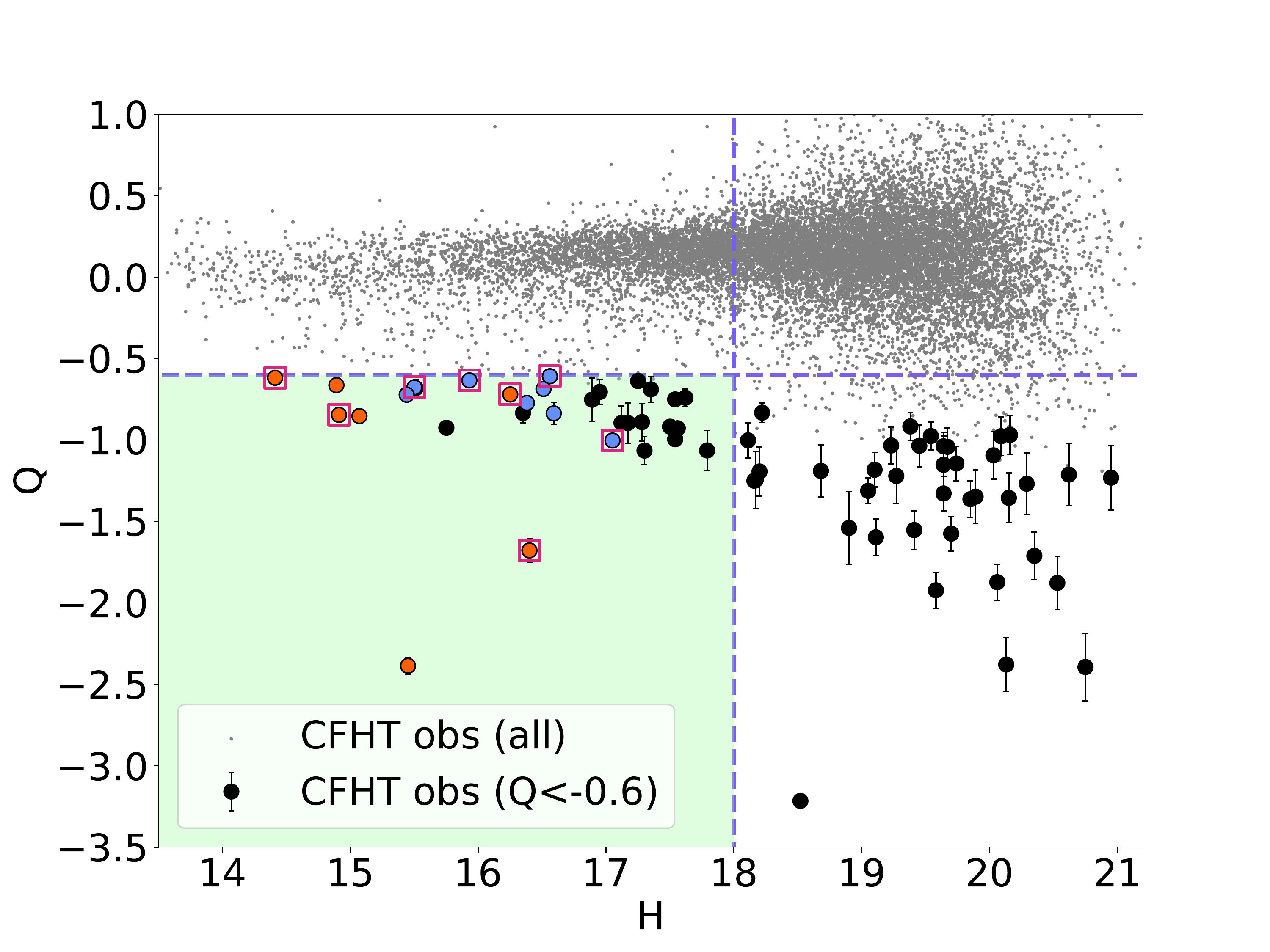}
    \caption{$H$ mag vs $Q$ for all objects in our Serpens South catalogue with photometric precision $< 0.1$ mag (grey). The black points show all of the objects that satisfy the photometric criteria $Q \le -(0.6 +Q_{\text{err}})$ if $H < 18$ mag or $Q \le -(0.6 +3 \times Q_{\text{err}})$ if $H > 18$ mag (69 targets). The orange points show objects followed up spectroscopically in \citet{jose20}; blue shows objects followed up spectroscopically in this work. Pink boxes indicate confirmed young, low-mass Serpens South candidates. The green box shows the region of parameter space suitable for spectroscopic follow-up with 4-m class telescopes. } 
    \label{fig:q_phot_ss}
\end{figure}

Figure \ref{fig:all_phot_ss} shows the $J$-$H$ colour vs $H$ magnitude for all objects in Serpens South with photometric uncertainty $< 0.1$mag, extracted from our CFHT images (a total of 15,276 objects, grey points without error bars in Figures \ref{fig:all_phot_ss} and \ref{fig:q_phot_ss}). Also shown in Figure \ref{fig:all_phot_ss} are young stellar objects (YSOs) found to be likely members of Serpens South by \citet{winston18} (yellow squares).

As discussed in Section \ref{sec:q}, objects with $Q < -0.6$ are likely to have spectral types $\ge$ M6. Figures \ref{fig:all_phot_ss} and \ref{fig:q_phot_ss} show all objects, in black, that satisfy the criteria $H < 18$ mag and $Q \le -(0.6 +Q_{\text{err}})$ or $H > 18$ mag and $Q \le -(0.6 +3 \times Q_{\text{err}})$ (where $Q_{\text{err}}$ is calculated using the standard propagation of errors). A more stringent cut on $Q_{\text{err}}$ is required for fainter objects, as their photometric errors are significantly larger, as is the likelihood of contamination by reddened background sources. This is effectively a 1$\sigma$/3$\sigma$ cut on the sample, and gives a total of 69 objects.

To identify targets for spectroscopic follow, we further examined the $Q < -0.6$ sample. We additionally required objects to be sufficiently bright for spectroscopic follow-up with a 4-m class telescope ($H < 18$ mag) - this was a total of 32 objects, highlighted by the green quadrant of Figure \ref{fig:q_phot_ss}. We then cross-matched these with the ALLWISE catalogue \citep{cutri13}. 
We expect targets with spectral types $\ge$ M6 to have intrinsic $W1-W2 > 0.1$ \citep{kirkpatrick11}, and thus required this condition to be met by all candidates. We also noted that any target with a very bright $W1$ magnitude is unlikely to be sub-stellar at the distance of Serpens. We used evolutionary models from \citet{baraffe15} to quantify this brightness cutoff (approximately equating the $L'$ band with $W1$, thus assuming solely photospheric flux). Considering an upper limit of 0.2M$_{\odot}$, and an object at the age (0.5 Myr) and distance (436 pc) of Serpens, we find that objects with masses above this limit will have a W1 magnitude brighter than $\approx$ 11 mag. This is considering objects without visual extinction - which is clearly unlikely for objects in Serpens (see Section \ref{sec:extinc}). Considering instead $A_{\text{V}}$ = 5 mag,  we find that objects with masses 0.2M$_{\odot}$ will have a $W1$ magnitude brighter than $\approx$ 12 mag. Consequently, we remove targets with $W1 < 12 $mag, or $W1-W2 < 0.1$ from our final sample. We also cross-matched with PS1 \citep{chambers16} to remove objects with bright optical counterparts (e.g. $z_{\text{PS1}}-J_{\text{CFHT}} \le 2$). Combining these photometric criteria, we obtain a final sample of 29 objects suitable for spectroscopic follow-up. CFHT and ALLWISE photometry for the 29 photometric candidates is reported in Table \ref{tab:phot_cfht}.

We chose to follow-up the brightest objects ($H < 17$ mag) spectroscopically. Seven bright targets were reported and discussed in \citet{jose20}, highlighted as orange points in Figures \ref{fig:serpens}, \ref{fig:all_phot_ss} and \ref{fig:q_phot_ss}. We report spectroscopic follow-up of eight additional bright targets (blue in Figures \ref{fig:serpens}, \ref{fig:all_phot_ss} and \ref{fig:q_phot_ss}). The pink squares highlight which of these targets were later confirmed as late-types. The remaining targets from Table \ref{tab:phot_cfht} are strong candidates for further spectroscopic follow-up.

\subsubsection{Serpens Core} \label{sec:phot_sc}

\begin{figure}
    \centering
    \includegraphics[width=\columnwidth]{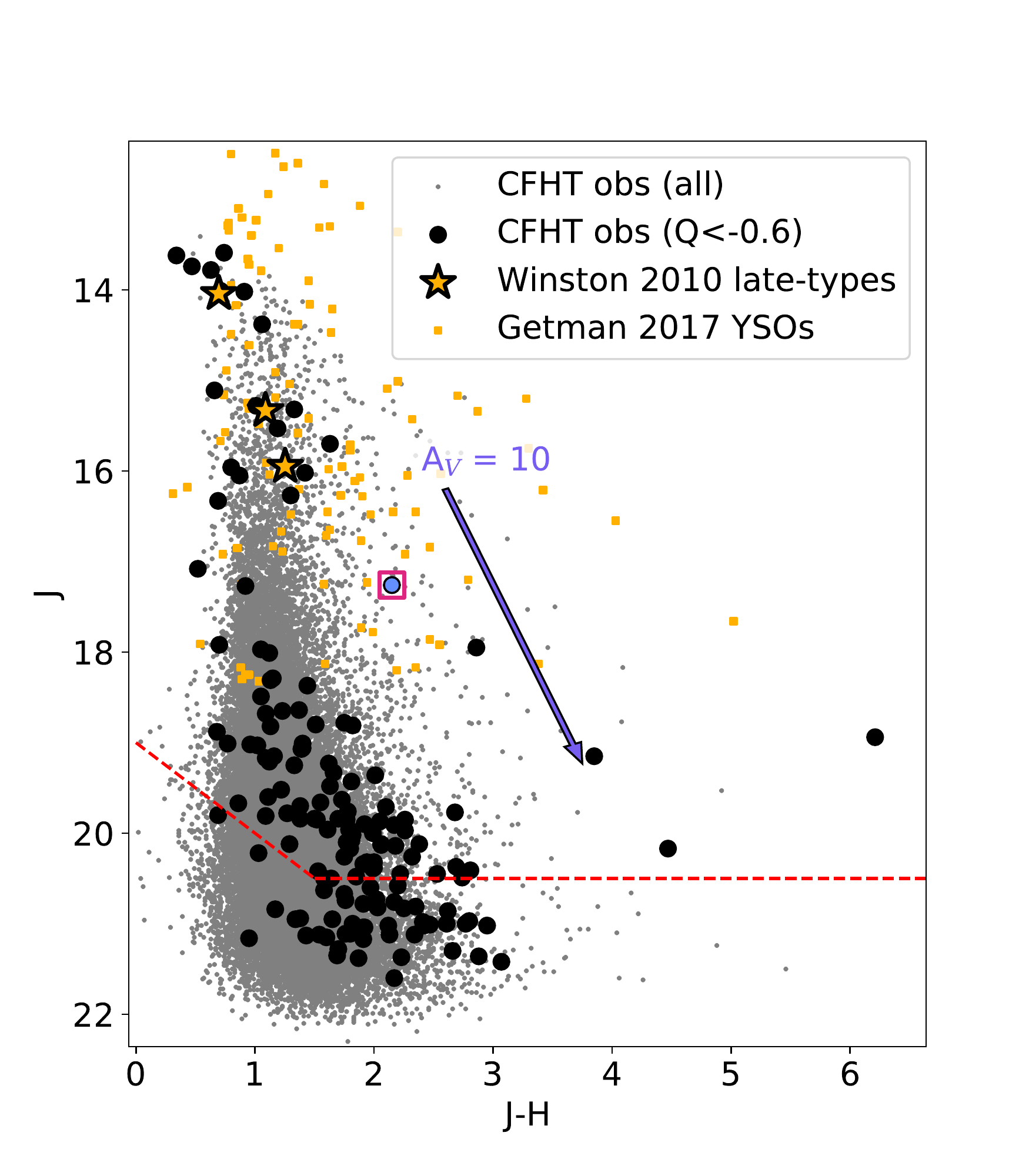}
    \caption{$J$-$H$ vs $J$ colour-magnitude diagram for Serpens Core objects with magnitude errors $< 0.1$ mag (grey). The black points show all of the objects that satisfy the photometric criterion $Q \le -(0.6 + 5 \times Q_{\text{err}})$ (151 targets). The blue point shows the target followed up spectroscopically in this work, SC182952+011618 - with a pink box indicating it was confirmed as a young, low-mass Serpens Core candidate member. Likely YSO members of Serpens South from \citet{getman17} are shown as yellow squares. Also shown are three late-type spectroscopically confirmed objects from \citet{winston10}, discussed in Section \ref{sec:phot}. The dashed red line indicates the completeness limit for the CFHT photometric survey, and the purple arrow shows the $A_{\text{V}} = 10$ extinction vector.}
    \label{fig:all_phot_sc}
\end{figure}

\begin{figure}
    \centering
    \includegraphics[width=\columnwidth]{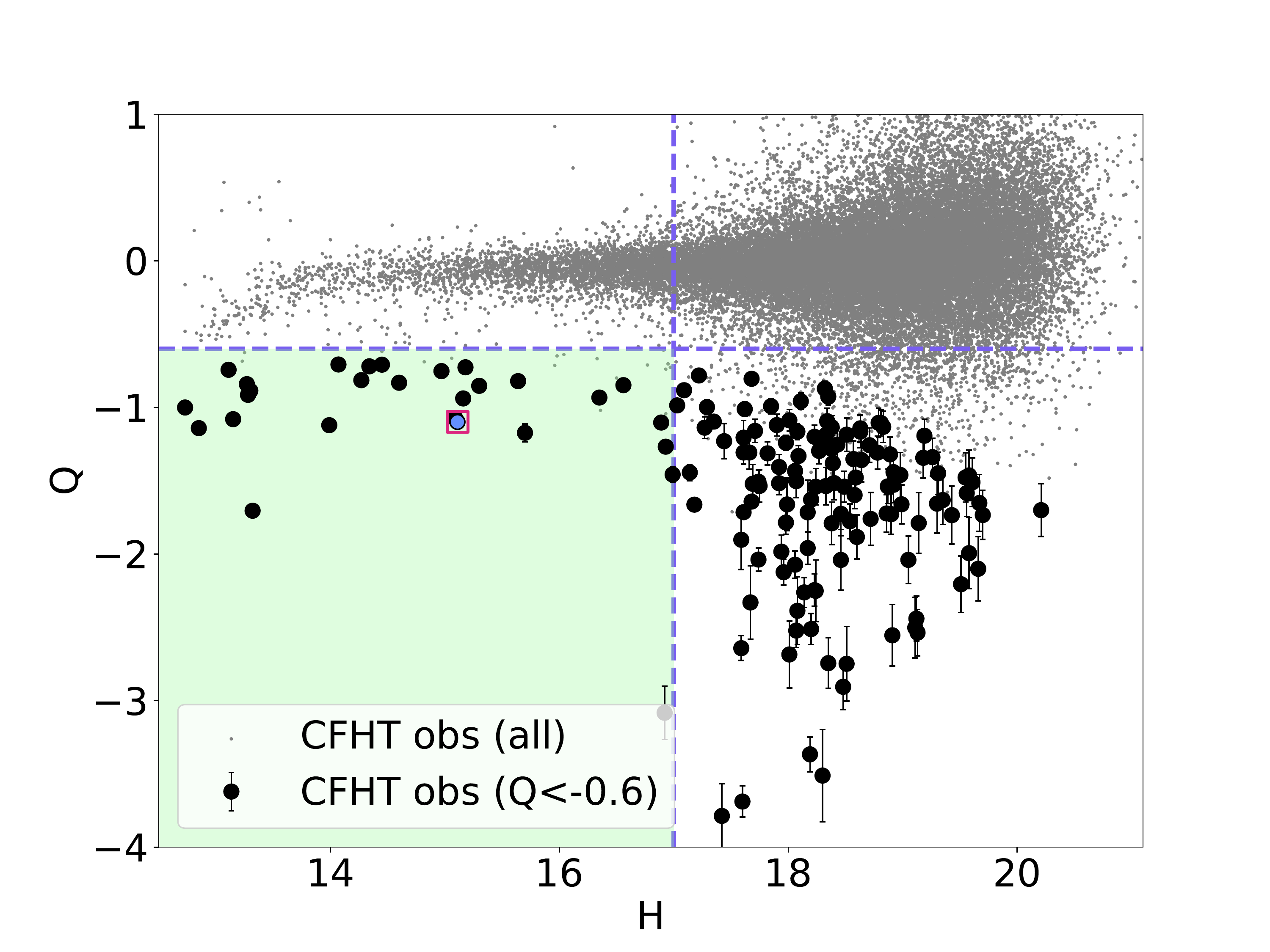}
    \caption{$H$ mag vs $Q$ for all objects in our Serpens Core catalogue with photometric precision $< 0.1$ mag (grey). The black points show all of the objects that satisfy the photometric criterion $Q \le -(0.6 + 5 \times Q_{\text{err}})$ (151 targets). The blue point shows the target followed up spectroscopically in this work - with a pink box indicating it was confirmed as a young, low-mass Serpens Core candidate. The green box shows the region of parameter space suitable for spectroscopic follow-up with 4-m class telescopes.}
    \label{fig:q_phot_sc}
\end{figure}

Objects with photometric uncertainty $ < 0.1$ mag in Serpens Core are shown in grey in Figures \ref{fig:all_phot_sc} and \ref{fig:q_phot_sc}, a total of 34,835 objects. The high density of sources in the core is seen in Figure \ref{fig:serpens}, and leads to a larger quantity of objects in the full photometric sample. As a result, we used a more stringent cut on $Q_{\text{err}}$ to create a subsample of objects considered for follow-up. We selected objects satisfying $Q \le -(0.6 +5 \times Q_{\text{err}})$, effectively a 5$\sigma$ cut. These are shown as the black points in Figures \ref{fig:all_phot_sc} and \ref{fig:q_phot_sc}, a total of 151 objects. Also shown in Figure \ref{fig:all_phot_ss} are likely YSO members of Serpens Core from \citet{getman17}.

To select targets for spectroscopic follow-up, we retain objects observable with a 4-m class telescope ($H$ = 17 mag for Serpens Core). We also removed objects with $Q < -2.0$ here, as previous experience showed that the majority of these objects with extremely low $Q$ values (in this magnitude range) would be contaminant objects with spurious photometry. More specifically, we expect $H$ vs $Q$ to follow a trend for a given cluster (i.e objects at the same distance with similar extinctions) -  a brighter object should have a higher mass and more positive value of $Q$ than a lower mass object. This can be seen in Figure \ref{fig:q_phot_ss}: the object with $H$ = 14.55, $Q$=-2.39 lies well below the general trend. Spectroscopic follow-up for this target was reported in \citet{jose20} (SS183038-021437 in Table \ref{tab:phot_cfht}) - and it was found to be a clear contaminant. After imposing this additional photometric cut, we retain a sample of 27 objects, shown in the green shaded quadrant of Figure \ref{fig:q_phot_sc}.

As in Serpens South, we applied WISE-based selection criteria. In addition to the criteria discussed above ($W1 > 12$ and $W1-W2 > 0.1$), we also required Serpens Core candidates to have WISE detections - i.e. WISE dropouts were removed from the sample. This gave us a sample of 7 candidates suitable for spectroscopic follow-up - 26$\%$ of the targets that meet all other photometric criteria also have WISE detections. CFHT photometry for these candidates is reported in Table \ref{tab:phot_cfht}. We obtained spectroscopic follow-up of one of these objects, which is highlighted in pink in Figures \ref{fig:all_phot_sc} and \ref{fig:q_phot_sc}.\\
SC182955+011034, SC182956+011218 and SC183005+011235, three of the six candidates we have not yet followed up, are spectroscopically characterised in \citet{winston10}. These targets have spectral types of M5, M8.5 and M9, respectively. They are shown for comparison to the W-band sample in Figure \ref{fig:all_phot_sc} as yellow stars.

\begin{table*}
    \centering 
    \begin{tabular}{lcccccccr}
        \hline
        Object ID & RA & Dec & [1.45] & $J_{\text{CFHT}}$ & $H_{\text{CFHT}}$ & $Q_{\text{CFHT}}$ & $W1$ & $W2$ \\
        & deg & deg & mag & mag & mag & & mag & mag \\
        \hline
        \multicolumn{5}{l}{{\it Serpens South candidates with spectroscopic follow-up:}} \\
        
        SS182917-020340$^{\dagger}$ & 277.3193 & -2.0610 & 17.25 $\pm$ 0.01 & 18.38 $\pm$ 0.01 & 16.25 $\pm$ 0.01 & -0.72$ \pm$ 0.02 & - & - \\  
        SS182918-020245$^{\dagger}$ & 277.3256 & -2.0458 & 15.47 $\pm$ 0.01 & 16.43 $\pm$ 0.01 & 14.91 $\pm$ 0.01 & -0.85 $\pm$ 0.01 & 12.39 $\pm$ 0.03 & 11.84 $\pm$ 0.03 \\ 
        SS182949-020308 & 277.4540 & -2.0522 & 17.43 $\pm$ 0.01 & 18.60 $\pm$ 0.01 & 16.38 $\pm$ 0.01 & -0.77 $\pm$ 0.01 & 14.58 $\pm$ 0.04 & 14.34 $\pm$ 0.07 \\
        SS182953-015639 & 277.4696 & -1.9442 & 16.59 $\pm$ 0.01 & 17.93 $\pm$ 0.01 & 15.50 $\pm$ 0.01 & -0.68 $\pm$ 0.01 & 13.40 $\pm$ 0.03 & 12.73 $\pm$ 0.03 \\
        SS182955-020416 & 277.4807 & -2.0712 & 17.57 $\pm$ 0.01 & 18.83 $\pm$ 0.01 & 16.56 $\pm$ 0.01 & -0.61 $\pm$ 0.01 & - & - \\
        SS182959-020335 & 277.4995 & -2.0598 & 18.23 $\pm$ 0.01 & 19.41 $\pm$ 0.02 & 17.05 $\pm$ 0.01 & -1.00 $\pm$ 0.04 & - & - \\
        SS183006-020219 & 277.5258 & -2.0387 & 18.68 $\pm$ 0.01 & 21.71 $\pm$ 0.06 & 16.59 $\pm$ 0.01 & -0.84 $\pm$ 0.07 & - & - \\
        SS183029-015409 & 277.6235 & -1.9026 & 17.16 $\pm$ 0.01 & 19.62 $\pm$ 0.02 & 15.44 $\pm$ 0.01 & -0.72 $\pm$ 0.02 & 12.18 $\pm$ 0.03 & 11.40 $\pm$ 0.02 \\
        SS183032-021028 & 277.6365 & -2.1745 & 16.77 $\pm$ 0.01 & 17.69 $\pm$ 0.01 & 15.93 $\pm$ 0.01 & -0.63 $\pm$ 0.01 & 13.75 $\pm$ 0.05 & 13.61 $\pm$ 0.08 \\
        SS183037-021411$^{\dagger}$ & 277.6545 & -2.2363 & 16.45 $\pm$ 0.01 & 18.15 $\pm$ 0.03 & 15.07 $\pm$ 0.01 & -0.85 $\pm$ 0.05 & 12.54 $\pm$ 0.03 & 12.25 $\pm$ 0.02 \\ 
        SS183037-020941 & 277.6566 & -2.1616 & 17.60 $\pm$ 0.01 & 18.93 $\pm$ 0.01 & 16.51 $\pm$ 0.01 & -0.69 $\pm$ 0.01 & - & - \\
        SS183038-021419$^{\dagger}$ & 277.6568 & -2.2386 & 17.95 $\pm$ 0.01 & 19.14 $\pm$ 0.04 & 16.40 $\pm$ 0.01 & -1.68 $\pm$ 0.07 & 12.86 $\pm$ 0.03 & 11.86 $\pm$ 0.02 \\ 
        SS183038-021437$^{\dagger}$ & 277.6570 & -2.2437 & 17.41 $\pm$ 0.01 & 18.65 $\pm$ 0.04 & 15.45 $\pm$ 0.01 & -2.39 $\pm$ 0.05 & 12.00 $\pm$ 0.03 & 11.70 $\pm$ 0.02 \\ 
        SS183044-020918$^{\dagger}$ & 277.6847 & -2.1551 & 15.29 $\pm$ 0.01 & 16.30 $\pm$ 0.01 & 14.41 $\pm$ 0.01 & -0.62 $\pm$ 0.01 & 12.28 $\pm$ 0.05 & 11.42 $\pm$ 0.04 \\ 
        SS183047-020133$^{\dagger}$ & 277.6943 & -2.0258 & 15.47 $\pm$ 0.01 & 15.88 $\pm$ 0.01 & 14.89 $\pm$ 0.01 & -0.66 $\pm$ 0.01 & 13.73 $\pm$ 0.11 & 15.15 $\pm$ 0.38 \\ 
         \\
         
        \multicolumn{5}{l}{{\it Other Serpens South photometric candidates:}} \\
    
        SS182912-021300 & 277.3011 & -2.2167 & 18.40 $\pm$ 0.03 & 19.78 $\pm$ 0.05 & 17.17 $\pm$ 0.04 & -0.90 $\pm$ 0.12 & - & - \\
        SS182917-020923 & 277.3217 & -2.1564 & 18.58 $\pm$ 0.01 & 19.66 $\pm$ 0.02 & 17.50 $\pm$ 0.01 & -0.92 $\pm$ 0.04 & - & - \\
        SS182938-015935 & 277.4112 & -1.9930 & 18.47 $\pm$ 0.00 & 19.44 $\pm$ 0.01 & 17.54 $\pm$ 0.01 & -0.75 $\pm$ 0.02 & - & - \\
        SS182942-015935 & 277.4275 & -1.9931 & 18.44 $\pm$ 0.00 & 19.11 $\pm$ 0.01 & 17.54 $\pm$ 0.01 & -1.00 $\pm$ 0.02 & - & -  \\
        SS182949-020558 & 277.4557 & -2.0995 & 18.13 $\pm$ 0.00 & 19.12 $\pm$ 0.01 & 17.25 $\pm$ 0.01 & -0.64 $\pm$ 0.02 & 13.88 $\pm$ 0.03 & 13.48 $\pm$ 0.05 \\
        SS182957-015409 & 277.4891 & -1.9027 & 19.02 $\pm$ 0.01 & 21.42 $\pm$ 0.07 & 17.35 $\pm$ 0.01 & -0.69 $\pm$ 0.08 & 13.96 $\pm$ 0.03 & 13.32$ \pm$ 0.03 \\
        SS182959-020917 & 277.4978 & -2.1547 & 19.28 $\pm$ 0.01 & 22.09 $\pm$ 0.11 & 17.28 $\pm$ 0.01 & -0.89 $\pm$ 0.12 & 13.47 $\pm$ 0.03 & 12.37 $\pm$ 0.03 \\
        SS183016-015728 & 277.5683 & -1.9578 & 19.15 $\pm$ 0.01 & 21.24 $\pm$ 0.04 & 17.62 $\pm$ 0.01 & -0.74 $\pm$ 0.05 & 15.36 $\pm$ 0.06 & 14.41 $\pm$ 0.07 \\
        SS183019-020130 & 277.5824 & -2.0251 & 18.64 $\pm$ 0.01 & 19.71 $\pm$ 0.01 & 17.56 $\pm$ 0.01 & -0.93 $\pm$ 0.04 & - & - \\
        SS183022-015315 & 277.5929 & -1.8875 & 19.22 $\pm$ 0.01 & 22.21 $\pm$ 0.10 & 17.12 $\pm$ 0.01 & -0.89 $\pm$ 0.11 & 13.37 $\pm$ 0.04 & 12.41 $\pm$ 0.03 \\
        SS183042-021334 & 277.6751 & -2.2263 & 18.05 $\pm$ 0.02 & 19.38 $\pm$ 0.05 & 16.95 $\pm$ 0.01 & -0.71 $\pm$ 0.08 & - & - \\
        SS183046-021202 & 277.6947 & -2.2007 & 18.54 $\pm$ 0.03 & 20.84 $\pm$ 0.10 & 16.89 $\pm$ 0.01 & -0.75 $\pm$ 0.13 & - & - \\
        SS183046-015651 & 277.6954 & -1.9476 & 18.60 $\pm$ 0.02 & 19.94 $\pm$ 0.06 & 17.30 $\pm$ 0.01 & -1.07 $\pm$ 0.08 & 13.49 $\pm$ 0.04 & 11.74 $\pm$ 0.04 \\
        SS183047-015834 & 277.6974 & -1.9761 & 19.03 $\pm$ 0.03 & 20.26 $\pm$ 0.08 & 17.79 $\pm$ 0.02 & -1.06 $\pm$ 0.12 & - & - \\
        
        \\
        \\
        \multicolumn{5}{l}{{\it Serpens Core candidates with spectroscopic follow-up:}}\\
        
        SC182952+011618 & 277.4679 & 1.2717 & 16.25 $\pm$ 0.01 & 17.26 $\pm$ 0.01 & 15.11 $\pm$ 0.01 & -1.10 $\pm$ 0.02 & 13.13 $\pm$ 0.03 & 12.56 $\pm$ 0.03\\
        \\
        
        \multicolumn{5}{l}{{\it Other Serpens Core photometric candidates: }} \\
        
        SC182955+011034$^{\star}$ & 277.4806 & 1.1761 & 13.86 $\pm$ 0.00 & 14.01 $\pm$ 0.00 & 13.30 $\pm$ 0.01 & -0.89 $\pm$ 0.02 & 12.62	$\pm$ 0.03 & 12.44 $\pm$ 0.03\\
        SC182956+010940 & 277.4837 & 1.1612 & 14.89 $\pm$ 0.00 & 15.70 $\pm$ 0.00 & 14.07 $\pm$ 0.01 & -0.71 $\pm$ 0.02 & 12.72	$\pm$ 0.03 & 12.36 $\pm$ 0.02\\
        SC182956+011218$^{\star}$ & 277.4848 & 1.2050 & 14.91 $\pm$ 0.00 & 15.28 $\pm$ 0.00 & 14.27 $\pm$ 0.01 & -0.81 $\pm$ 0.02 & 13.11	$\pm$ 0.03 & 12.62 $\pm$ 0.03\\
        SC183005+011235$^{\star}$ & 277.5209 & 1.2099 & 15.39 $\pm$ 0.00 & 16.02 $\pm$ 0.00 & 14.60 $\pm$ 0.01 & -0.83 $\pm$ 0.02 & 12.74	$\pm$ 0.05 & 12.14 $\pm$ 0.05\\
        SC183008+010830 & 277.5364 & 1.1418 & 14.85 $\pm$ 0.00 & 15.32 $\pm$ 0.00 & 13.99 $\pm$ 0.01 & -1.12 $\pm$ 0.02 & 12.85	$\pm$ 0.03 & 12.43 $\pm$ 0.03\\
        SC183016+013307 & 277.5708 & 1.2187 & 15.69 $\pm$ 0.00 & 16.27 $\pm$ 0.01 & 14.97 $\pm$ 0.01 & -0.75 $\pm$ 0.02 & 13.41	$\pm$ 0.03 & 13.03 $\pm$ 0.03\\
        \hline
        \multicolumn{3}{l}{\footnotesize$^{\dagger}$ Objects first reported in \citet{jose20}.} \\
        \multicolumn{3}{l}{\footnotesize$^{\star}$ Spectroscopically characterised in \citet{winston10}.}
    \end{tabular}
    \caption{ Magnitudes in 1.45$\mu$m, $J$ and $H$ from CHFT for Serpens South and Serpens Core objects that meet the criteria from Sections \ref{sec:phot_ss} \& \ref{sec:phot_sc}. $Q$-value calculated from these, as detailed in Section \ref{sec:q}. Also given are ALLWISE magnitudes for detected sources \citep{cutri13}.}
    \label{tab:phot_cfht}
\end{table*}

\subsection{Spectroscopic Follow-up} \label{sec:spec}

\begin{figure}
    \centering
    \includegraphics[width=\columnwidth]{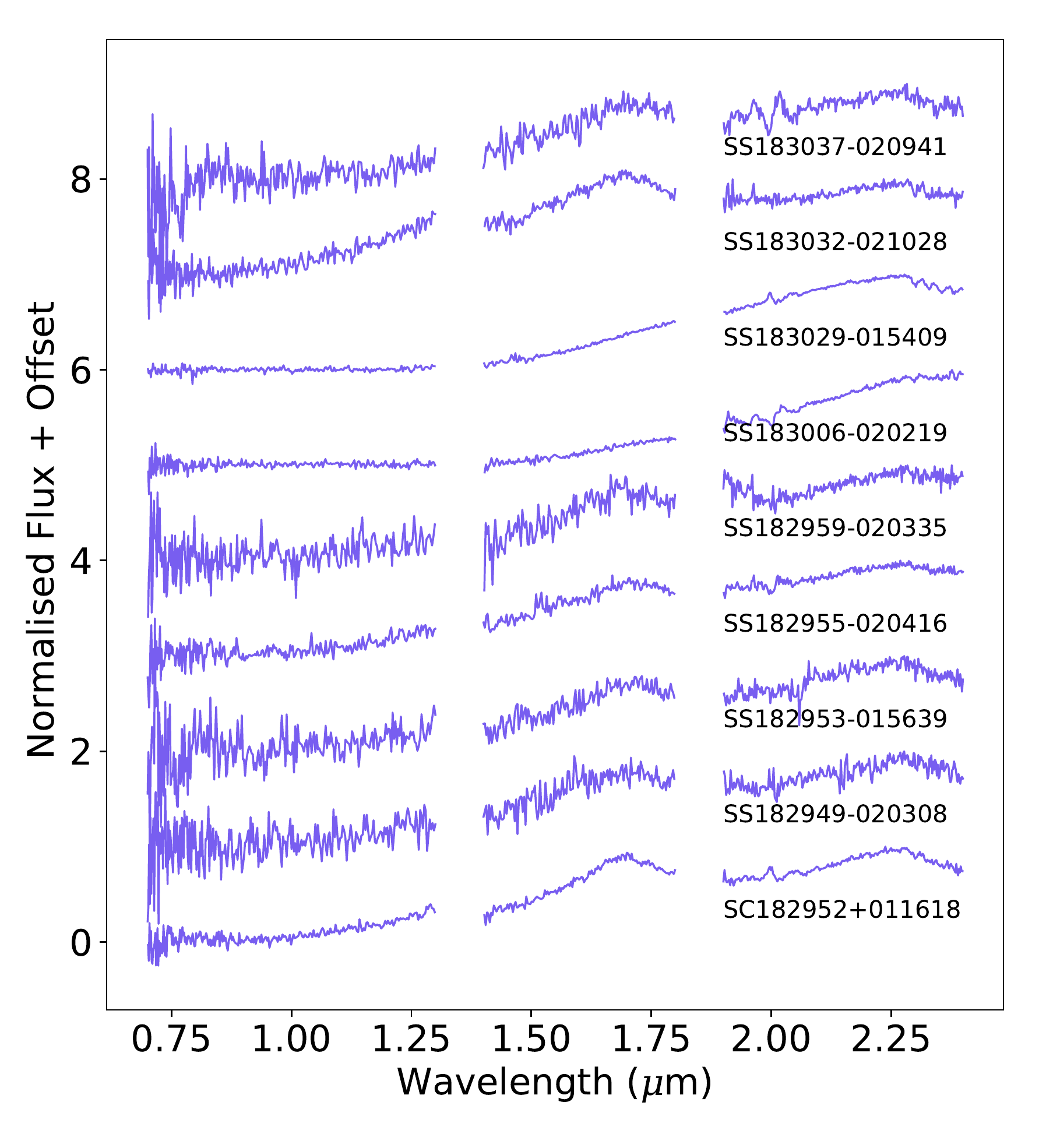}
    \caption{IRTF SpeX spectra of Serpens South and Serpens Core objects ($J-$, $H-$ and $K-$bands) reported in this paper (before dereddening), normalised to the $K$-band peak.}
    \label{fig:all_spec}
\end{figure}

We obtained spectra of each of the objects identified in Section \ref{sec:selection}. Spectroscopy can confirm water absorption, and be used to characterise the spectral type and age of a target. The full sample of 9 spectra obtained is shown in Figure \ref{fig:all_spec}. Each is normalised to the peak of the $K-$band flux.

We observed 8 targets in Serpens South and 1 target in Serpens Core using the SpeX spectrograph on the IRTF \citep{rayner03}, on 5 nights between July 2017 and June 2018. The coordinates and object ID for each observation are given in Table \ref{tab:phot_hst}. For each observation, a standard ABBA nodding pattern was used to obtain sky and target spectra, with total integration times ranging from 1430 - 3820s. The observations were undertaken in prism mode using either the 0.5" or 0.8" slit, depending on the observing conditions, corresponding to an average resolving power R $\sim$ 100. When observing the objects, flat field frames were taken before each set of observations. Wavelength calibration was achieved using Argon lines, determined by observing an argon lamp, again prior to each set of observations. We used {\tt Spextool} \citep[V4.1][]{cushing04}, an IDL-based data reduction package, to extract the spectra of the target object from the data, and to combine the individual frames for each target into one spectrum. Bad pixels and other variable effects were pruned from the spectra, which were then scaled to a common median flux level. After visual inspection, the spectra were median combined into a single file. 
Additionally, an A0 standard star was observed within an hour either before or after each science observation. We used one of four standard stars for each of the target observations: either HD174240, HD172792, HD182299 or HD167163. The median combined spectra of the standard star was then used for telluric correction of the science targets. By comparing this modelled spectra to the A0 V spectra and modifying it for any observed differences, the intrinsic and observed spectra of the standard can be used to determine the function necessary for correcting the target spectra \citep{vacca03}.

\subsection{High Resolution Follow-up Observations} \label{sec:hst}

We obtained HST WFC3 optical and near-IR imaging (GO 15628, PI Biller) for 6 objects identified as candidate members of Serpens South \citep[three from this work, three from][]{jose20} in order to search for low-mass companions to these objects. These are identified in Table \ref{tab:phot_hst} where we present their HST fluxes and $Q_{\text{HST}}$-values. Through spectral fitting analysis (see Section \ref{sec:spectypes}), we have confirmed 5 of these as bonafide late-type candidate Serpens members, and classified one as a late-M field-age contaminant.

We observed each object using the IR detector with the F127M and F139M filters, and using the UVIS detector with the F850LP filter. IR imaging was obtained using a 3-point dither pattern with exposure times of 197s. UVIS imaging was obtained using a 3-point dither pattern with 275s exposures, as well as the 2K2C sub-aperture and a 10s flash to correct for charge transfer inefficiency. 
The combination of F127M and F139M is analogous to the W-band technique - we expect to see objects showing water-absorption to be visible in F127M and drop out significantly in F139M. The reddening insensitive index, $Q_{\text{HST}}$, is also used here, calculated specifically for this combination of HST filters using Eq. \ref{eq:q_hst}. As discussed in Section \ref{sec:q_hst}, objects with late-M, L \& T spectral types will have $Q_{\text{HST}} > 1$, while background stars will have $Q_{\text{HST}}$ $\approx$ 0.  
The magnitudes, fluxes and $Q_{\text{HST}}$ values for each object are given in Table \ref{tab:phot_hst} ($Q_{\text{HST}}$ is calculated using the fluxes presented). Two targets were not detected in F850LP, and as a result no $Q_{\text{HST}}$ value is reported.

\begin{landscape}
\begin{table}
    \renewcommand{\arraystretch}{1.3}
    \centering
    \begin{tabular}{lccccccccccr}
        \hline
        Object ID & RA & Dec & F850LP & F127M & F139M & $F_{\text{F850LP}}$ & $F_{\text{F127M}}$ & $F_{\text{F139M}}$ & $Q_{\text{HST}}$ & SpT & Conf.?  \\
        & deg & deg & mag & mag & mag & erg cm$^{-2}$ s$^{-1}$ $\text{\normalfont\AA}^{-1}$ & erg cm$^{-2}$ s$^{-1}$ $\text{\normalfont\AA}^{-1}$ & erg cm$^{-2}$ s$^{-1}$ $\text{\normalfont\AA}^{-1}$ & & & \\
        \hline
        \multicolumn{3}{l}{{\it Serpens South (this work):}} \\
        SS182949-020308 & 277.4540 & -2.0522 & - & 18.332$^{+0.017}_{-0.009}$ & 17.864$^{+0.016}_{-0.008}$ & - & (1.243 $\pm$ 0.010)$\times10^{-17}$ & (1.458 $\pm$ 0.011)$\times10^{-17}$ & - & Contaminant (late-M) & N \\ 
        SS182953-015639 & 277.4696 & -1.9442 & - & 17.411$^{+0.010}_{-0.005}$ & 17.116$^{+0.010}_{-0.005}$ & - & (2.224 $\pm$ 0.012)$\times10^{-17}$ & (2.904 $\pm$ 0.013)$\times10^{-17}$& - & M7-LO & Y \\
        SS182955-020416 & 277.4807 & -2.0712 & - & - & - & - & - & - & - & M4-M6.5 & Y \\
        SS182959-020335 & 277.4995 & -2.0598 & - & - & - & - & - & - & - & M5-LO & Y \\
        SS183006-020219 & 277.5258 & -2.0387 & - & - & - & - & - & - & - & Contaminant (early-M) & N \\
        SS183029-015409 & 277.6235 & -1.9026 & - & - & - & - & - & - & - & Contaminant (early-M) & N \\
        SS183032-021028 & 277.6365 & -2.1745 & 20.374$^{+0.058}_{-0.120}$ & 17.495$^{+0.012}_{-0.009}$ & 17.180$^{+0.012}_{-0.006}$ & (5.442 $\pm$ 0.300)$\times10^{-18}$ & (2.686 $\pm$ 0.015)$\times10^{-17}$ & (2.737 $\pm$ 0.015)$\times10^{-17}$ & 1.619 & M5-M6.5 & Y \\
        SS183037-020941 & 277.6566 & -2.1616 & & & & - & - & - & - & Contaminant (early-M)  & N \\
        \\
        \multicolumn{3}{l}{{\it Serpens Core (this work):}}\\
        SC182952+011618 & 277.4679 & +1.2717 & - & - & - & - & - & - & - &M7-M9 & Y \\
        \\
        \multicolumn{3}{l}{{\it Serpens South \citep{jose20}:}} \\
        SS182917-020340 & 277.3193 & -2.0610 & 22.040$^{+0.208}_{-0.100}$ & 18.200$^{+0.017}_{-0.009}$ & 17.802$^{+0.016}_{-0.008}$ &(1.169 $\pm$ 0.111)$\times10^{-18}$ & (1.392 $\pm$ 0.011)$\times10^{-17}$ & (1.544 $\pm$ 0.011)$\times10^{-17}$ & 2.035 & M4-M7 & Y \\
        SS182918-020245 & 277.3256 & -2.0458 & 19.645$^{+0.028}_{-0.014}$ & 16.317$^{+0.006}_{-0.003}$ & 16.901$^{+0.006}_{-0.003}$ &(1.062 $\pm$ 0.014)$\times10^{-17}$& (7.949 $\pm$ 0.022)$\times10^{-17}$ & (7.464 $\pm$ 0.021)$\times10^{-17}$ & 2.583 & M5-M8 & Y \\
        SS183037-021411 & 277.6545 & -2.2363 & - & - & - & - & - & - & - & Contaminant (early-M) & N \\
        SS183038-021419 & 277.6568 & -2.2386 & - & - & - & - & - & - & - & M3-M6 & Y \\
        SS183038-021437 & 277.6570 & -2.2437 & - & - & - & - & - & - & - & Contaminant (early-M) & N \\
        SS183044-020918 & 277.6847 & -2.1551 & 20.184$^{+0.045}_{-0.022}$ & 16.132$^{+0.005}_{-0.003}$ & 15.768$^{+0.005}_{-0.003}$ &(6.463 $\pm$ 0.133)$\times10^{-18}$ & (9.429 $\pm$ 0.024)$\times10^{-18}$ & (1.005 $\pm$ 0.002)$\times10^{-16}$ & 2.508 & M7-M9 & Y \\
        SS183047-020133 & 277.6943 & -2.0258 & - & - & - & - & - & - & - & Contaminant (early-M) & N \\

        \hline
        
    \end{tabular}
    \caption{ W-band Serpens survey objects that were followed-up spectroscopically, and subsequently characterised. HST filter magnitudes and fluxes are given for objects that were imaged and detected, as well as the HST $Q$-value. We report a spectral type, if one was determined from spectral fitting, and highlight the objects that were confirmed as late-type brown dwarfs in this work and in \citet{jose20} (Y in 'Conf.?' column).}
    \label{tab:phot_hst}
\end{table}
\end{landscape}

\section{Effects of Extinction} \label{sec:extinc}

\begin{figure*}
    \centering
    \includegraphics[width=\textwidth]{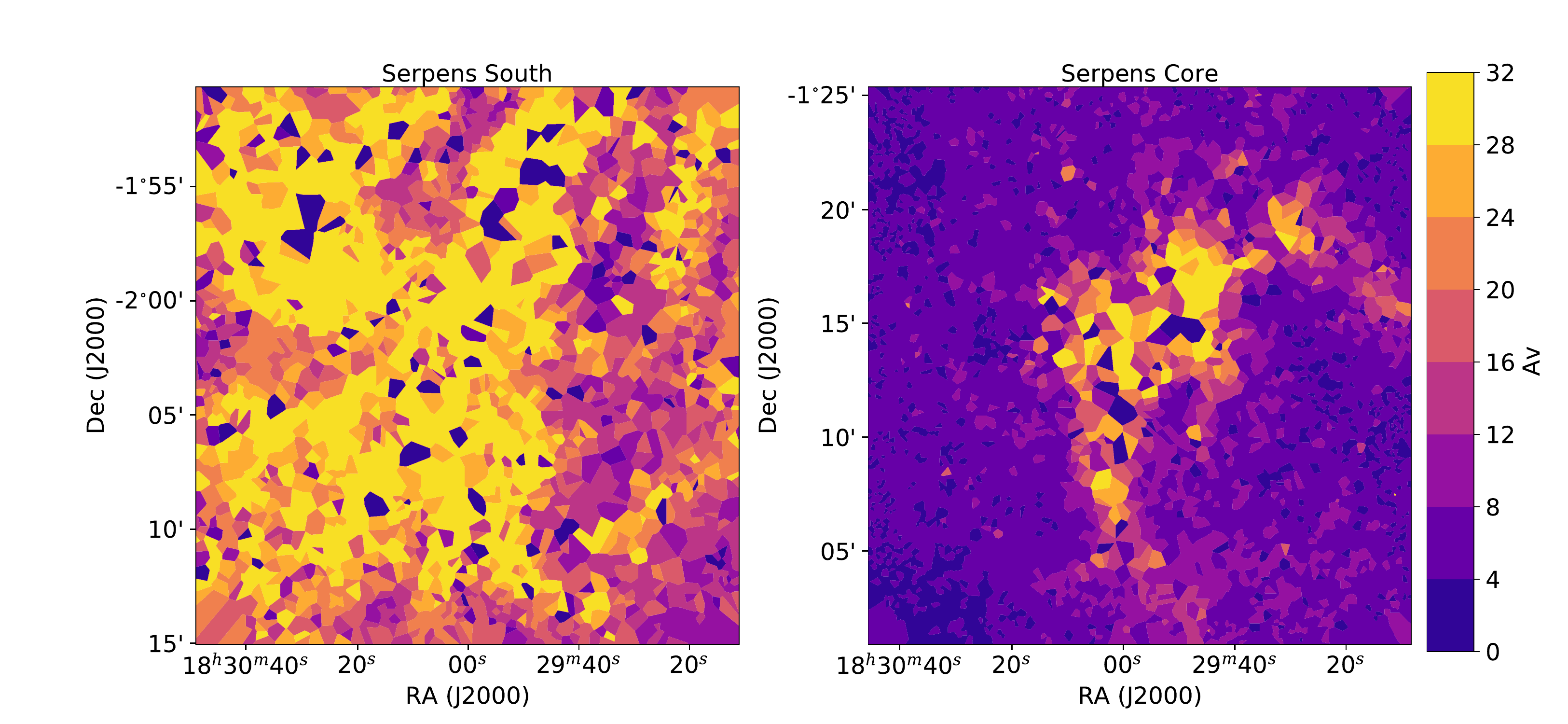}
    \caption{Visual extinction ($A_{\text{V}}$) maps for Serpens South and Serpens Core, generated using the photometric catalogues from CFHT WIRCam. Nearest neighbour interpolation is used to fill the parameter space between coordinates with observed values of $A_{\text{V}}$.}
    \label{fig:av_map}
\end{figure*}

The visual extinction ( $A_{\text{V}}$) along the Serpens line of sight is high. We used our CFHT photometric catalogues of Serpens Core and Serpens South to produce extinction maps of both regions, shown in Figure \ref{fig:av_map}. An SED fit is performed for each detected object in the field, using photometry from the three observed bands and any available in literature catalogues (e.g 2MASS, PanSTARRS), with visual extinction as an additional fitting parameter.  The best fit $A_{\text{V}}$ is stored for every object. To create Figure \ref{fig:av_map}, we used nearest-neighbour interpolation\footnote{using the {\tt scipy NearestNDInterpolator} function.} to find an estimate of the spatial distribution of the dust attenuation across both full fields. 
We found an average value of $A_{\text{V,mean}} = 21.3$ mag, and a range of $A_{\text{V}} = 0-31$ mag (the maximum value of extinction considered) for Serpens South. As can be seen from the map in Figure \ref{fig:av_map}, high extinction dominates, but localised, lower-extinction regions are distributed uniformly across the area of sky that we imaged. We also found an average value of $A_{\text{V,mean}} = 6.3$ mag and a range of $A_{\text{V}} = 0-31$ mag for Serpens Core. However, here the spatial distribution of high and low values is not uniform - extinction in the center of the imaged region (where the Core is located) is very high, and $A_{\text{V}}$ drops rapidly to lower values elsewhere. 
These values are comparable to those found in the 2MASS extinction analysis of the region, discussed in Section \ref{sec:serpens}. Both analyses show higher average and maximum values for Serpens South, and consistently high values across both subclusters.

As described in \citet{allers20}, the W-band method was originally designed and optimised for star-forming regions with far lower interstellar extinction levels (e.g Taurus). Consequently, we investigated the effect that the significantly higher extinctions of the two Serpens regions would have on the effectiveness of our technique. Starting from our photometric observations of Serpens South, we examined the location of true and false positives on the $H$ vs $Q$ diagram.

\begin{figure}
    \centering
    \includegraphics[width=\columnwidth]{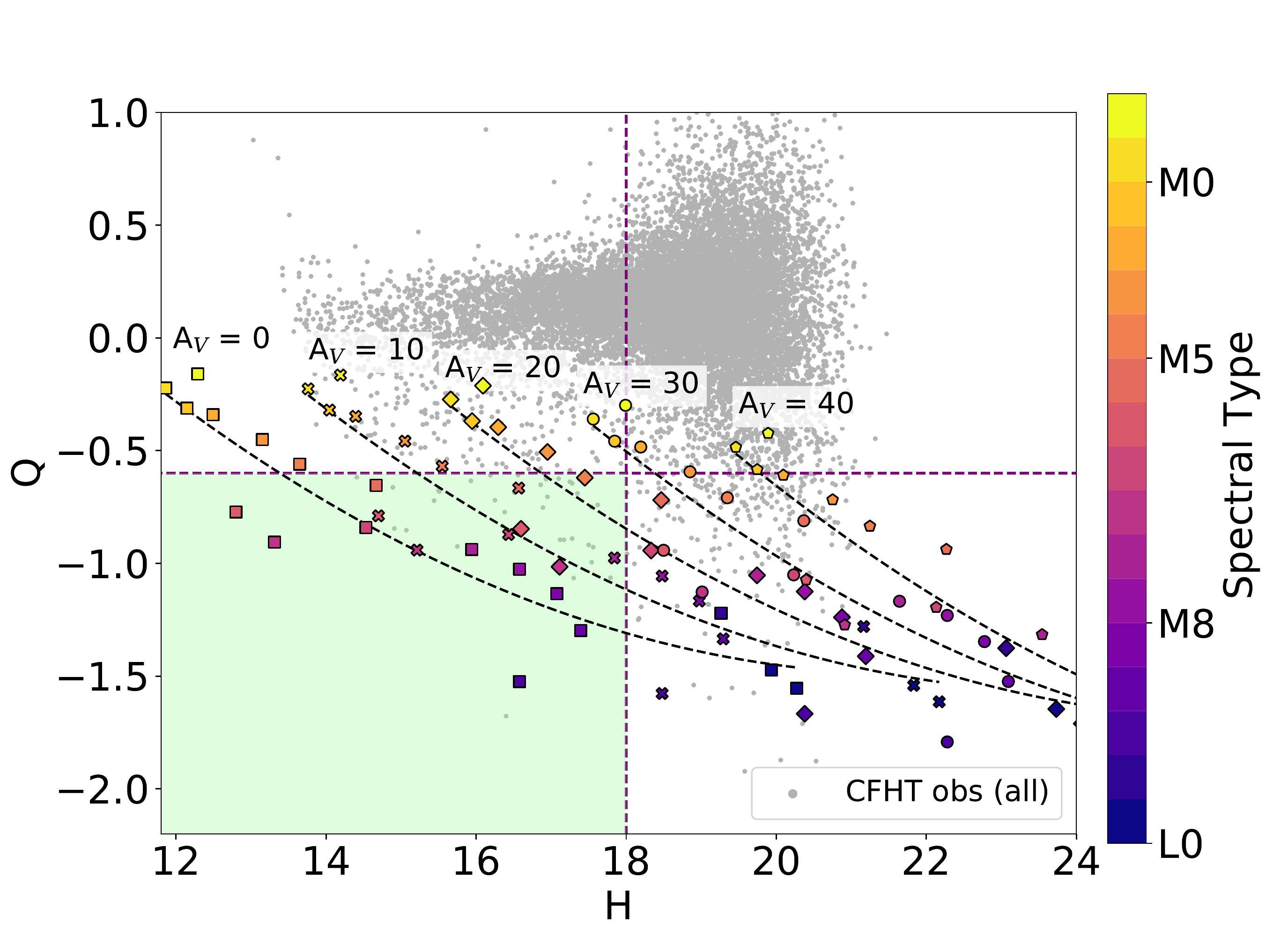}
    \caption{$H$ mag vs $Q$ for objects used to construct the spectral standards described in \citet{luhman17}. Sequences of objects with SpT = M0--L4 are reddened by varying amounts between $A_{\text{V}}$ = 0--40 mag. SpT is indicated by colour. The green box shows the region of parameter space suitable for spectroscopic follow-up with 4-m class telescopes.}
    \label{fig:q_h_l17}
\end{figure}

First, we considered the population of objects that the W-band filter was designed to find: young, late-type members of nearby star-forming regions. \citet{luhman17} (henceforth \citetalias{luhman17}) propose a set of near-IR spectral standards for young, late-type objects, which we adopted here as suitable template spectra. We calculated synthetic photometry in $J$, $H$ and $W$, as well as $Q$, for the standards with SpT = M0-L4. We repeated this calculation multiple times for each object, placing them in Serpens South each time ($d$ = 436 pc) and reddening them by $A_{\text{V}}$ = 0, 10, 20, 30 and 40 mag. These sequences are shown as the coloured markers in Figure \ref{fig:q_h_l17}. 
This analysis demonstrates that using the W-band filter to look at Serpens truly pushes the $Q$-index to its limits. When considering the highest extinction levels, we see that $Q$ can no longer be considered `reddening-insensitive' - it is significantly altered from the low-$A_{\text{V}}$ levels. However, Figure \ref{fig:q_h_l17} does show that we largely avoid early-M contaminants in the green shaded quadrant (which shows the area of parameter space suitable for spectroscopic follow-up with 4-m class telescopes in Serpens South).  When considering fainter magnitudes ($H > 18$), we find that highly-reddened early-mid Ms may have $Q < -0.6$, which will be an important consideration for further follow-up with larger telescopes. Additionally, late Ms and early Ls can be pushed into the $H > 18$ regime if highly attenuated by dust, meaning many may have been missed in this portion of follow-up.  \\

We also explored the effect of reddened field objects along the cluster line-of-sight on the success of our technique. As discussed, the $W$-band method and $Q$-index are optimised for separating late-type brown dwarfs from earlier spectral type objects. However, we cannot always distinguish between young late-type members of star-forming regions and older, field brown dwarfs, with simple photometry alone. Consequently, it is likely a lack of low surface gravity (youthful) features in follow-up spectra will indicate that some selected objects are actually field-age brown dwarfs.

\begin{figure*}
    \centering
    \includegraphics[width=\textwidth]{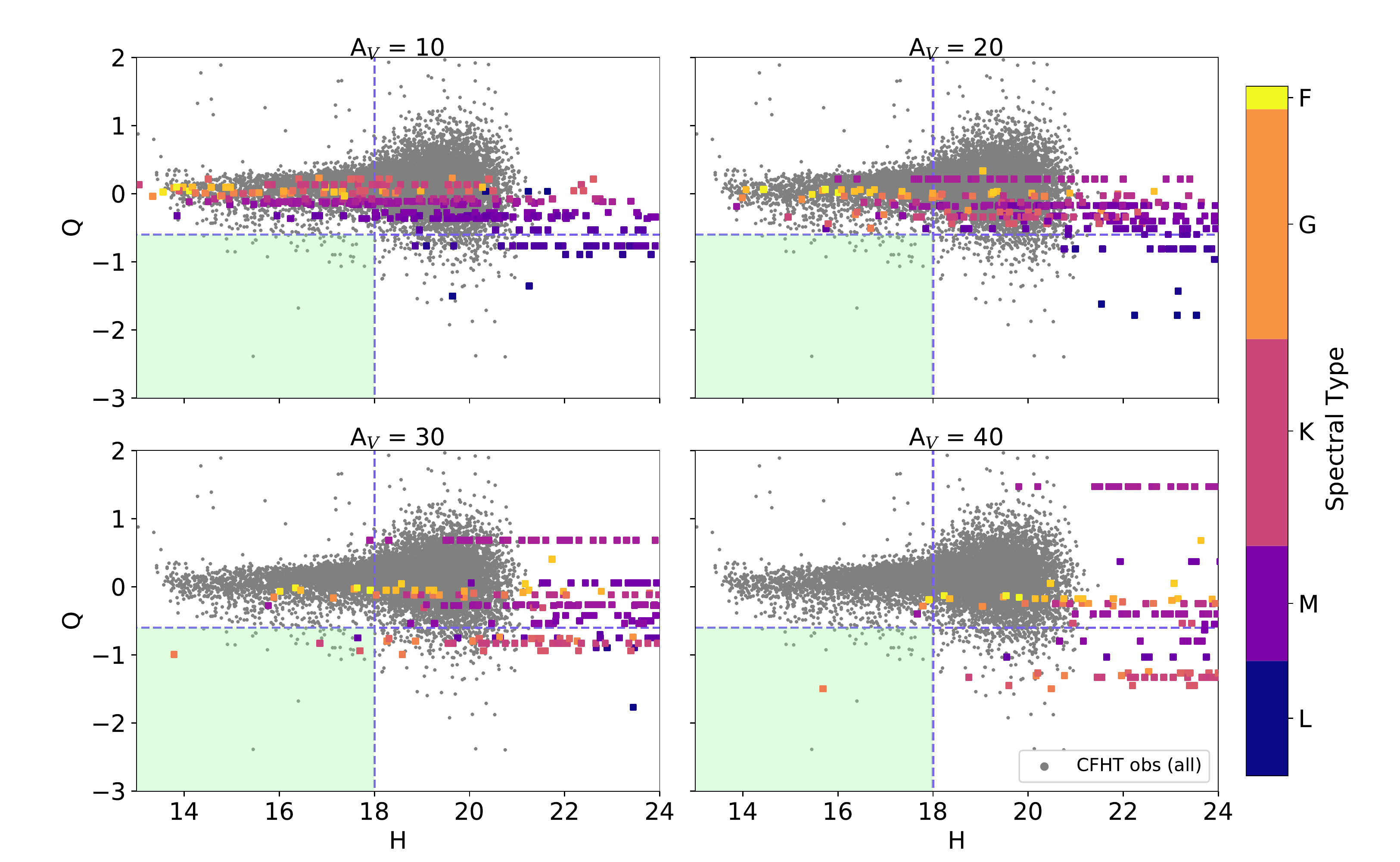}
    \caption{$H$ vs $Q$ for objects generated using the Trilegal galactic model \citep{girardi12} centred on Serpens South (RA = 277.5,Dec = -2.25). Spectral type is indicated by colour. The grey points are all objects in our Serpens South CFHT catalogue with photometric precision $< 0.1$ mag. The green boxes shows the region of parameter space suitable for spectroscopic follow-up with 4-m class telescopes.}
    \label{fig:q_h_tl}
\end{figure*}

We used the Trilegal galactic model \citep{girardi12} to obtain a population of objects along the Serpens South line-of-sight. We simulated a field of view of 0.1225 sq. deg, centred on RA = 277.5$^{\circ}$, Dec = -2.25$^{\circ}$. The CFHT MegaCam+WIRCam system was chosen for output magnitudes, and we used the Kroupa IMF \citep{kroupa01}, including binaries. This resulted in a population of 1,543 objects along the Serpens South line-of-sight.
Each object generated by the Trilegal model has an associated effective temperature. To convert these to spectral types, we used Mamajek's `Modern Mean Dwarf Stellar Color and Effective Temperature Sequence' \footnote{\hyperlink{}{https$://$www.pas.rochester.edu/$~$emamajek/EEM\_dwarf\_UBVIJHK\_colors\_Teff.txt}}
(described in part in \citet{pecaut13}).
We associated each object with the appropriate spectral type from the IRTF spectral library to obtain a template spectrum \citep{rayner09}, and then used CFHT filter information to calculate the $Q$-index and associated apparent magnitudes for each member of the population. We calculated synthetic photometry for a range of visual extinction values, $A_{\text{V}}$ = 10, 20, 30 and 40 mag, and plotted each resulting population on a $H$-mag vs $Q$ plot, as shown in Figure \ref{fig:q_h_tl}. 

Figure \ref{fig:q_h_tl} demonstrates that the majority of field objects with $H < 18$ retain $Q > -0.6$, despite extreme reddening, and as such would not be false positives in our survey. However, there are a handful of objects with $A_{\text{V}}$ = 30 mag that fall into the $Q < -0.6$, $H < 18$ region, and as such would be considered for follow-up. On closer inspection, these objects are M, K, G stars with extinction-altered colours. As a result, there is a possibility of including a small numbers of significantly earlier-type stars in a spectroscopic follow-up sample (as well as field-age brown dwarfs with M or later spectral types). Therefore, it is important to check for signs of youth when characterising our late-type discoveries spectroscopically, and determine whether they are young, candidate cluster members, or field-age background contaminants.

\section{Characterisation} \label{sec:results}

\subsection{Spectral Fitting} \label{sec:spectypes}

\begin{figure*}
    \centering
    \includegraphics[width=\textwidth]{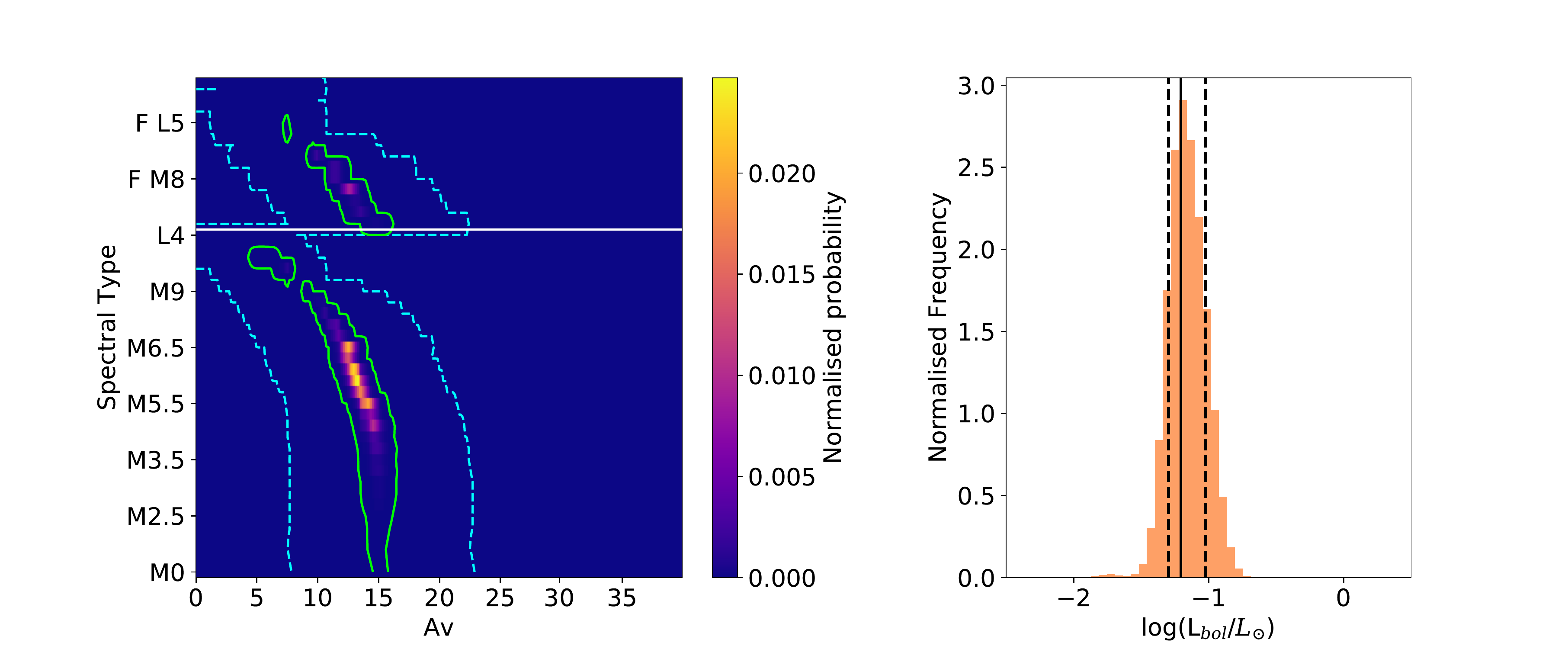}
    \caption{Left: Normalised probability map for SS183032-021028. Contours show 1$\sigma$ (blue, dashed) and 2$\sigma$ (green, solid) levels, respectively. Field and young standard solutions are separated by the white line (field denoted by F). Right: Histogram of log($L_{\text{bol}}/L_{\odot})$ solutions for SS183032-021028, derived from the probability map. Also shown is the peak value (solid line), and the 68\% (or 1$\sigma$) Bayesian credible intervals (dashed lines), which reflect the asymmetry of the distribution. }
    \label{fig:prob_map}
\end{figure*}

As discussed in Section \ref{sec:intro}, one must consider the spectral type, visual extinction and age in combination to be able to accurately characterise an object.  The procedure used for characterising targets in the W-band survey is explained in detail in \citet{jose20} and Albert et al. (in preparation).
We used a grid-based approach to compare the Serpens spectra with spectral standards. We used a library of standards compiled from two different sources: spectra of field brown dwarfs from the SpeX spectral library \citep{cushing05,rayner09}, and spectra of young brown dwarfs from \citetalias{luhman17}. Each standard belongs to one of three age groups: VL-G (very low gravity, $\approx$ 1 Myr), INT-G (intermediate gravity, $\approx$ 10 Myr) and FLD-G (field gravity, for much older field stars). These age classifications are based on the work of \citet{allers13}. For each standard we evaluated a $\chi^{2}$ 'goodness-of-fit', with the data reddened by $A_{\text{v}}$ values from 0 to 40 mag using the reddening law from \citet{fitzpatrick99}. We investigated the effect of the choice of extinction law on our best fit parameters in \citet{jose20}, and found it to have a minimal effect on our results. \\
The signal-to-noise (S/N) in the $J$-band portion of some of the spectra (1.07-1.4 $\mu$m) was not sufficient for a reliable fit, due to the high extinction of the region. As a result, some of the spectral types presented here were determined using a fit to the $H$- (1.4 - 1.8 $\mu$m) and $K$- (1.9 - 2.3 $\mu$ m) bands only.
Having obtained a $\chi^{2}$ value for each grid point of $A_{\text{v}}$ vs SpT, we plotted normalised probability maps of the parameter space, in order to evaluate the best-fit parameter combination. An example of such a map is shown in Figure \ref{fig:prob_map}, for SS183032-021028. Results for the younger and field age standards are distinguished by the horizontal white line. We used the 1$\sigma$ contour in $A_{\text{v}}$-SpT space to inform the best fit spectral type and extinction combination. $A_{\text{v}}$ vs SpT probability maps for SS182953-015639, S182955-020416, SS182959-020335 and SC182952+011618 (the other low-mass candidate Serpens members reported in this work, see below) are given in Figures \ref{fig:props_SS182953}, \ref{fig:props_SS182955}, \ref{fig:props_SS182959} $\&$ \ref{fig:props_SC182952} in Appendix \ref{sec:app_plots}.
The spectral types (if constrained) for each Serpens object are given in Table \ref{tab:phot_hst}, along with spectral types of the seven targets discussed in \citet{jose20}. Extinctions and ages for late-type discoveries are given in Table \ref{tab:phys_prop}.
Of the 9 objects in this portion of spectroscopic follow-up of Serpens South and Serpens Core, we report spectral types for five objects (discussed further in Section \ref{sec:spec_lates}). From rest of the sample reported in this paper, three objects (SS183029-015409,  SS183037-020941 and SS183006-020219) are classified as clear early-M contaminants. We also find one late-M, field age object, SS182949-020308. 
We used visual inspection of the spectra to confirm by eye whether our classifications from this analysis were reasonable. The spectral features of young and field age late M and L type objects differ. Low surface gravity (i.e. young) objects have a distinctly 'peaky' $H$-band shape \citep{lucas00,allers07,allers13}. We see this feature (to varying degrees) in each of our late-type detections. Additional indicators of low surface gravity are present in the $J-$,$H-$ and $K-$ bands, but we have insufficient S/N to use these for classification. Similarly, we do not see any features that can be used to distinguish between early M (M0-M4) objects. As a result, the targets classified as early M objects could have spectral types ranging from M0-M4.

\subsection{New Detections} \label{sec:spec_lates}

\begin{figure}
    \centering
    \includegraphics[width=\columnwidth]{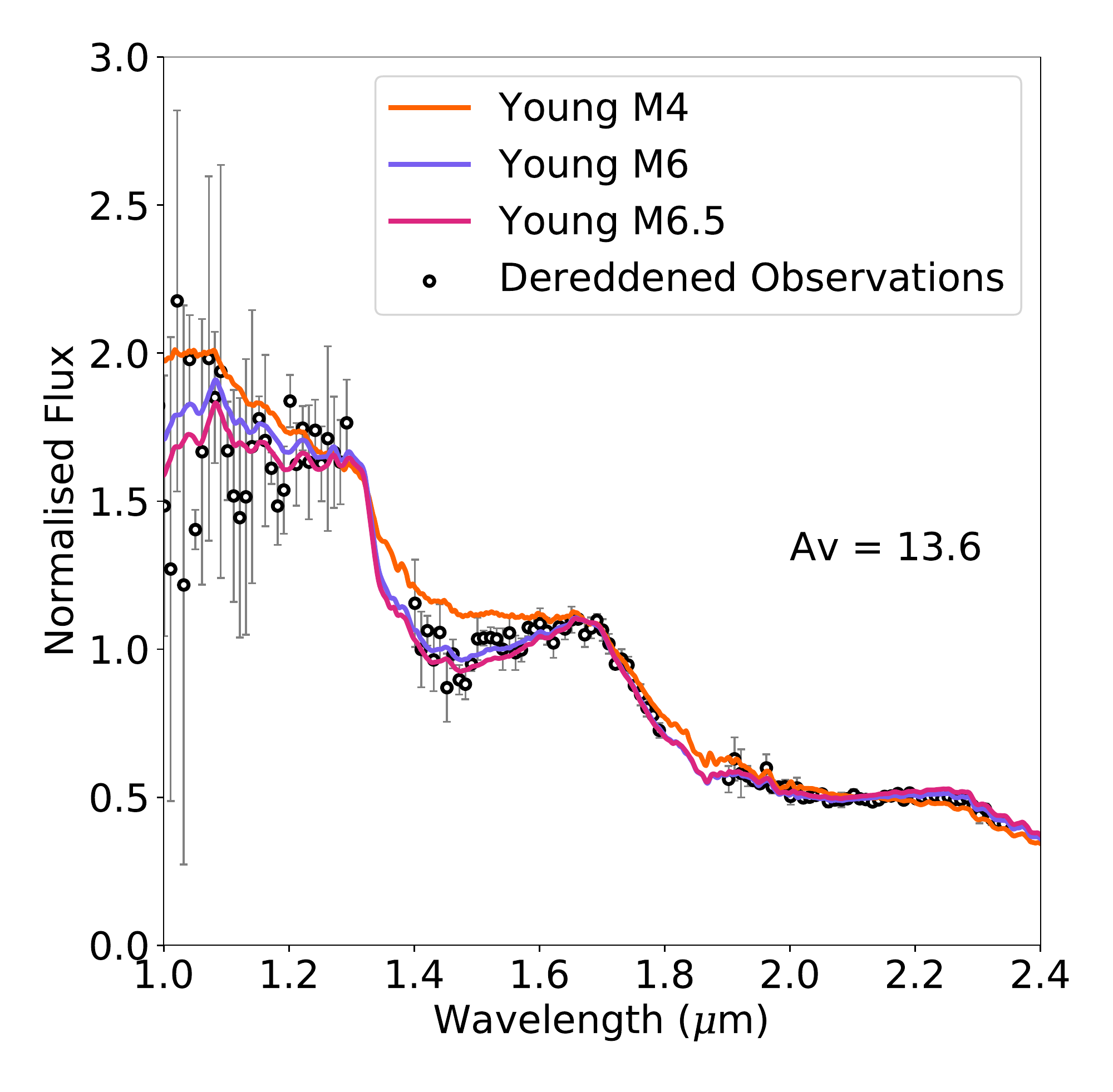}
    \caption{Spectrum of SS183032-021028, dereddened by the best fit $A_{\text{V}}$ (black open circles). Spectral data is compared to best fitting M6 spectral template from \citetalias{luhman17} (purple), and the earliest (M4, orange) and latest (M6.5, pink) templates as informed by 1$\sigma$ errors. }
    \label{fig:spec_SS183032}
\end{figure}

\begin{figure}
    \centering
    \includegraphics[width=\columnwidth]{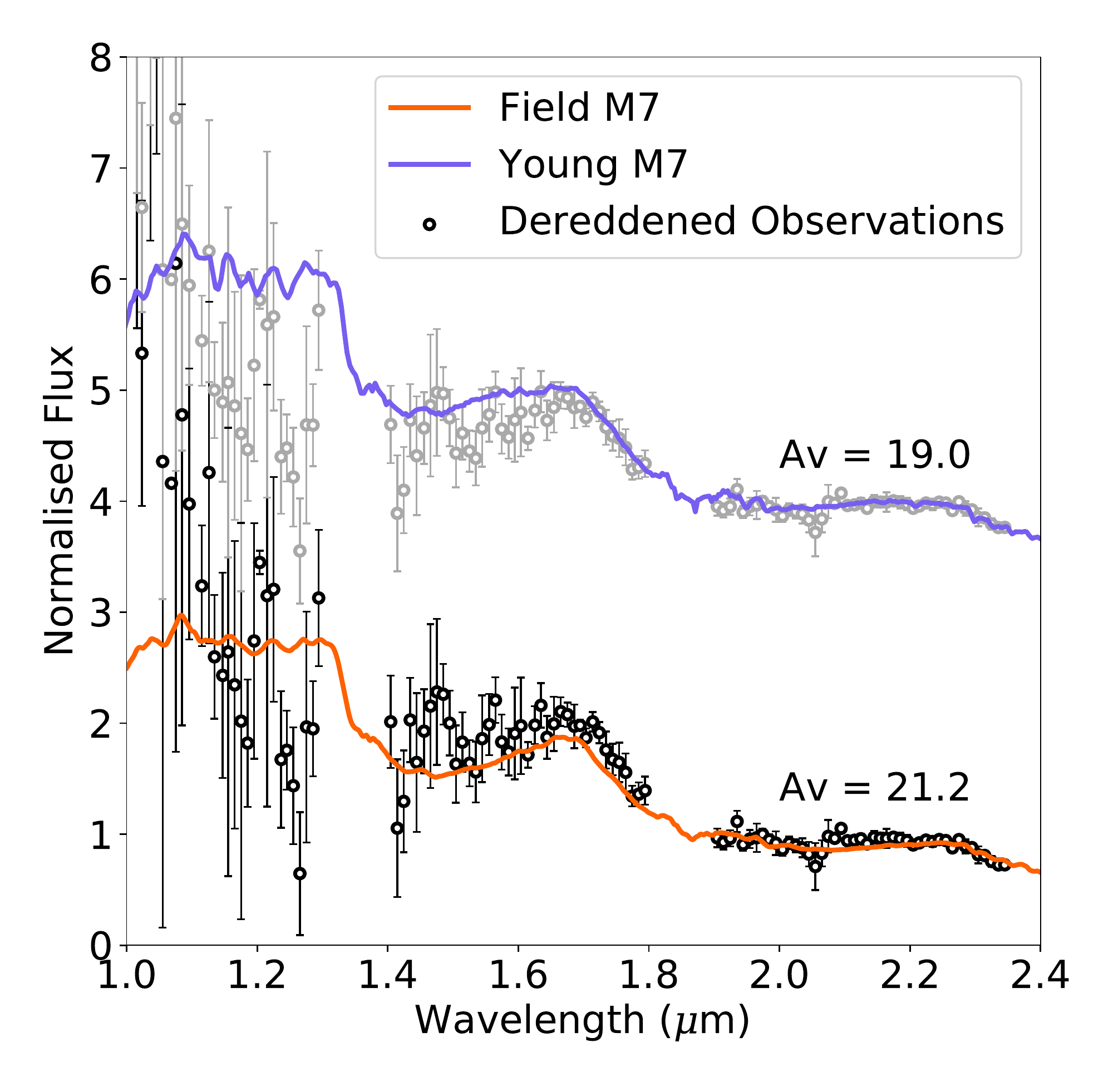}
    \caption{Spectrum of SS182953-015639, dereddened by the best fit $A_{\text{V}}$ (black open circles), compared to a young M7 template fit (upper) and a field M7 template fit (lower).}
    \label{fig:spec_SS182953}
\end{figure}

\begin{figure}
    \centering
    \includegraphics[width=\columnwidth]{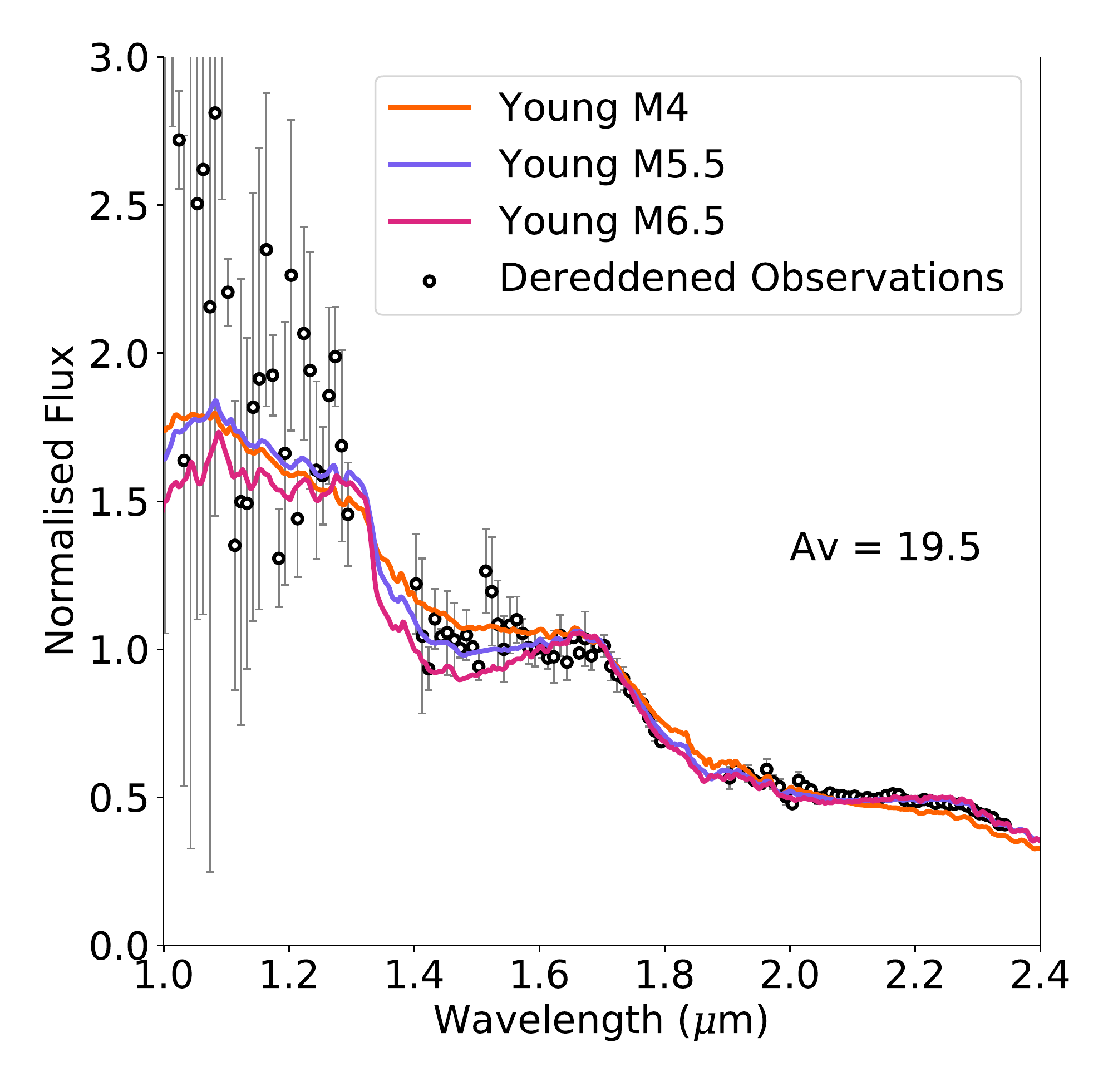}
    \caption{Spectrum of SS182955-020416, dereddened by the best fit $A_{\text{V}}$ (black open circles). Spectral data is compared to best fitting M5.5 spectral template from \citetalias{luhman17} (purple), and the earliest (M4, orange) and latest (M6.5, pink) templates as informed by 1$\sigma$ errors.}
    \label{fig:spec_SS182955}
\end{figure}

\begin{figure}
    \centering
    \includegraphics[width=\columnwidth]{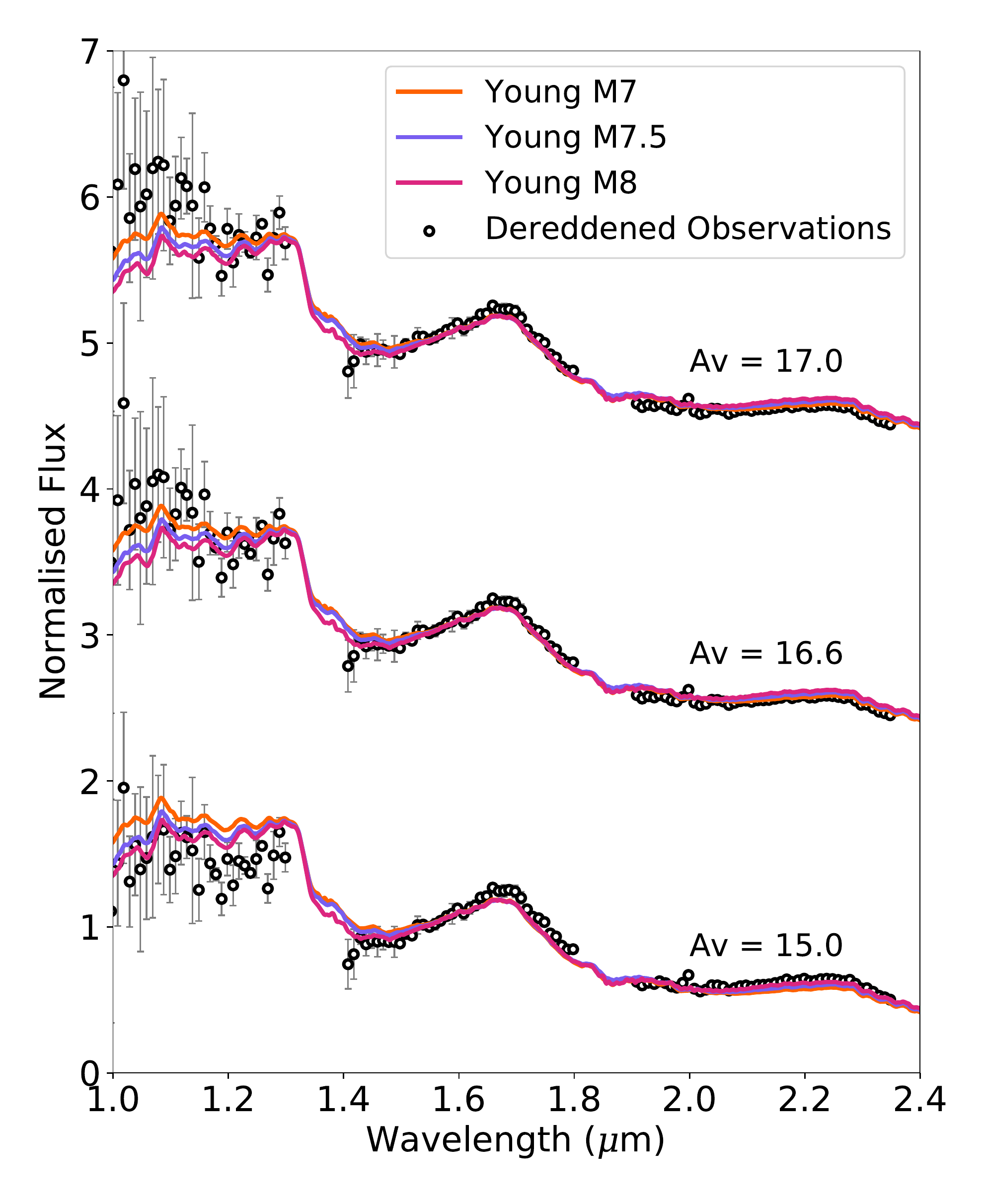}
    \caption{Spectrum of SC182952+011618 (black open circles). Middle: data dereddened by the best fitting $A_{\text{v}}$, compared to the 1$\sigma$ range of \citetalias{luhman17} spectral templates determined from the probability maps (best fit = M7.5). Upper and lower plots show the same template spectra, with the minimum and maximum bounds of $A_{\text{v}}$, compared with the dereddened data.}
    \label{fig:spec_sc182952}
\end{figure}

\begin{figure}
    \centering
    \includegraphics[width=\columnwidth]{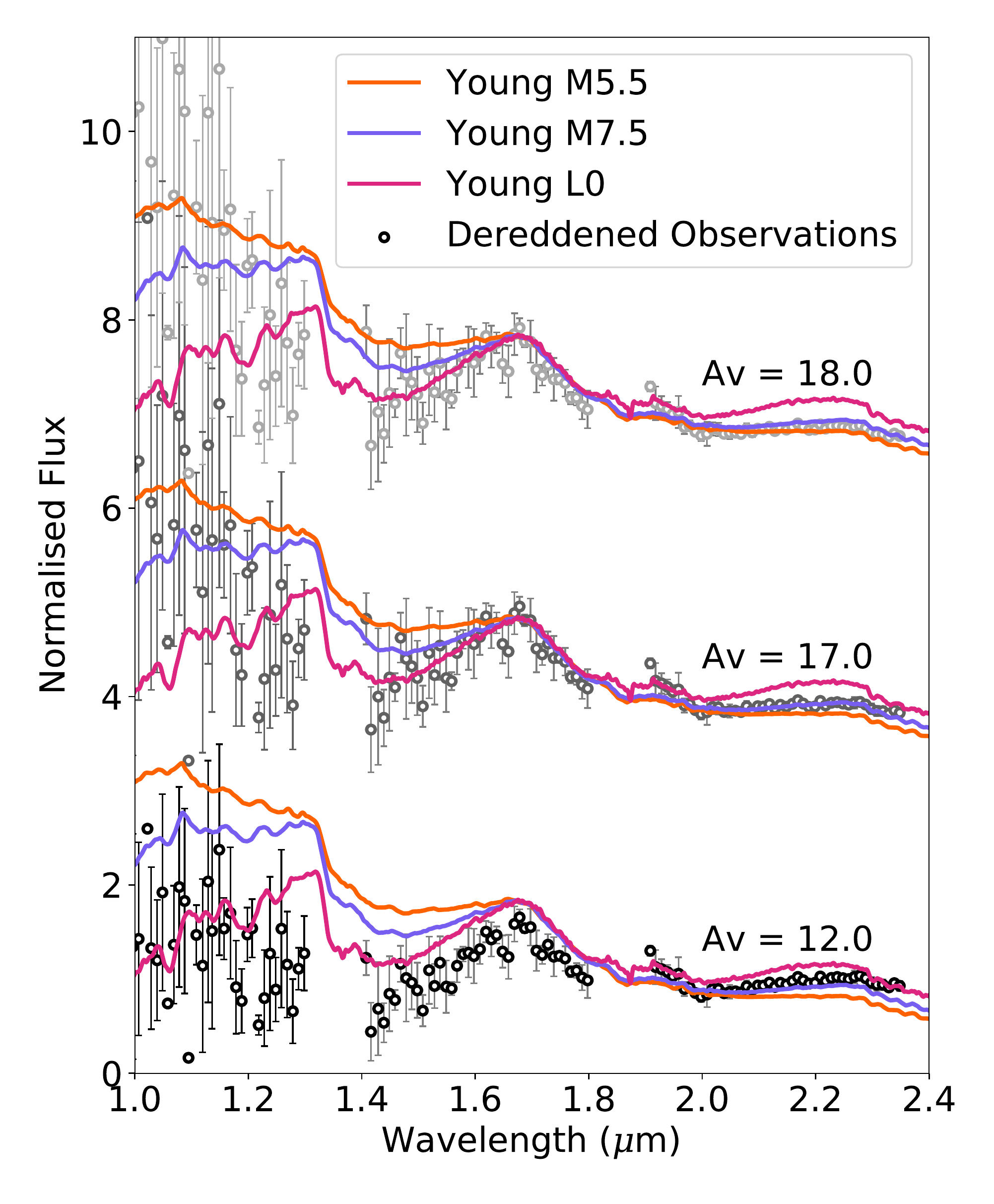}
    \caption{Spectrum of SS182959-020335 (black open circles). Middle: data dereddened by the best fitting $A_{\text{v}}$, compared to the 1$\sigma$ range of \citetalias{luhman17} spectral templates determined from the probability maps (best fit = M7.5). Upper and lower plots show the same template spectra, with the minimum and maximum bounds of $A_{\text{v}}$, compared with the dereddened data.}
    \label{fig:spec_ss182959}
\end{figure}

We find five new candidate Serpens members: four in Serpens South and one in Serpens Core. SC182952+011626 and SS182959-020335 are found to have best fitting spectral types of M8 and M7.5 respectively, and are discussed further below. SS183032-021028 (spectrum shown in Figure \ref{fig:spec_SS183032}) is found to be best fit by a young, M6 standard, placing it at the boundary of sub-stellar objects in Serpens South. SS182953-015639 (spectrum shown in Figure \ref{fig:spec_SS182953}) is best fit by an M7 standard spectrum, but we see clear bimodality between young and field age templates. This can be seen in the normalised probability map shown in Figure \ref{fig:props_SS182953}. We are confident in the late spectral type of this object, but cannot report a confident age as the spectral fitting results do not strongly favour one population. Constraining the age of this object, and thus whether it is a likely Serpens South member, will require additional spectroscopy to improve the S/N of possible youthful features.
The final likely young Serpens South candidate member, SS182955-020416, has a best fit spectral type of M5, in the range M4.5-M6, likely above the boundary of sub-stellar objects in Serpens South (spectrum shown in Figure \ref{fig:spec_SS182955}). We classify this target as a young object (despite bimodality in age from spectral fitting, it appears in multiple YSO catalogues - see Section \ref{sec:yso}), and as a result report this target as a Serpens South candidate member.

\subsubsection{SC182952+011618}

SC182952+011618 (Serpens Core) has a best fit spectral type of M7.5, in the range M7-M8, with a best fit $A_{\text{v}}$ of 16.6 mag, in the range of 15-17 mag. We quote a range in $A_{\text{v}}$ as we found that our errors were often asymmetric. The spectral type was determined by fitting standard templates to the full spectrum ($J$-,$H$- and $K$- bands).
The spectrum of SC182952+011618 is shown in Figure \ref{fig:spec_sc182952}: the peaky $H$-band feature is clear, as we would expect to see for a young, late-type object. In this figure, we also demonstrate the spread in spectral types and extinctions that fit well to this data, informed by the parameter ranges given in Table \ref{tab:phys_prop}.

\subsubsection{SS182959-020335}

SS182959-020335 (Serpens South) has a best fit spectral type of M7.5, in the range M5-L0, and best fit $A_{\text{v}}$ of 20.2 mag, with a range of 12-18 mag. Again, we can see that the best-fit young templates in Figure \ref{fig:spec_ss182959} reproduce the shape of the $H$-band well. However, the clear spread in the $H$-band data demonstrates the need for a wider range of possible spectral type fits for this object, compared to SC182952+011618. Additionally, the spectral type for this object was determined only using the $H$- and $K$- bands, as the $J$-band suffers from significantly lower S/N. Constraining the spectral type range further will require additional observation of the $J$- and $H$- bands.

\begin{table*}
    \renewcommand{\arraystretch}{1.3}
    \centering
    \caption{Physical properties of the new candidate Serpens members discovered in the W-band Serpens surveys. Adopted spectral types and extinctions are given, as well as the population according to the near-IR gravity classification in \citet{allers13} (VLG = very low gravity, INTG-G = intermediate gravity). Objects that appear in YSO catalogues are indicated: G = appears as YSO in \citet{getman17}, D = appears as YSO in \citet{dunham15}, P = appears as YSO in \citet{povich13}, G-NOD = listed as `No disk' in \citet{getman17}. The bolometric correction needed to calculate log($L_{\text{bol}}/L_{\odot})$ is taken from either \citet{filippazzo15} or \citet{herczeg15}, depending on the spectral type of the object. The target number given shows the numbering used in Figure \ref{fig:hr}. Mass estimates shown are approximate values derived from the positions of targets in Figure \ref{fig:hr}. }
    \begin{tabular}{lccccccccccr}
        \hline
        Target No. & Object ID & RA & Dec & SpT & A$_{\text{V,best}}$ & A$_{\text{V,range}}$ & Age & log($L_{\text{bol}}/L_{\odot})$ & $T_{\text{eff}}$ & Mass & YSO? \\
        & & (deg) & (deg) & & & & & & (K) & (M$_{\odot}$) &  \\
        \hline
        \multicolumn{3}{l}{{\it Serpens South (this work):}} \\
        1 & SS182953-015639 & 277.4696 & -1.9442 & M6--L0 & - & 10--23 & - & -0.91$^{+0.47}_{-0.30}$ & 2720 $\pm$ 200 & 0.07--0.1 &... \\ 
        2 & SS182955-020416 & 277.4807 & -2.0712 & M4--M6.5 & 19.5 & 17--21 & INT-G & -0.88$^{+0.13}_{-0.19}$ & 2920 $\pm$ 165 & 0.1--0.15 & G,D,P \\ 
        3 & SS182959-020335 & 277.4807 & -2.0712 & M5--LO & 17.0 & 12--18 & VL-G & -1.47$^{+0.52}_{-0.31}$ & 2770 $\pm$ 200 & 0.05--0.09 & G,P  \\
        4 & SS183032-021028 & 277.6365 & -2.1745 & M5--M6.5 & 13.6 & 12--15 & INT-G &  -1.21$^{+0.18}_{-0.09}$ & 2860 $\pm$ 120 & 0.07--0.1 & ... \\
        \\
        \multicolumn{3}{l}{{\it Serpens Core (this work):}} \\
        5 & SC182952+011618 & 277.4679 & +1.2717 & M7--M8 & 16.6 & 15--17 & VL-G & -0.61$^{+0.08}_{-0.17}$ & 2720 $\pm$ 50 & 0.07--0.12 & G-NOD\\
        \\
        \multicolumn{3}{l}{{\it Serpens South \citep{jose20}:}} \\
        6 & SS182917-020340 & 277.3193 & -2.0610 & M4--M7 & 20.0 & 17--22 & VL-G & -0.78 $\pm$ 0.28  & 2980 $\pm$ 210 & $\sim$0.1 & G,D,P \\ 
        7 & SS182918-020245 & 277.3256 & -2.0458 & M5--M8 & 11.9 & 10--13 & VL-G & -0.91 $\pm$ 0.17  & 2825 $\pm$ 155 & 0.05--0.08 & G,D,P \\ 
        8 & SS183038-021419 & 277.6568 & -2.2386 & M3--M6 & 21.9 & 18--24 & VL-G & -0.89 $\pm$ 0.34 & 3135 $\pm$ 275 & $\sim$0.1 &G,D,P \\ 
        9 & SS183044-020918 & 277.6847 & -2.1551 & M7--M9 & 13.6 & 12--15 & VL-G & -0.67 $\pm$ 0.09  & 2670 $\pm$ 100 & 0.05--0.08 & G,D,P \\ 
        \hline
    \end{tabular}
    \label{tab:phys_prop}
\end{table*}

\subsection{YSO Catalogues} \label{sec:yso}

\citet{jose20} use the presence of a Serpens candidate member in catalogues of young stellar objects (YSOs) as an independent indication of youth and Serpens membership. In all cases in \citet{jose20}, the late-type objects identified spectroscopically appeared in the MYStIX IR-Excess Source Catalogue \citep[][hereafter P13]{povich13}, SFiNCs Xray-Infrared Catalogue \citep[][hereafter G17]{getman17} and the Gould Belt Survey \citep[][hereafter D15]{dunham15}. This was used as strong evidence to break the population degeneracy found in the $\chi^{2}$ maps for some of these objects, and classify them as young. However, for the objects discussed in this paper, the evidence is less conclusive. In Table \ref{tab:phys_prop}, the column labelled `YSO?' indicates whether an object is identified as a YSO in these surveys. SS182953-015639 and SS183032-021028 do not appear in any of the catalogues. SS182959-020335 is listed as a young stellar object in both \citetalias{getman17} and \citetalias{povich13}, and SS182955-020416 appears in both of these and in \citetalias{dunham15}.
SC182952+011618 does not appear in \citetalias{dunham15}, and is flagged as `NO-DISK' in \citetalias{getman17}, meaning this object is identified as diskless by this survey (\citetalias{povich13} does not cover Serpens Core). This classification is based on the shape of the IR SED, as well as other properties of IR and X-Ray photometry \citep[see][]{getman17}. This is not conclusive proof against the youth of this object, but does strongly indicate that the object does not host an accreting disk. \citet{winston18} report a sub-sample of 18 diskless objects in Serpens South. They argue that whilst they clearly lack disks, other indicators of youth imply these objects are likely young cluster members, that may have rapidly lost their disks due to external environmental factors (e.g stripping by nearby massive stars, tidal disruption). They report that 30-53\% of X-ray sources discovered in Serpens South are diskless \citep[comparable to 48$\pm$11\% for Serpens Core,][]{winston07} - consequently, we consider it likely that SC182952+011618 is one of these diskless, young cluster members. 

In general, the information gathered about these objects from YSO catalogues is somewhat inconclusive. For SS182959-020335, SS183032-021028 and SC182952+011618, spectral type fitting unambiguously classifies these objects as young, based on their spectral features. However, this is not the case for SS182953-015639 and SS182955-020416, where we see bimodality in the SpT-$A_{\text{V}}$ $\chi^{2}$ maps. The latter appears in all 3 YSO surveys considered here, demonstrating additional evidence of youth and supporting the young spectral solution for this target, which we henceforth adopt. The lack of information in YSO catalogues for SS182953-015639 means we are unable to break the degeneracy between the young and field solutions, and we cannot confidently assign SS182953-015639 to a population.

\subsection{Physical Parameters} \label{sec:params}

We used our CFHT photometry and SpT-$A_{\text{V}}$ probability maps to calculate the bolometric magnitude and luminosity of each candidate Serpens member in our sample. We performed a Monte Carlo analysis, using 500,000 model objects distributed proportionally in the SpT-$A_{\text{V}}$ probability map. For each simulated object in each SpT-$A_{\text{V}}$ probability bin, we calculated a bolometric magnitude (Eq. \ref{eq:bolmag}) by sampling Gaussian distributions of distance modulus, $d_{\text{mod}}$ and apparent magnitude, $m_{\text{J}}$. 

\begin{equation}
    \centering
    M_{\text{bol}} = m_{\text{J}} - d_{\text{mod}} - A_{\text{J}} + BC_{\text{J}}
    \label{eq:bolmag}
\end{equation}

The Gaussian distribution of $m_{\text{J}}$ is constructed using the CFHT apparent magnitude and error, as listed in Table \ref{tab:phot_cfht}. The Gaussian distribution of $d_{\text{mod}}$ is constructed using the assumed value of $8.2 \pm 0.18$ mag, calculated using the Serpens distance given in \citet{ortiz17}. In Eq. \ref{eq:bolmag} we also use the conversion factor from $A_{\text{V}}$ to $A_{\text{J}}$, taken to be 0.282 \citep{cardelli89}, and the bolometric correction, $BC_{\text{J}}$. The bolometric correction is taken from two sources: for spectral type M7 or later, the relevant polynomial described in \citet{filippazzo15} is used; if the spectral type is earlier than M7, the bolometric corrections given in \citet{herczeg15} are used. 
To convert from bolometric magnitude to bolometric luminosity, we use Eq. \ref{eq:bollum}, where the solar bolometric magnitude is taken to be 4.73 mags:

\begin{equation}
    \centering
    \text{log}_{10}(L_{\text{bol}}/L_{\odot}) = - (M_{\text{bol}} - M_{\text{bol},\odot}) / 2.5
    \label{eq:bollum}
\end{equation}

\begin{figure}
    \centering
    \includegraphics[width=\columnwidth]{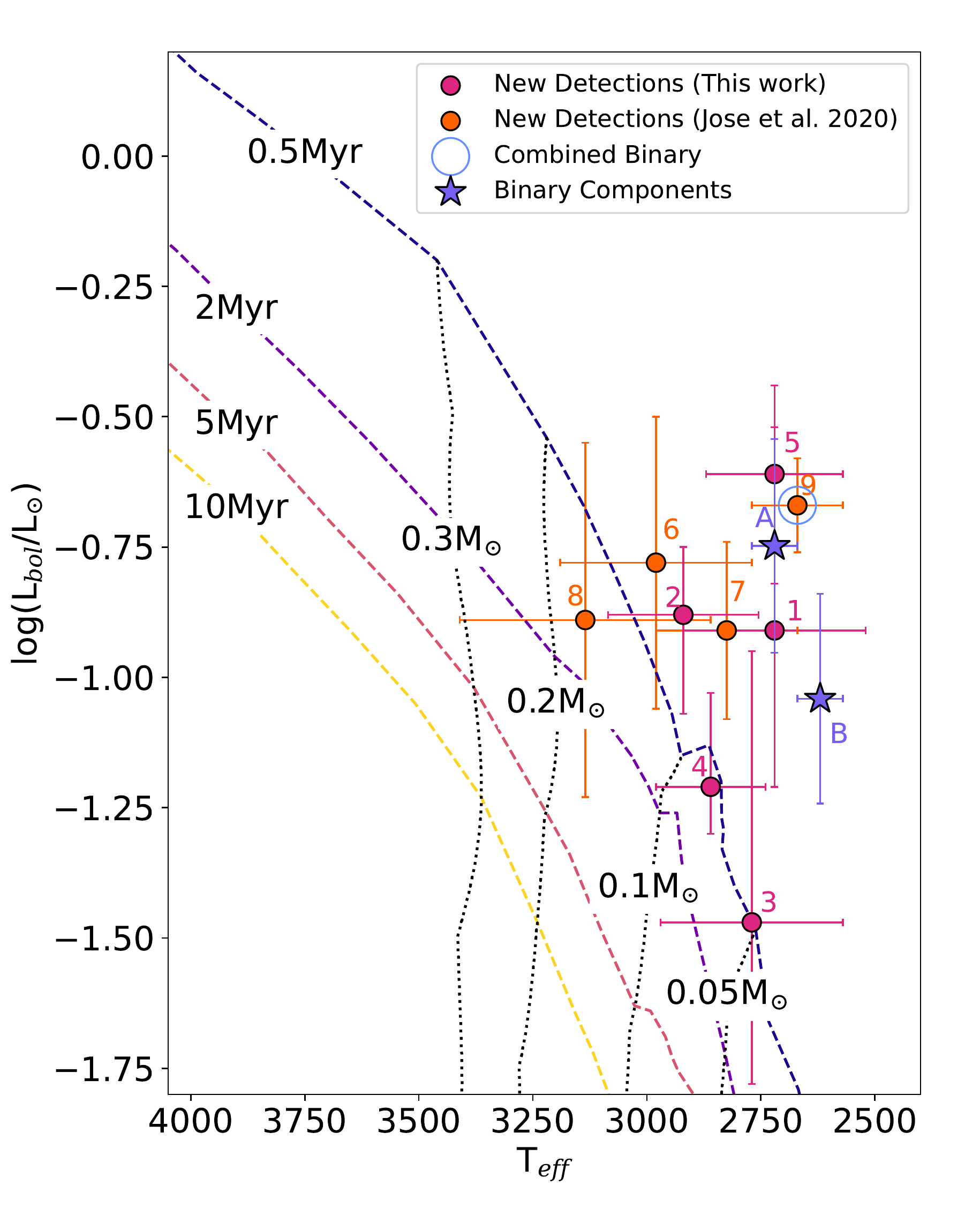}
    \caption{H-R diagram for the five candidate Serpens members identified in this paper (pink points), 4 in Serpens South (1,2,3,4) and 1 in Serpens Core (5). Also shown are the four low-mass candidate members presented in \citet{jose20} (orange points, 6,7,8,9). Object 9 is the newly discovered binary system (blue circle): purple stars show the binary components (A and B). Isochrones (dashed lines) and evolutionary tracks (dotted lines) shown are taken from \citet{baraffe15}.}
    \label{fig:hr}
\end{figure}

Having obtained a bolometric luminosity ($L_{\text{bol}}/L_{\odot}$) for each model object in each bin, we use this data to plot a histogram of log($L_{\text{bol}}/L_{\odot}$) values. The values given in Table \ref{tab:phys_prop} are the peak positions of these histograms, with errors derived from the 68\% Bayesian credible intervals of the distribution, which reflect the asymmetry seen for every object. An example histogram is shown in Figure \ref{fig:prob_map} for SS183032-021028. Log($L_{\text{bol}}/L_{\odot}$) histograms are shown in Figures \ref{fig:props_SS182953}, \ref{fig:props_SS182955}, \ref{fig:props_SS182959} $\&$ \ref{fig:props_SC182952} in Appendix \ref{sec:app_plots} for the remaining Serpens candidate members reported in this work. In some cases, we saw bimodal distributions in log($L_{\text{bol}}/L_{\odot}$): the parameter values in these cases are taken from the most prominent peak in the distribution. 
We made one key assumption when modelling the luminosity distribution across the full range of Av and SpT: we only included the section of the parameter space corresponding to young template fits. This is due to the inclusion of the distance modulus in the bolometric magnitude calculation, as the adopted distance modulus is only valid for actual members of Serpens. As a result, we assume that all 5 objects are young (with strong evidence of this being true for all targets except SS182953-015639), and calculate a luminosity range that would be valid in this case. 
Figure \ref{fig:hr} shows the 5 new detections presented in this paper plotted on a HR-diagram (numbering from Table \ref{tab:phys_prop}), also showing age and mass isochrones \citep[taken from~][]{baraffe15} to give context for other objects of these masses and ages. The effective temperatures ($T_{\text{eff}}$) plotted here and given in Table \ref{tab:phys_prop} are estimated using our derived spectral types and the relations given in Table 5 of \citet{herczeg14}. The 4 young, late-type candidate Serpens members presented in \citet{jose20} are also shown, with physical parameters given in Table \ref{tab:phys_prop}.

\subsection{HST Imaging}

As discussed in Section \ref{sec:hst}, to search for low-mass companions of our new Serpens candidate members, we obtained IR and UVIS (UV and visible) imaging of 6 targets in Serpens South, using Wide Field Camera 3 (WFC3) on the Hubble Space Telescope (HST).  

\subsubsection{New Binary Detection: SS183044-020918} \label{sec:S1}

\begin{figure}
    \centering
    \includegraphics[width=\columnwidth]{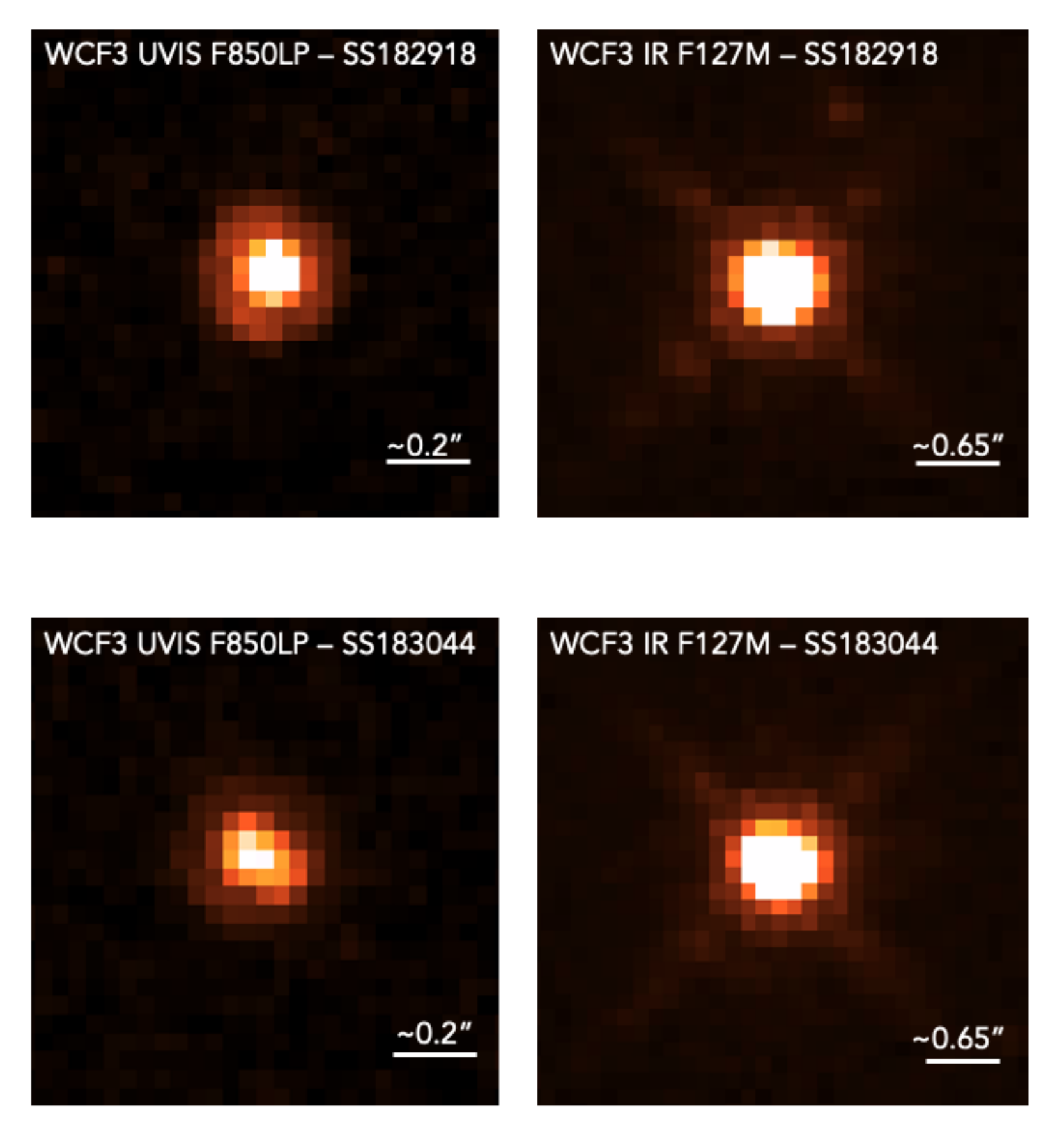}
    \caption{Upper panels: for comparison, F850LP (upper left) and F127M (upper right) images of SS182918-020245, a non-binary target from the Serpens HST program. Lower panels: the newly discovered low-mass binary, SS183044-020918. The binary components are resolved in F850LP (lower left), and unresolved in F127M (lower right), due to the larger pixel scale of the HST IR channel.}
    \label{fig:binary}
\end{figure}

The F850LP image of target SS183044-020918 appears notably elongated compared to other objects in the field. The components appear marginally resolved in the F850LP data, shown in Figure \ref{fig:binary}, along with the unresolved components in the IR filter F127M - the IR channel of has a much coarser pixel scale than the UVIS. We also show the F127M and F850LP images for SS182918-020245, to highlight the elliptical appearance of SS183044-020918 in F850LP. This elongation is clearly caused by two distinct components: best explained by a binary system, as we show below. \\

\begin{figure*}
    \centering
    \includegraphics[width=\textwidth]{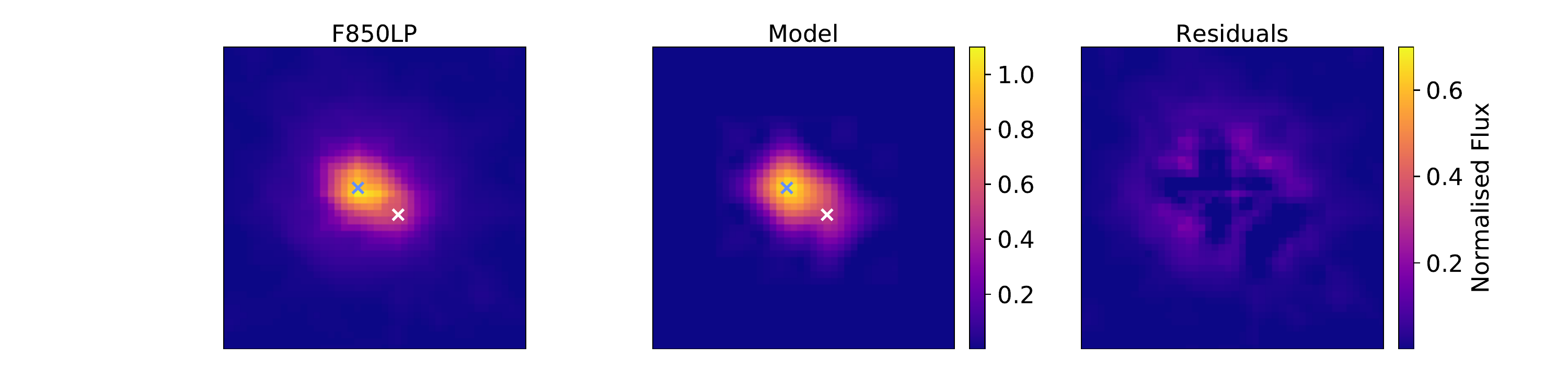}
    \caption{Best-fitting PSF subtraction model for both components of SS183044-020918, using ePSF method. Coloured crosses indicate the best fit component positions. Left panel shows raw F850LP data. Centre panel shows model imaged constructed using best fit parameters from MCMC analysis. Right panel shows residuals from model fit.}
    \label{fig:binary_sub}
\end{figure*}

To determine the separation and the relative flux contributions of each binary component, we built an effective point spread function (ePSF) using the other target stars from the HST survey. After removing some stars due to contamination or bright pixels, we retained a sample of 28 stars to build the ePSF. These were centred and scaled by the {\tt EPSFBuilder} algorithm available in the python package {\tt photutils}, and combined into a single ePSF. We then ran a Markov chain Monte Carlo (MCMC) analysis for parameter estimation, using the python package {\tt emcee} \citep{foreman13}. At each step of the MCMC chain, an artificial image is constructed using two scaled ePSFs, with the coordinates of both components as model parameters, as well as the relative flux ratio. We minimise over $\chi^{2}$ to obtain a best-fitting model image and binary component parameters, as well as associated errors. The best fit model and fit residuals are shown in Figure \ref{fig:binary_sub}.

We calculated the binary separation using the best-fit component positions from this analysis, and calculated the magnitude difference of the two components from the best-fit flux ratio. We find that the components are not equal magnitude: SS183044-020918B is 0.76$\pm$0.06 magnitudes fainter than SS183044-020918A in the F850LP filter. All data calculated using this model fitting is given in Table \ref{tab:binary_prop}, including errors in each fit parameter. The secondary is at a separation of 0.092$\pm$0.008" from the primary, which corresponds to 41.9$\pm$3.6 AU in Serpens South. This distance was calculated using the F850LP images only, as it is less than the pixel scale of the IR channel detector (0.13"), and thus the two components are unresolved in F127M and F139M. 

\citet{jose20} present a combined spectrum of SS183044-020918, which was best fit with a young standard of spectral type M7-M9, and an $A_{\text{V}}$ in the range 12-15 mag. We explored whether a combination of multiple spectral type templates (one for each binary component) provides a better fit than a single component template. We again used the young spectral templates from \citetalias{luhman17}, combined in pairs and fitted to the observed spectrum, whilst varying the $A_{\text{V}}$ between the previously-found best fitting range. For the primary, we considered M7-M9 as reasonable spectral types, and for the secondary, we used the broader range of M7-L2. We only considered solutions where the secondary was of a later spectral type than the primary - informed by the flux ratio in F850LP. The resulting best-fit spectral types for each component are given in Table \ref{tab:binary_prop}. We again give a range of spectral types - as we saw that different combinations of templates gave similarly good answers, with slightly varying $A_{\text{V}}$ values. We find that SS183044-020918A has a best fit spectral type of M7-M8, and SS183044-020918B has a best fit spectral type of M8-M9. Figure \ref{fig:binary_spec} shows the overall best fit of M8+M9 with an $A_{\text{V}}$ of 12.9$\pm$0.3 mag, slightly lower than the previously reported value of 13.6 \citep{jose20}.

We used the flux ratio of the components, and the photometry and spectroscopy obtained from both the W-band and HST surveys, to determine physical properties. These are detailed in Table \ref{tab:binary_prop}. The $J$ and $H$ mags were calculated using $\Delta$mag = 0.76, derived from the best fit to the F850LP image and the CFHT photometry for the combined components. As the binary is only resolved in F850LP, we must assume that the flux ratio between the two components is similar in every filter, and calculate properties based on this assumption. The similar spectral types of both components support this assumption, indicating that $\Delta$mag will likely be similar in different filters. As in Section \ref{sec:params}, to calculate the bolometric luminosity of each component, the bolometric correction $BC_{\text{J}}$ was taken from \citet{filippazzo15}. We use the extinction determined from our combined standard best-fit, $A_{\text{V}}$ = 12.9. $T_{\text{eff}}$ for each component is estimated using Table 5 of \citet{herczeg15}. If we plot the two binary on Figure {\ref{fig:hr} (purple stars), we find that SS183044+020918A is consistent with a mass of $\sim$ 0.08-0.1M$_{\odot}$, and SS183044+020918B is consistent with a lower mass of $\sim$ 0.05-0.07M$_{\odot}$, truly pushing into the substellar regime. As seen with the other reported discoveries in Section \ref{sec:results}, the two binary components lie above the 0.5Myr isochrone, and as such the masses determined by extended the evolutionary models beyond this age have large associated errors.

\subsubsection{Likelihood of Binarity}

At a distance of $> 400$ pc, the proper motion of objects in Serpens South is too low to allow common proper motion confirmation of this candidate on a timescale $< 10$ years. Instead, we estimate the likelihood that SS183044-020918B is a background object using the 2MASS survey and Eq. 1 from \citet{brandner00}:

\begin{equation}
    P(\Theta,m) = 1 - e^{-\pi \rho(m) \Theta^{2}}
    \label{eq:brandner}
\end{equation}

where $\Theta$ is defined as the angular distance from the target considered, and $\rho$(m) is defined as `the cumulative surface density of background sources down to a limiting magnitude m' \citep{brandner00}. A 2MASS query within a 1$^{\circ}$ radius of the primary returns 102,537 sources with $J$ magnitudes of 16.5 or brighter. We use this as the surface density of background sources $\rho$(m), with an angular distance $\Theta$ equal to the binary separation. This returns a probability of $ P(\Theta,m) \approx 2.1\times10^{-4}$ of finding a background source with a brightness at least that of the secondary within 0.096" of the primary. This calculation does not account for the size of the survey, or the colour information of the sources, both of which give us additional insight into the nature of the secondary. Thus, we are confident that the binary we report is indeed a binary, and not a Serpens South member aligned with an unrelated background star.

Further evidence of the binarity of this system can be found in the spectral data. As detailed above, our combined spectrum is best fit by a young M8+M9 combined template. This strongly implies the secondary has a similar or later spectral type to the primary, as does the F850LP flux ratio. When considering this in conjunction with the calculation of the likelihood of the secondary being an unrelated foreground / background object, the probability of a similarly late and young background object in chance alignment with the primary is, as stated above, very small.

\begin{figure}
    \centering
    \includegraphics[width=\columnwidth]{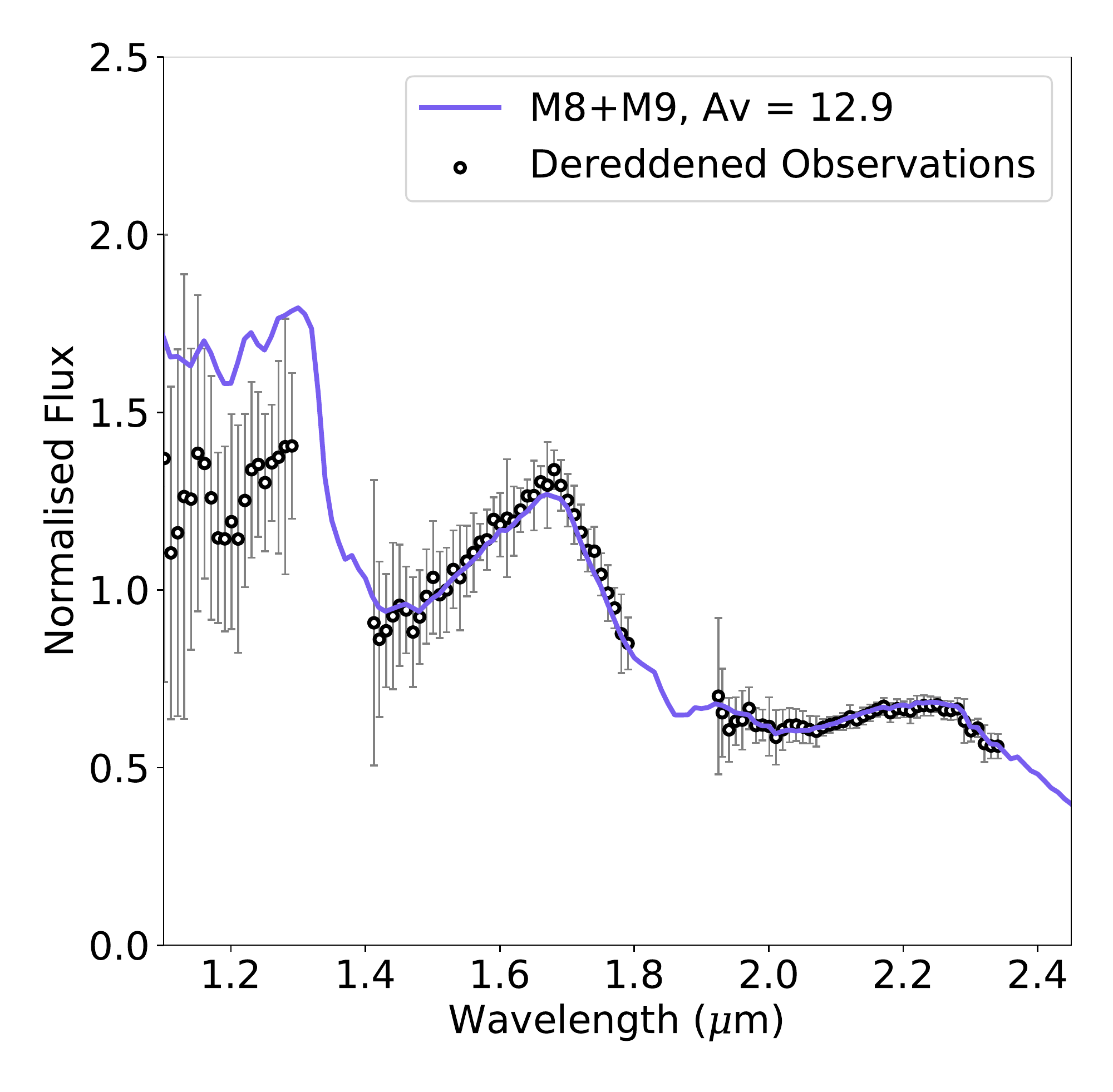}
    \caption{Spectrum of SS183044-020918, first reported in \citet{jose20}. Data is shown in black open circles, and is dereddened by the best fit $A_{\text{V}}$ = 12.9. The best fit combined template spectrum, M8+M9, is shown in purple.}
    \label{fig:binary_spec}
\end{figure}

\begin{table}
    \centering
    \caption{Properties of the binary system, SS183044-020918. }
    \begin{tabular}{lcr}
        \hline
        Property & Primary & Secondary \\
        \hline
        Distance & \multicolumn{2}{|c|}{436.0 $\pm$ 9.2 pc$^{a}$}  \\
        Age & \multicolumn{2}{|c|}{0.5 Myr$^{b}$}  \\
        Separation & \multicolumn{2}{|c|}{ 0.092 $\pm$ 0.008" (41.9 $\pm$ 3.6 AU)} \\
        $\Delta$mag (F850LP) & \multicolumn{2}{|c|}{0.76 $\pm$ 0.06} \\
        $z$(mag)$^{\dagger}$ & 21.46 & 22.22 \\
        $J$(mag)$^{\dagger}$ & 16.74 & 17.50 \\
        $H$(mag)$^{\dagger}$ & 14.85 & 15.61 \\
        $A_{\text{V}}$(mag) & \multicolumn{2}{|c|}{12.9 $\pm$ 0.3} \\
        Spectral Type & M7-M8 & M8-M9 \\
        log($L/L_{\odot}$)$^{\dagger}$ & -0.748 $\pm$ 0.205 & -1.041 $\pm$ 0.210 \\
        $T_{\text{eff}}$ & 2720 $\pm$ 50 K & 2620 $\pm$ 50 K\\
        Estimated mass$^{\dagger}$ & 0.08--0.1~M$_{\odot}$ & 0.05--0.07~M$_{\odot}$ \\
        \hline
        \multicolumn{3}{l}{\footnotesize$^{a}$ Distance from \citet{ortiz17,ortiz18}.} \\
        \multicolumn{3}{l}{\footnotesize$^{b}$ Age from \citet{gutermuth08}.} \\
        \multicolumn{3}{l}{\footnotesize$^{\dagger}$ Assuming F850LP $\Delta$mag is true for all wavelengths}\\
    \end{tabular}
    \label{tab:binary_prop}
\end{table}

\subsection{Other Serpens South HST observations}

We examined the 6 Serpens South HST datasets for binary systems, close-in and wide companions, and do not report any additional candidate companions. To perform this examination, we used a two-fold approach. We visually inspected each image for close companions, looking for obvious elongation or close neighbouring sources. This initial inspection led to the identification of SS183044+020918 as a binary, but of no other candidates. Next, we performed PSF subtraction on each of the images, using two parallel techniques. 

The first method uses an ePSF constructed from other target stars in the survey, as detailed above for the binary analysis. We used this ePSF in conjunction with PSF subtraction algorithms available in {\tt photutils}. More specifically, the {\tt IterativelySubtractedPSFPhotometry} function was used to subtract the target-constructed ePSF from each of the objects, leaving a residual image with the starlight removed. If there were any close in companions around the targets objects, previously obscured by the objects themselves, these should be visible in the ePSF-subtracted residuals. 

To confirm the results of the aforementioned method to detect possible Serpens companions, we made use of tools from the Vortex Image Processing package \citep[VIP;][]{gomezgonzalez17} to perform a PSF subtraction. VIP has an implementation of the Reference Differential Imaging method \citep[RDI;][]{lagrange09,lafreniere09,soummer12} where a reference PSF is constructed using similar but distinct PSFs to subtract from the science PSF. This is combined with an implementation of Principal Component Analysis \citep{soummer12} to reduce the dimensionality of the data. We build our reference library by using other stellar PSFs in the Serpens cluster data in addition to previously observed data from the Ophiuchus and Taurus associations, all observed with HST in the F850LP filter. We then divide this into nine separate "sublibraries" based on which subpixel the PSF peak position falls on: where the subpixels are each pixel divided into nine equal squares. To determine the peak of the science PSF, a Gaussian fit is made to the Serpens target PSF. This peak defines which sublibrary will be used for the PSF subtraction, as the reference sublibrary will be made up only of PSFs whose peak falls in the same subpixel as the target's peak.
Using this method with 5 principal components, we made a redetection for the binary companion to SS183044+020918, shown in Figure \ref{fig:vip_binary_im} in Appendix \ref{sec:app_binary}. We also performed this analysis on the other Serpens objects, for which we made no further detections, agreeing with the results of the method above. \\

We also examined the images for wide companions, and investigated the nature of other objects nearby in the field. We examined each object with 3 criteria in mind: 1) were there any nearby (within $\sim$ 20") sources with similar or larger $Q_{\text{HST}}$ values than the target, 2) were there any nearby sources (within $\sim$ 3") with blue F127M - F139M colours and 3) were there any nearby F139M dropouts (i.e >10$\sigma$ detection in F127M and no detection in F139M). For all of the Serpens images, there were no nearby sources that satisfied these criteria, ruling out finding any wide-companions with these diagnostics. 

\subsection{Contrast Curves}

We present contrast curves for our HST observations in Figure \ref{fig:contrast}. To generate contrast curves, we added a simulated planet (the ePSF used in Section \ref{sec:S1}, scaled and cropped) into the F127M images of the five datasets with no detections, at a specified position and magnitude difference (relative to the primary star). We varied the planet-star separation and magnitude difference to cover a grid ranging from 0.13 - 2.0" separation and from 0 - 8 $\Delta$mags. To obtain the contrast value plotted for each separation, we performed PSF subtraction (as in Section \ref{sec:S1}) on the image containing the injected planet. We then calculated the signal in an ePSF-sized aperture placed over the location of the injected planet in the PSF-subtracted image. To calculate the noise level, we used the same aperture at the position of the injected planet, but in a PSF-subtracted image where no planet was added. For each separation, the magnitude difference that yielded a signal-to-noise ratio (S/N) $> 5$ was adopted as the contrast at that radius. \\
The achieved contrast is similar for all targets at separations $\lesssim 0.6"$ from the primary star. The contrasts diverge at separations greater than this, and reach differing levels at separations $\gtrsim 1.0"$  (corresponding to $\approx$ 10 pixels in the F127M images). In Figure \ref{fig:contrast}, we also present the minimum detectable apparent magnitude of a planet around each target, to account for the variation in contrast caused by the differing intrinsic brightness of each target. From this, we can see that the performance for each target is actually quite similar, with a mean value reaching $\sim m=23$ at a separation of $\sim$1.0" from the primary. We present contrast curves in F127M, as two targets were not detected in F850LP. We also used VIP and the technique discussed previously to calculate contrast curves, and find good agreement between the two methods for all targets.

\begin{figure*}
    \centering
    \includegraphics[width=\textwidth]{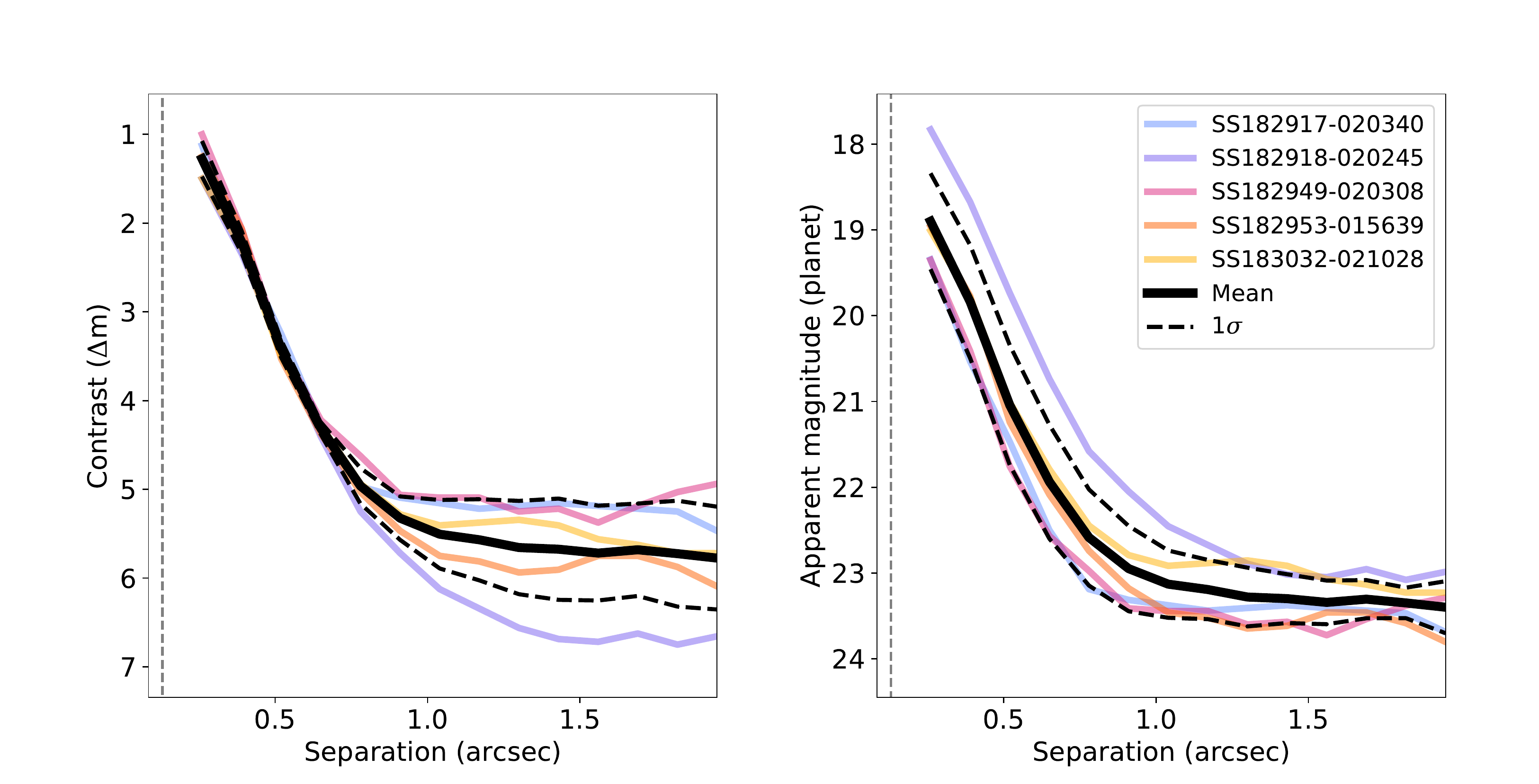}
    \caption{Left: Contrast curves in F127M for the five Serpens objects with no detected companions. Contrast values were calculated by injecting synthetic planets into the images and checking the magnitude at which these are no longer detectable. Dashed line indicates the pixel scale of the F127M HST images (0.13"). Right: Minimum detectable apparent magnitude for planets around each target (coloured dashed lines). Solid black line shows the mean value for all 5 targets at each separation, dashed lines show 1$\sigma$ error on this value.}
    \label{fig:contrast}
\end{figure*}

\section{Discussion} \label{sec:disc}

The W-band technique has proven highly effective in identifying ultracool dwarfs in nearby star-forming regions. By looking at Serpens South and Serpens Core, we have pushed its effectiveness to its limits, yet still demonstrated its usefulness by finding more low-mass, likely members of these subclusters.  The number of new detections presented in this paper is small, but adds to the growing number of objects creating a statistically significant low-mass sample. We cannot provide additional constrains on the form of the low-mass end of the IMF with just the small sample presented in this work.  However as further low-mass members of Serpens are discovered, the possibility for a meaningful, focused investigation of the IMF in this region increases. 

\subsection{Mass estimates of new detections}

In Section \ref{sec:params}, we calculated physical properties of the 5 mid-late M, candidate Serpens members reported in this portion of the W-band Serpens survey. Figure \ref{fig:hr} shows these objects on a HR diagram, as well as the four candidates reported in \citet{jose20}, and isochrones and evolutionary tracks from \citet{baraffe15}.

SS183032-021028 (target 4, numbering given in Table \ref{tab:phys_prop}) lies effectively on the 0.5 Myr isochrone \citep[the approximate age of Serpens South,][]{gutermuth08}, and has an effective temperature consistent with a mass of approximately 0.07--0.1~M$_{\odot}$. SS182955-020416 (target 2) lies close the 0.5 Myr isochrone, but likely has a higher mass, approximately 0.1--0.15~M$_{\odot}$. 

Similarly, SS182959-020335 (target 3) also sits on the 0.5 Myr isochrone, and likely has the lowest mass of the 5 Serpens candidate members. We report a mass interval of 0.05--0.09~M$_{\odot}$ for this object based on its position in Figure \ref{fig:hr} - although the mass range encompassed by the error bars indicates that this target could be consistent with a mass lower than this. However, the evolutionary tracks shown here do not extend below 0.05~M$_{\odot}$, and so we cannot confidently report a lower limit that goes beyond the coverage of the models used for characterisation.

SS182953-015639 (target 1) lies above the youngest available isochrone (0.5 Myr). Again, since this is beyond the range of the evolutionary models, estimating a mass requires extending the model trends to the position of SS182953-015639 - leading to a mass of 0.07--0.1~M$_{\odot}$, which should be treated as very approximate. In previous sections, we discussed the likely age of SS182953-015639, and were unable to constrain it to either a young or field-age population. The calculation of $L_{\text{bol}}/L_{\odot}$ used here assumes that the best solution is a young template - field-age solutions are not considered. Thus, we cannot use Figure \ref{fig:hr} as strong proof of a young age for SS182953-015639, as the assumption of youth is used in the calculations of the parameters plotted. \\

One of the objects reported in \citet{jose20}, SS183044-020918 (target 9), sits high above the 0.5 Myr isochrone. As discussed above in Section \ref{sec:results}, we report that this object is actually an unresolved binary system, confirmed using multi-filter HST imaging. This explains the anomalously high luminosity seen in this analysis, as we are actually measuring the combined properties of two unresolved objects.
Similar to SS183044-020918, SC182952+010116 (target 5) also sits high above the 0.5 Myr isochrone. It is consistent with a mass of $\approx$ 0.07--0.12~M$_{\odot}$, although again, extending the evolutionary tracks beyond the provided values creates a large margin of error. The similarity in physical properties between this target and the newly-discovered binary system suggest that binarity could also explain the anomalously high $L_{\text{bol}}/L_{\odot}$ of SC182952-010116.

SS182918-020245 (target 7) also sits above the youngest isochrone in Figure \ref{fig:hr}. \citet{jose20} found signatures of ongoing accretion in the spectrum of this object. The models shown in Figure \ref{fig:hr} \citep{baraffe15} assume photospheric emission as the primary contributing factor to luminosity - consequently, if an object is also actively accreting, these models are not best suited for characterising its properties. The positioning of many of the Serpens discoveries above the 0.5 Myr isochrone is discussed in further detail in \citet{jose20}.
Ongoing accretion could explain the properties of the two targets from this work that are inconsistent with the 0.5 Myr isochrone, SS182953-015639 (target 1) and SC182952+010116 (target  5).  However, the IRTF SpeX spectra obtained for these objects have insufficient resolution to look for the accretion signatures discussed in \citet{jose20}, and as such we cannot identify accretion as the driving factor for their high luminosities. \\

The five late-type candidate Serpens members discussed in this work all have masses consistent with 0.05--0.15~M$_{\odot}$ (using isochrones from \citet{baraffe15}). Mass estimates are given in Table \ref{tab:phys_prop}, including objects from \citet{jose20}. If the boundary between low-mass stellar objects and substellar brown dwarfs is taken to be $\sim$0.075~M$_{\odot}$ \citep{reid99}, SS182953-015639, SS182959-020335, SS183032-021028 and SC182952+011618 could all be characterised as substellar, although the large error bars on their derived masses mean that this characterisation remains uncertain. It should be noted that the mass estimates given in \citet{jose20} (and in Table \ref{tab:phys_prop}) are derived using solely the $T_{\text{eff}}$ of each object, which differs slightly from the method used in this work (where luminosities and effective temperatures are used in combination).

\subsection{New binary discovery}

In Section \ref{sec:S1}, we present the properties of SS183044-020918, derived from analysis of IR and UVIS HST images. As discussed here, the mass estimate of 0.05--0.07~M$_\odot$ derived for SS183044-020918B makes it the lowest mass object discovered in the W-band/HST survey of Serpens, and also the most likely sub-stellar candidate member.
With a primary mass of 0.08--0.10~M$_\odot$, the more massive component of this binary is likely a very low-mass star, with a mass lying just above the hydrogen-burning limit. Such component masses, with a measured projected separation of $\sim$40~AU, are not unusual for late-type M dwarfs, which tend to have more similar component masses and tighter binary separations compared to more massive stars \citep{bergfors10}.

Although low-mass stars and brown dwarfs are preferentially seen in compact binaries, a number of wide low-mass systems have been discovered with separations of several tens to hundreds of AU, both in young associations and in the field population. For example, the young Taurus binaries CFHT-Tau-7 (with masses of 0.07 and 0.06~M$_\odot$), CFHT-Tau-17 (0.10 and 0.06~M$_\odot$) and CFHT-Tau-18 (0.10 and 0.06~M$_\odot$) have orbital separations of 32, 82 and 31~AU, respectively \citep{konopacky07}. In the field, systems in similar configurations include DENIS J220002.05$-$303832.9AB \citep{burgasser06}, a 0.085$+$0.083~M$_\odot$ binary with a 38-AU separation, or 2MASS J15500845$+$1455180AB \citep{burgasser09}, which consists of 0.070 and 0.067-M$_\odot$ L dwarfs separated by 30~AU. While not very common, such wide systems are thus known to exist at various evolutionary stages. In fact, systems with comparable masses to SS183044-020918 are even known with separations larger than 100~AU, such as the 2--4 Gyr old 2MASS J0130$-$4445 binary (0.083$+$0.070~M$_\odot$, 130 AU; \citealp{dhital11}). We thus consider that SS183044-020918 is not abnormally wide for a binary with a very low-mass stellar host.

An alternative way to compare binary properties is in terms of gravitational binding energy. We calculate upper and lower limits on the binding energy of the SS183044-020918 binary system, based on the ranges of mass and separation given in Table \ref{tab:binary_prop}. Using masses of 0.1M~$_\odot$ and 0.07~M$_\odot$ ($M_{\text{T,upper}} \approx 178$~M$_{\text{J}}$), and the minimum derived separation of 38.3~AU, we obtain an upper limit on binding energy, $E_{\text{b,upper}} = 3.2 \times 10^{42}$~ergs. Alternatively, using masses of 0.08~M$_\odot$ and 0.05~M$_\odot$ ($M_{\text{T,lower}} \approx 136$~M$_{\text{J}}$), and the maximum derived separation of 45.5~AU, we obtain a lower limit on binding, $E_{\text{b,lower}} = 1.8 \times 10^{42}$~ergs. \citet{fontanive20} present an up-to-date summary of binding energy vs total mass in their Figure 4, showing low-mass binaries in the field and in young associations (see \citealp{faherty20} for a full compilation). Considering the limits on binding energy and total mass of SS183044-020918 derived here, this new binary system agrees well with previous measurements of both field and young association systems, consistent with our conclusions above. Given that systems with similar architectures are observed from young star-forming regions to the old field population, our newly-discovered Serpens binary is likely to be stable to dynamical evolutionary processes and to survive to field ages.

\section{Conclusions} \label{sec:concl}

We present results from a multi-technique survey of Serpens South and Serpens Core, using photometric, spectroscopic and high-resolution imaging data to hunt for the lowest mass, youngest members of these regions. We have identified five likely-young low-mass candidate members of Serpens South and Serpens Core, adding to the four detections reported in \citet{jose20}. We find that four of these objects have spectral types later than $\sim$M5, and have effective temperatures and luminosities consistent with $\lesssim 0.12$M$_{\odot}$. Additional evidence suggests that three of these objects are highly likely to be Serpens members, but this cannot be confirmed with proper motion follow-up on reasonable timescales for the Serpens region. We also provide an update on one of the detections from \citet{jose20}, SS183044-020918, which we have found through a HST imaging program to be a binary system, resolved in visible light. We cannot confirm the binarity of the two components via proper motion analysis, but show that the likelihood of chance alignment with a background star is very small. We find the binary components to have likely spectral types of M7--M8 and M8--M9, derived from a combined IR spectrum. Calculating bolometric luminosities using the difference in magnitudes between the components, and our CFHT photometry, we find that the secondary has a mass of 0.05-0.07M$_{\odot}$, making it one of the lowest mass candidate members of the Serpens South star-forming region.

\section*{Acknowledgements}

Based on observations obtained with WIRCam, a joint project of CFHT, Taiwan, Korea, Canada, and France, at the Canada–France–Hawaii Telescope (CFHT) which is operated by the National Research Council (NRC) of Canada, the Institut National des Sciences de l’Univers of the Centre National de la Recherche Scientifique of France, and the University of Hawaii. \\
Based on observations made with the NASA/ESA Hubble Space Telescope, which is operated by the Association of Universities for Research in Astronomy, Inc., under NASA contract NAS5-26555\\
Visiting Astronomer at the Infrared Telescope Facility, which is operated by the University of
Hawaii under contract NNH14CK55B with the National Aeronautics and Space Administration. The authors wish to recognise and acknowledge the very significant cultural role and reverence that the summit of Maunakea has always had within the indigenous Hawaiian community. We are most fortunate to have the opportunity to conduct observations from this mountain.\\
This research has benefited from the SpeX Prism Spectral Libraries, maintained by Adam Burgasser at \url{http://pono.ucsd.edu/~adam/browndwarfs/spexprism/}.\\
This publication makes use of data products from the Wide-field Infrared Survey Explorer, which is a joint project of the University of California, Los Angeles, and the Jet Propulsion Laboratory/California Institute of Technology, funded by the National Aeronautics and Space Administration.\\
This research has made use of the SIMBAD database, operated at CDS, Strasbourg, France, and NASA's Astrophysics Data System.\\
This research made use of Photutils, an Astropy package for
detection and photometry of astronomical sources\\

\section*{Data Availability}

The data underlying this article will be shared on reasonable request to the corresponding author.




\bibliographystyle{mnras}
\bibliography{paper.bib} 

\begin{thebibliography}{}
\makeatletter
\relax
\def\mn@urlcharsother{\let\do\@makeother \do\$\do\&\do\#\do\^\do\_\do\%\do\~}
\def\mn@doi{\begingroup\mn@urlcharsother \@ifnextchar [ {\mn@doi@}
  {\mn@doi@[]}}
\def\mn@doi@[#1]#2{\def\@tempa{#1}\ifx\@tempa\@empty \href
  {http://dx.doi.org/#2} {doi:#2}\else \href {http://dx.doi.org/#2} {#1}\fi
  \endgroup}
\def\mn@eprint#1#2{\mn@eprint@#1:#2::\@nil}
\def\mn@eprint@arXiv#1{\href {http://arxiv.org/abs/#1} {{\tt arXiv:#1}}}
\def\mn@eprint@dblp#1{\href {http://dblp.uni-trier.de/rec/bibtex/#1.xml}
  {dblp:#1}}
\def\mn@eprint@#1:#2:#3:#4\@nil{\def\@tempa {#1}\def\@tempb {#2}\def\@tempc
  {#3}\ifx \@tempc \@empty \let \@tempc \@tempb \let \@tempb \@tempa \fi \ifx
  \@tempb \@empty \def\@tempb {arXiv}\fi \@ifundefined
  {mn@eprint@\@tempb}{\@tempb:\@tempc}{\expandafter \expandafter \csname
  mn@eprint@\@tempb\endcsname \expandafter{\@tempc}}}

\bibitem[\protect\citeauthoryear{{Allers} \& {Liu}}{{Allers} \&
  {Liu}}{2013}]{allers13}
{Allers} K.~N.,  {Liu} M.~C.,  2013, \mn@doi [\apj]
  {10.1088/0004-637X/772/2/79}, \href
  {https://ui.adsabs.harvard.edu/abs/2013ApJ...772...79A} {772, 79}

\bibitem[\protect\citeauthoryear{{Allers} \& {Liu}}{{Allers} \&
  {Liu}}{2020}]{allers20}
{Allers} K.~N.,  {Liu} M.~C.,  2020, \mn@doi [\pasp]
  {10.1088/1538-3873/aba811}, \href
  {https://ui.adsabs.harvard.edu/abs/2020PASP..132j4401A} {132, 104401}

\bibitem[\protect\citeauthoryear{{Allers} et~al.,}{{Allers}
  et~al.}{2007}]{allers07}
{Allers} K.~N.,  et~al., 2007, \mn@doi [\apj] {10.1086/510845}, \href
  {https://ui.adsabs.harvard.edu/abs/2007ApJ...657..511A} {657, 511}

\bibitem[\protect\citeauthoryear{{Alves de Oliveira}, {Moraux}, {Bouvier},
  {Bouy}, {Marmo}  \& {Albert}}{{Alves de Oliveira} et~al.}{2010}]{alves10}
{Alves de Oliveira} C.,  {Moraux} E.,  {Bouvier} J.,  {Bouy} H.,  {Marmo} C.,
  {Albert} L.,  2010, \mn@doi [\aap] {10.1051/0004-6361/200913900}, \href
  {https://ui.adsabs.harvard.edu/abs/2010A&A...515A..75A} {515, A75}

\bibitem[\protect\citeauthoryear{{Bally}, {Walawender}, {Johnstone}, {Kirk}  \&
  {Goodman}}{{Bally} et~al.}{2008}]{bally08}
{Bally} J.,  {Walawender} J.,  {Johnstone} D.,  {Kirk} H.,   {Goodman} A.,
  2008, {The Perseus Cloud}.
p.~308

\bibitem[\protect\citeauthoryear{{Baraffe}, {Homeier}, {Allard}  \&
  {Chabrier}}{{Baraffe} et~al.}{2015}]{baraffe15}
{Baraffe} I.,  {Homeier} D.,  {Allard} F.,   {Chabrier} G.,  2015, \mn@doi
  [\aap] {10.1051/0004-6361/201425481}, \href
  {https://ui.adsabs.harvard.edu/abs/2015A&A...577A..42B} {577, A42}

\bibitem[\protect\citeauthoryear{{Bergfors} et~al.,}{{Bergfors}
  et~al.}{2010}]{bergfors10}
{Bergfors} C.,  et~al., 2010, \mn@doi [\aap] {10.1051/0004-6361/201014114},
  \href {https://ui.adsabs.harvard.edu/abs/2010A&A...520A..54B} {520, A54}

\bibitem[\protect\citeauthoryear{{Brandner} et~al.,}{{Brandner}
  et~al.}{2000}]{brandner00}
{Brandner} W.,  et~al., 2000, \mn@doi [\aj] {10.1086/301483}, \href
  {https://ui.adsabs.harvard.edu/abs/2000AJ....120..950B} {120, 950}

\bibitem[\protect\citeauthoryear{{Brice{\~n}o}, {Luhman}, {Hartmann},
  {Stauffer}  \& {Kirkpatrick}}{{Brice{\~n}o} et~al.}{2002}]{briceno02}
{Brice{\~n}o} C.,  {Luhman} K.~L.,  {Hartmann} L.,  {Stauffer} J.~R.,
  {Kirkpatrick} J.~D.,  2002, \mn@doi [\apj] {10.1086/343127}, \href
  {http://adsabs.harvard.edu/abs/2002ApJ...580..317B} {580, 317}

\bibitem[\protect\citeauthoryear{{Burgasser} \& {McElwain}}{{Burgasser} \&
  {McElwain}}{2006}]{burgasser06}
{Burgasser} A.~J.,  {McElwain} M.~W.,  2006, \mn@doi [\aj] {10.1086/499042},
  \href {https://ui.adsabs.harvard.edu/abs/2006AJ....131.1007B} {131, 1007}

\bibitem[\protect\citeauthoryear{{Burgasser}, {Dhital}  \& {West}}{{Burgasser}
  et~al.}{2009}]{burgasser09}
{Burgasser} A.~J.,  {Dhital} S.,   {West} A.~A.,  2009, \mn@doi [\aj]
  {10.1088/0004-6256/138/6/1563}, \href
  {https://ui.adsabs.harvard.edu/abs/2009AJ....138.1563B} {138, 1563}

\bibitem[\protect\citeauthoryear{{Cardelli}, {Clayton}  \& {Mathis}}{{Cardelli}
  et~al.}{1989}]{cardelli89}
{Cardelli} J.~A.,  {Clayton} G.~C.,   {Mathis} J.~S.,  1989, \mn@doi [\apj]
  {10.1086/167900}, \href
  {https://ui.adsabs.harvard.edu/abs/1989ApJ...345..245C} {345, 245}

\bibitem[\protect\citeauthoryear{{Chambers}, {Magnier}, {Metcalfe},
  {Flewelling}, {Huber}  et~al.}{{Chambers} et~al.}{2016}]{chambers16}
{Chambers} K.~C.,  {Magnier} E.~A.,  {Metcalfe} N.,  {Flewelling} H.~A.,
  {Huber} M.~E.,   et~al., 2016, arXiv e-prints, \href
  {https://ui.adsabs.harvard.edu/abs/2016arXiv161205560C} {p. arXiv:1612.05560}

\bibitem[\protect\citeauthoryear{{Cushing}, {Vacca}  \& {Rayner}}{{Cushing}
  et~al.}{2004}]{cushing04}
{Cushing} M.~C.,  {Vacca} W.~D.,   {Rayner} J.~T.,  2004, \mn@doi [\pasp]
  {10.1086/382907}, \href {http://adsabs.harvard.edu/abs/2004PASP..116..362C}
  {116, 362}

\bibitem[\protect\citeauthoryear{{Cushing}, {Rayner}  \& {Vacca}}{{Cushing}
  et~al.}{2005}]{cushing05}
{Cushing} M.~C.,  {Rayner} J.~T.,   {Vacca} W.~D.,  2005, \mn@doi [\apj]
  {10.1086/428040}, \href
  {https://ui.adsabs.harvard.edu/abs/2005ApJ...623.1115C} {623, 1115}

\bibitem[\protect\citeauthoryear{{Cushing} et~al.,}{{Cushing}
  et~al.}{2008}]{cushing08}
{Cushing} M.~C.,  et~al., 2008, \mn@doi [\apj] {10.1086/526489}, \href
  {https://ui.adsabs.harvard.edu/abs/2008ApJ...678.1372C} {678, 1372}

\bibitem[\protect\citeauthoryear{{Cutri}, {others}  \& {others}}{{Cutri}
  et~al.}{2013}]{cutri13}
{Cutri} R.~M.,  {others}  {others} 2013, VizieR Online Data Catalog, \href
  {https://ui.adsabs.harvard.edu/abs/2014yCat.2328....0C} {p. II/328}

\bibitem[\protect\citeauthoryear{{Dhital}, {Burgasser}, {Looper}  \&
  {Stassun}}{{Dhital} et~al.}{2011}]{dhital11}
{Dhital} S.,  {Burgasser} A.~J.,  {Looper} D.~L.,   {Stassun} K.~G.,  2011,
  \mn@doi [\aj] {10.1088/0004-6256/141/1/7}, \href
  {https://ui.adsabs.harvard.edu/abs/2011AJ....141....7D} {141, 7}

\bibitem[\protect\citeauthoryear{{Dunham} et~al.,}{{Dunham}
  et~al.}{2015}]{dunham15}
{Dunham} M.~M.,  et~al., 2015, \mn@doi [\apjs] {10.1088/0067-0049/220/1/11},
  \href {https://ui.adsabs.harvard.edu/abs/2015ApJS..220...11D} {220, 11}

\bibitem[\protect\citeauthoryear{{Eiroa}, {Djupvik}  \& {Casali}}{{Eiroa}
  et~al.}{2008}]{eiroa08}
{Eiroa} C.,  {Djupvik} A.~A.,   {Casali} M.~M.,  2008, {The Serpens Molecular
  Cloud}.
p.~693

\bibitem[\protect\citeauthoryear{{Faherty} et~al.,}{{Faherty}
  et~al.}{2020}]{faherty20}
{Faherty} J.~K.,  et~al., 2020, \mn@doi [\apj] {10.3847/1538-4357/ab5303},
  \href {https://ui.adsabs.harvard.edu/abs/2020ApJ...889..176F} {889, 176}

\bibitem[\protect\citeauthoryear{{Filippazzo}, {Rice}, {Faherty}, {Cruz}, {Van
  Gordon}  \& {Looper}}{{Filippazzo} et~al.}{2015}]{filippazzo15}
{Filippazzo} J.~C.,  {Rice} E.~L.,  {Faherty} J.,  {Cruz} K.~L.,  {Van Gordon}
  M.~M.,   {Looper} D.~L.,  2015, \mn@doi [\apj] {10.1088/0004-637X/810/2/158},
  \href {https://ui.adsabs.harvard.edu/abs/2015ApJ...810..158F} {810, 158}

\bibitem[\protect\citeauthoryear{{Fitzpatrick}}{{Fitzpatrick}}{1999}]{fitzpatrick99}
{Fitzpatrick} E.~L.,  1999, \mn@doi [\pasp] {10.1086/316293}, \href
  {http://adsabs.harvard.edu/abs/1999PASP..111...63F} {111, 63}

\bibitem[\protect\citeauthoryear{{Fontanive} et~al.,}{{Fontanive}
  et~al.}{2020}]{fontanive20}
{Fontanive} C.,  et~al., 2020, \mn@doi [\apjl] {10.3847/2041-8213/abcaf8},
  \href {https://ui.adsabs.harvard.edu/abs/2020ApJ...905L..14F} {905, L14}

\bibitem[\protect\citeauthoryear{{Foreman-Mackey}, {Hogg}, {Lang}  \&
  {Goodman}}{{Foreman-Mackey} et~al.}{2013}]{foreman13}
{Foreman-Mackey} D.,  {Hogg} D.~W.,  {Lang} D.,   {Goodman} J.,  2013, \mn@doi
  [PASP] {10.1086/670067}, \href
  {http://adsabs.harvard.edu/abs/2013PASP..125..306F} {125, 306}

\bibitem[\protect\citeauthoryear{{Gaia Collaboration} et~al.}{{Gaia
  Collaboration} et~al.}{2018}]{gaia18}
{Gaia Collaboration} et~al., 2018, \mn@doi [\aap]
  {10.1051/0004-6361/201833051}, \href
  {http://adsabs.harvard.edu/abs/2018A%26A...616A...1G} {616, A1}

\bibitem[\protect\citeauthoryear{{Gennaro} et~al.,}{{Gennaro}
  et~al.}{2018}]{gennaro18}
{Gennaro} M.,  et~al., 2018, \mn@doi [\apj] {10.3847/1538-4357/aaa973}, \href
  {https://ui.adsabs.harvard.edu/abs/2018ApJ...855...20G} {855, 20}

\bibitem[\protect\citeauthoryear{{Getman}, {Broos}, {Kuhn}, {Feigelson},
  {Richert}, {Ota}, {Bate}  \& {Garmire}}{{Getman} et~al.}{2017}]{getman17}
{Getman} K.~V.,  {Broos} P.~S.,  {Kuhn} M.~A.,  {Feigelson} E.~D.,  {Richert}
  A. J.~W.,  {Ota} Y.,  {Bate} M.~R.,   {Garmire} G.~P.,  2017, \mn@doi [\apjs]
  {10.3847/1538-4365/229/2/28}, \href
  {https://ui.adsabs.harvard.edu/abs/2017ApJS..229...28G} {229, 28}

\bibitem[\protect\citeauthoryear{{Girardi} et~al.,}{{Girardi}
  et~al.}{2012}]{girardi12}
{Girardi} L.,  et~al., 2012, \mn@doi [Astrophysics and Space Science
  Proceedings] {10.1007/978-3-642-18418-5_17}, \href
  {https://ui.adsabs.harvard.edu/abs/2012ASSP...26..165G} {26, 165}

\bibitem[\protect\citeauthoryear{{Gomez Gonzalez} et~al.,}{{Gomez Gonzalez}
  et~al.}{2017}]{gomezgonzalez17}
{Gomez Gonzalez} C.~A.,  et~al., 2017, \mn@doi [\aj]
  {10.3847/1538-3881/aa73d7}, \href
  {http://adsabs.harvard.edu/abs/2017AJ....154....7G} {154, 7}

\bibitem[\protect\citeauthoryear{{Gutermuth} et~al.,}{{Gutermuth}
  et~al.}{2008}]{gutermuth08}
{Gutermuth} R.~A.,  et~al., 2008, \mn@doi [\apjl] {10.1086/528710}, \href
  {https://ui.adsabs.harvard.edu/abs/2008ApJ...673L.151G} {673, L151}

\bibitem[\protect\citeauthoryear{{Herbst}}{{Herbst}}{2008}]{herbst08}
{Herbst} W.,  2008, {Star Formation in IC 348}.
p.~372

\bibitem[\protect\citeauthoryear{{Herczeg} \& {Hillenbrand}}{{Herczeg} \&
  {Hillenbrand}}{2014}]{herczeg14}
{Herczeg} G.~J.,  {Hillenbrand} L.~A.,  2014, \mn@doi [\apj]
  {10.1088/0004-637X/786/2/97}, \href
  {https://ui.adsabs.harvard.edu/abs/2014ApJ...786...97H} {786, 97}

\bibitem[\protect\citeauthoryear{{Herczeg} \& {Hillenbrand}}{{Herczeg} \&
  {Hillenbrand}}{2015}]{herczeg15}
{Herczeg} G.~J.,  {Hillenbrand} L.~A.,  2015, \mn@doi [\apj]
  {10.1088/0004-637X/808/1/23}, \href
  {https://ui.adsabs.harvard.edu/abs/2015ApJ...808...23H} {808, 23}

\bibitem[\protect\citeauthoryear{{Herczeg} et~al.,}{{Herczeg}
  et~al.}{2019}]{herczeg19}
{Herczeg} G.~J.,  et~al., 2019, \mn@doi [\apj] {10.3847/1538-4357/ab1d67},
  \href {https://ui.adsabs.harvard.edu/abs/2019ApJ...878..111H} {878, 111}

\bibitem[\protect\citeauthoryear{{Hosek}, {Lu}, {Anderson}, {Najarro}, {Ghez},
  {Morris}, {Clarkson}  \& {Albers}}{{Hosek} et~al.}{2019}]{hosek19}
{Hosek} Matthew~W. J.,  {Lu} J.~R.,  {Anderson} J.,  {Najarro} F.,  {Ghez}
  A.~M.,  {Morris} M.~R.,  {Clarkson} W.~I.,   {Albers} S.~M.,  2019, \mn@doi
  [\apj] {10.3847/1538-4357/aaef90}, \href
  {https://ui.adsabs.harvard.edu/abs/2019ApJ...870...44H} {870, 44}

\bibitem[\protect\citeauthoryear{{Jose} et~al.,}{{Jose} et~al.}{2020}]{jose20}
{Jose} J.,  et~al., 2020, arXiv e-prints, \href
  {https://ui.adsabs.harvard.edu/abs/2020arXiv200202473J} {p. arXiv:2002.02473}

\bibitem[\protect\citeauthoryear{{Kenyon}, {G{\'o}mez}  \& {Whitney}}{{Kenyon}
  et~al.}{2008}]{kenyon08}
{Kenyon} S.~J.,  {G{\'o}mez} M.,   {Whitney} B.~A.,  2008, {Low Mass Star
  Formation in the Taurus-Auriga Clouds}.
p.~405

\bibitem[\protect\citeauthoryear{{Kirkpatrick}, {Looper}, {Burgasser},
  {Schurr}, {Cutri}  et~al.}{{Kirkpatrick} et~al.}{2010}]{kirkpatrick10}
{Kirkpatrick} J.~D.,  {Looper} D.~L.,  {Burgasser} A.~J.,  {Schurr} S.~D.,
  {Cutri} R.~M.,   et~al., 2010, \mn@doi [\apjs] {10.1088/0067-0049/190/1/100},
  \href {http://adsabs.harvard.edu/abs/2010ApJS..190..100K} {190, 100}

\bibitem[\protect\citeauthoryear{{Kirkpatrick} et~al.,}{{Kirkpatrick}
  et~al.}{2011}]{kirkpatrick11}
{Kirkpatrick} J.~D.,  et~al., 2011, \mn@doi [\apjs]
  {10.1088/0067-0049/197/2/19}, \href
  {https://ui.adsabs.harvard.edu/abs/2011ApJS..197...19K} {197, 19}

\bibitem[\protect\citeauthoryear{{Kirkpatrick} et~al.,}{{Kirkpatrick}
  et~al.}{2019}]{kirkpatrick18}
{Kirkpatrick} J.~D.,  et~al., 2019, \mn@doi [\apjs] {10.3847/1538-4365/aaf6af},
  \href {https://ui.adsabs.harvard.edu/abs/2019ApJS..240...19K} {240, 19}

\bibitem[\protect\citeauthoryear{{Klotz}, {Caux}, {Monin}  \& {Lodieu}}{{Klotz}
  et~al.}{2004}]{klotz04}
{Klotz} A.,  {Caux} E.,  {Monin} J.~L.,   {Lodieu} N.,  2004, \mn@doi [\aap]
  {10.1051/0004-6361:20041081}, \href
  {https://ui.adsabs.harvard.edu/abs/2004A&A...425..927K} {425, 927}

\bibitem[\protect\citeauthoryear{{Knapp} et~al.,}{{Knapp}
  et~al.}{2004}]{knapp04}
{Knapp} G.~R.,  et~al., 2004, \mn@doi [\aj] {10.1086/420707}, \href
  {https://ui.adsabs.harvard.edu/abs/2004AJ....127.3553K} {127, 3553}

\bibitem[\protect\citeauthoryear{{Konopacky}, {Ghez}, {Rice}  \&
  {Duch{\^e}ne}}{{Konopacky} et~al.}{2007}]{konopacky07}
{Konopacky} Q.~M.,  {Ghez} A.~M.,  {Rice} E.~L.,   {Duch{\^e}ne} G.,  2007,
  \mn@doi [\apj] {10.1086/518360}, \href
  {https://ui.adsabs.harvard.edu/abs/2007ApJ...663..394K} {663, 394}

\bibitem[\protect\citeauthoryear{{Kroupa}}{{Kroupa}}{2001}]{kroupa01}
{Kroupa} P.,  2001, \mn@doi [\mnras] {10.1046/j.1365-8711.2001.04022.x}, \href
  {https://ui.adsabs.harvard.edu/abs/2001MNRAS.322..231K} {322, 231}

\bibitem[\protect\citeauthoryear{{Lafreni{\`e}re}, {Marois}, {Doyon}  \&
  {Barman}}{{Lafreni{\`e}re} et~al.}{2009}]{lafreniere09}
{Lafreni{\`e}re} D.,  {Marois} C.,  {Doyon} R.,   {Barman} T.,  2009, \mn@doi
  [\apjl] {10.1088/0004-637X/694/2/L148}, \href
  {https://ui.adsabs.harvard.edu/abs/2009ApJ...694L.148L} {694, L148}

\bibitem[\protect\citeauthoryear{{Lagrange} et~al.,}{{Lagrange}
  et~al.}{2009}]{lagrange09}
{Lagrange} A.~M.,  et~al., 2009, \mn@doi [\aap] {10.1051/0004-6361:200811325},
  \href {https://ui.adsabs.harvard.edu/abs/2009A&A...493L..21L} {493, L21}

\bibitem[\protect\citeauthoryear{{Larson}}{{Larson}}{1992}]{larson92}
{Larson} R.~B.,  1992, \mn@doi [\mnras] {10.1093/mnras/256.3.641}, \href
  {https://ui.adsabs.harvard.edu/abs/1992MNRAS.256..641L} {256, 641}

\bibitem[\protect\citeauthoryear{{Lodieu}, {Caux}, {Monin}  \&
  {Klotz}}{{Lodieu} et~al.}{2002}]{lodieu02}
{Lodieu} N.,  {Caux} E.,  {Monin} J.~L.,   {Klotz} A.,  2002, \mn@doi [\aap]
  {10.1051/0004-6361:20020034}, \href
  {https://ui.adsabs.harvard.edu/abs/2002A&A...383L..15L} {383, L15}

\bibitem[\protect\citeauthoryear{{Lodieu}, {Zapatero Osorio}, {B{\'e}jar}  \&
  {Pe{\~n}a Ram{\'\i}rez}}{{Lodieu} et~al.}{2018}]{lodieu18}
{Lodieu} N.,  {Zapatero Osorio} M.~R.,  {B{\'e}jar} V.~J.~S.,   {Pe{\~n}a
  Ram{\'\i}rez} K.,  2018, \mn@doi [\mnras] {10.1093/mnras/stx2279}, \href
  {https://ui.adsabs.harvard.edu/abs/2018MNRAS.473.2020L} {473, 2020}

\bibitem[\protect\citeauthoryear{{Lu}, {Do}, {Ghez}, {Morris}, {Yelda}  \&
  {Matthews}}{{Lu} et~al.}{2013}]{lu13}
{Lu} J.~R.,  {Do} T.,  {Ghez} A.~M.,  {Morris} M.~R.,  {Yelda} S.,   {Matthews}
  K.,  2013, \mn@doi [\apj] {10.1088/0004-637X/764/2/155}, \href
  {https://ui.adsabs.harvard.edu/abs/2013ApJ...764..155L} {764, 155}

\bibitem[\protect\citeauthoryear{{Lucas} \& {Roche}}{{Lucas} \&
  {Roche}}{2000}]{lucas00}
{Lucas} P.~W.,  {Roche} P.~F.,  2000, \mn@doi [\mnras]
  {10.1046/j.1365-8711.2000.03515.x}, \href
  {https://ui.adsabs.harvard.edu/abs/2000MNRAS.314..858L} {314, 858}

\bibitem[\protect\citeauthoryear{{Luhman}}{{Luhman}}{1999}]{luhman99}
{Luhman} K.~L.,  1999, \mn@doi [\apj] {10.1086/307902}, \href
  {https://ui.adsabs.harvard.edu/abs/1999ApJ...525..466L} {525, 466}

\bibitem[\protect\citeauthoryear{{Luhman}}{{Luhman}}{2008}]{luhman08}
{Luhman} K.~L.,  2008, {Chamaeleon}.
p.~169

\bibitem[\protect\citeauthoryear{{Luhman}, {Mamajek}, {Shukla}  \&
  {Loutrel}}{{Luhman} et~al.}{2017}]{luhman17}
{Luhman} K.~L.,  {Mamajek} E.~E.,  {Shukla} S.~J.,   {Loutrel} N.~P.,  2017,
  \mn@doi [\aj] {10.3847/1538-3881/153/1/46}, \href
  {http://adsabs.harvard.edu/abs/2017AJ....153...46L} {153, 46}

\bibitem[\protect\citeauthoryear{{Muench}, {Lada}, {Luhman}, {Muzerolle}  \&
  {Young}}{{Muench} et~al.}{2007}]{muench07}
{Muench} A.~A.,  {Lada} C.~J.,  {Luhman} K.~L.,  {Muzerolle} J.,   {Young} E.,
  2007, \mn@doi [\aj] {10.1086/518560}, \href
  {https://ui.adsabs.harvard.edu/abs/2007AJ....134..411M} {134, 411}

\bibitem[\protect\citeauthoryear{{Najita}, {Tiede}  \& {Carr}}{{Najita}
  et~al.}{2000}]{najita00}
{Najita} J.~R.,  {Tiede} G.~P.,   {Carr} J.~S.,  2000, \mn@doi [\apj]
  {10.1086/309477}, \href
  {https://ui.adsabs.harvard.edu/abs/2000ApJ...541..977N} {541, 977}

\bibitem[\protect\citeauthoryear{{Ortiz-Le{\'o}n} et~al.,}{{Ortiz-Le{\'o}n}
  et~al.}{2017}]{ortiz17}
{Ortiz-Le{\'o}n} G.~N.,  et~al., 2017, \mn@doi [\apj]
  {10.3847/1538-4357/834/2/143}, \href
  {https://ui.adsabs.harvard.edu/abs/2017ApJ...834..143O} {834, 143}

\bibitem[\protect\citeauthoryear{{Ortiz-Le{\'o}n} et~al.,}{{Ortiz-Le{\'o}n}
  et~al.}{2018}]{ortiz18}
{Ortiz-Le{\'o}n} G.~N.,  et~al., 2018, \mn@doi [\apjl]
  {10.3847/2041-8213/aaf6ad}, \href
  {https://ui.adsabs.harvard.edu/abs/2018ApJ...869L..33O} {869, L33}

\bibitem[\protect\citeauthoryear{{Pecaut} \& {Mamajek}}{{Pecaut} \&
  {Mamajek}}{2013}]{pecaut13}
{Pecaut} M.~J.,  {Mamajek} E.~E.,  2013, \mn@doi [\apjs]
  {10.1088/0067-0049/208/1/9}, \href
  {https://ui.adsabs.harvard.edu/abs/2013ApJS..208....9P} {208, 9}

\bibitem[\protect\citeauthoryear{{Povich} et~al.,}{{Povich}
  et~al.}{2013}]{povich13}
{Povich} M.~S.,  et~al., 2013, \mn@doi [\apjs] {10.1088/0067-0049/209/2/31},
  \href {https://ui.adsabs.harvard.edu/abs/2013ApJS..209...31P} {209, 31}

\bibitem[\protect\citeauthoryear{{Puget} et~al.,}{{Puget}
  et~al.}{2004}]{puget04}
{Puget} P.,  et~al., 2004, in {Moorwood} A.~F.~M.,  {Iye} M.,  eds,  \procspie
  Vol. 5492, Ground-based Instrumentation for Astronomy. pp 978--987,
  \mn@doi{10.1117/12.551097}

\bibitem[\protect\citeauthoryear{{Rayner}, {Toomey}, {Onaka}, {Denault},
  {Stahlberger}  et~al.}{{Rayner} et~al.}{2003}]{rayner03}
{Rayner} J.~T.,  {Toomey} D.~W.,  {Onaka} P.~M.,  {Denault} A.~J.,
  {Stahlberger} W.~E.,   et~al., 2003, \mn@doi [\pasp] {10.1086/367745}, \href
  {http://adsabs.harvard.edu/abs/2003PASP..115..362R} {115, 362}

\bibitem[\protect\citeauthoryear{{Rayner}, {Cushing}  \& {Vacca}}{{Rayner}
  et~al.}{2009}]{rayner09}
{Rayner} J.~T.,  {Cushing} M.~C.,   {Vacca} W.~D.,  2009, \mn@doi [\apjs]
  {10.1088/0067-0049/185/2/289}, \href
  {https://ui.adsabs.harvard.edu/abs/2009ApJS..185..289R} {185, 289}

\bibitem[\protect\citeauthoryear{{Rebull}, {Padgett}, {McCabe}, {Hillenbrand},
  {Stapelfeldt}  et~al.}{{Rebull} et~al.}{2010}]{rebull10}
{Rebull} L.~M.,  {Padgett} D.~L.,  {McCabe} C.-E.,  {Hillenbrand} L.~A.,
  {Stapelfeldt} K.~R.,   et~al., 2010, \mn@doi [\apjs]
  {10.1088/0067-0049/186/2/259}, \href
  {http://adsabs.harvard.edu/abs/2010ApJS..186..259R} {186, 259}

\bibitem[\protect\citeauthoryear{{Reid} et~al.,}{{Reid} et~al.}{1999}]{reid99}
{Reid} I.~N.,  et~al., 1999, \mn@doi [\apj] {10.1086/307589}, \href
  {https://ui.adsabs.harvard.edu/abs/1999ApJ...521..613R} {521, 613}

\bibitem[\protect\citeauthoryear{{Saumon}, {Marley}, {Cushing}, {Leggett},
  {Roellig}, {Lodders}  \& {Freedman}}{{Saumon} et~al.}{2006}]{saumon06}
{Saumon} D.,  {Marley} M.~S.,  {Cushing} M.~C.,  {Leggett} S.~K.,  {Roellig}
  T.~L.,  {Lodders} K.,   {Freedman} R.~S.,  2006, \mn@doi [\apj]
  {10.1086/505419}, \href
  {https://ui.adsabs.harvard.edu/abs/2006ApJ...647..552S} {647, 552}

\bibitem[\protect\citeauthoryear{{Skrutskie} et~al.,}{{Skrutskie}
  et~al.}{2006}]{skrutskie06}
{Skrutskie} M.~F.,  et~al., 2006, \mn@doi [\aj] {10.1086/498708}, \href
  {https://ui.adsabs.harvard.edu/abs/2006AJ....131.1163S} {131, 1163}

\bibitem[\protect\citeauthoryear{{Soummer}, {Pueyo}  \& {Larkin}}{{Soummer}
  et~al.}{2012}]{soummer12}
{Soummer} R.,  {Pueyo} L.,   {Larkin} J.,  2012, \mn@doi [\apj]
  {10.1088/2041-8205/755/2/L28}, \href
  {https://ui.adsabs.harvard.edu/abs/2012ApJ...755L..28S} {755, L28}

\bibitem[\protect\citeauthoryear{{Spezzi}, {Alves de Oliveira}, {Moraux},
  {Bouvier}, {Winston}  et~al.}{{Spezzi} et~al.}{2012}]{spezzi12}
{Spezzi} L.,  {Alves de Oliveira} C.,  {Moraux} E.,  {Bouvier} J.,  {Winston}
  E.,   et~al., 2012, \mn@doi [\aap] {10.1051/0004-6361/201219559}, \href
  {http://adsabs.harvard.edu/abs/2012A%26A...545A.105S} {545, A105}

\bibitem[\protect\citeauthoryear{{Vacca}, {Cushing}  \& {Rayner}}{{Vacca}
  et~al.}{2003}]{vacca03}
{Vacca} W.~D.,  {Cushing} M.~C.,   {Rayner} J.~T.,  2003, \mn@doi [\pasp]
  {10.1086/346193}, \href {http://adsabs.harvard.edu/abs/2003PASP..115..389V}
  {115, 389}

\bibitem[\protect\citeauthoryear{{Whitworth}, {Bate}, {Nordlund}, {Reipurth}
  \& {Zinnecker}}{{Whitworth} et~al.}{2007}]{whitworth07}
{Whitworth} A.,  {Bate} M.~R.,  {Nordlund} {\r{A}}.,  {Reipurth} B.,
  {Zinnecker} H.,  2007, in {Reipurth} B.,  {Jewitt} D.,   {Keil} K.,  eds,
  Protostars and Planets V. p.~459

\bibitem[\protect\citeauthoryear{{Winston} et~al.,}{{Winston}
  et~al.}{2007}]{winston07}
{Winston} E.,  et~al., 2007, \mn@doi [\apj] {10.1086/521384}, \href
  {https://ui.adsabs.harvard.edu/abs/2007ApJ...669..493W} {669, 493}

\bibitem[\protect\citeauthoryear{{Winston} et~al.,}{{Winston}
  et~al.}{2010}]{winston10}
{Winston} E.,  et~al., 2010, \mn@doi [\aj] {10.1088/0004-6256/140/1/266}, \href
  {https://ui.adsabs.harvard.edu/abs/2010AJ....140..266W} {140, 266}

\bibitem[\protect\citeauthoryear{{Winston}, {Wolk}, {Gutermuth}  \&
  {Bourke}}{{Winston} et~al.}{2018}]{winston18}
{Winston} E.,  {Wolk} S.~J.,  {Gutermuth} R.,   {Bourke} T.~L.,  2018, \mn@doi
  [\aj] {10.3847/1538-3881/aabe82}, \href
  {https://ui.adsabs.harvard.edu/abs/2018AJ....155..241W} {155, 241}

\bibitem[\protect\citeauthoryear{{Wright} et~al.,}{{Wright}
  et~al.}{2010}]{wright10}
{Wright} E.~L.,  et~al., 2010, \mn@doi [\aj] {10.1088/0004-6256/140/6/1868},
  \href {https://ui.adsabs.harvard.edu/abs/2010AJ....140.1868W} {140, 1868}

\bibitem[\protect\citeauthoryear{{Zapatero Osorio}, {B{\'e}jar}  \& {Pe{\~n}a
  Ram{\'\i}rez}}{{Zapatero Osorio} et~al.}{2017}]{zapatero17}
{Zapatero Osorio} M.~R.,  {B{\'e}jar} V.~J.~S.,   {Pe{\~n}a Ram{\'\i}rez} K.,
  2017, \mn@doi [\apj] {10.3847/1538-4357/aa70ec}, \href
  {https://ui.adsabs.harvard.edu/abs/2017ApJ...842...65Z} {842, 65}

\bibitem[\protect\citeauthoryear{{van Dokkum} \& {Conroy}}{{van Dokkum} \&
  {Conroy}}{2010}]{dokkum10}
{van Dokkum} P.~G.,  {Conroy} C.,  2010, \mn@doi [\nat] {10.1038/nature09578},
  \href {https://ui.adsabs.harvard.edu/abs/2010Natur.468..940V} {468, 940}

\makeatother
\end{thebibliography}


\appendix

\section{Physical properties from spectral fitting} \label{sec:app_plots}

\begin{figure*}
    \centering
    \includegraphics[width=\textwidth]{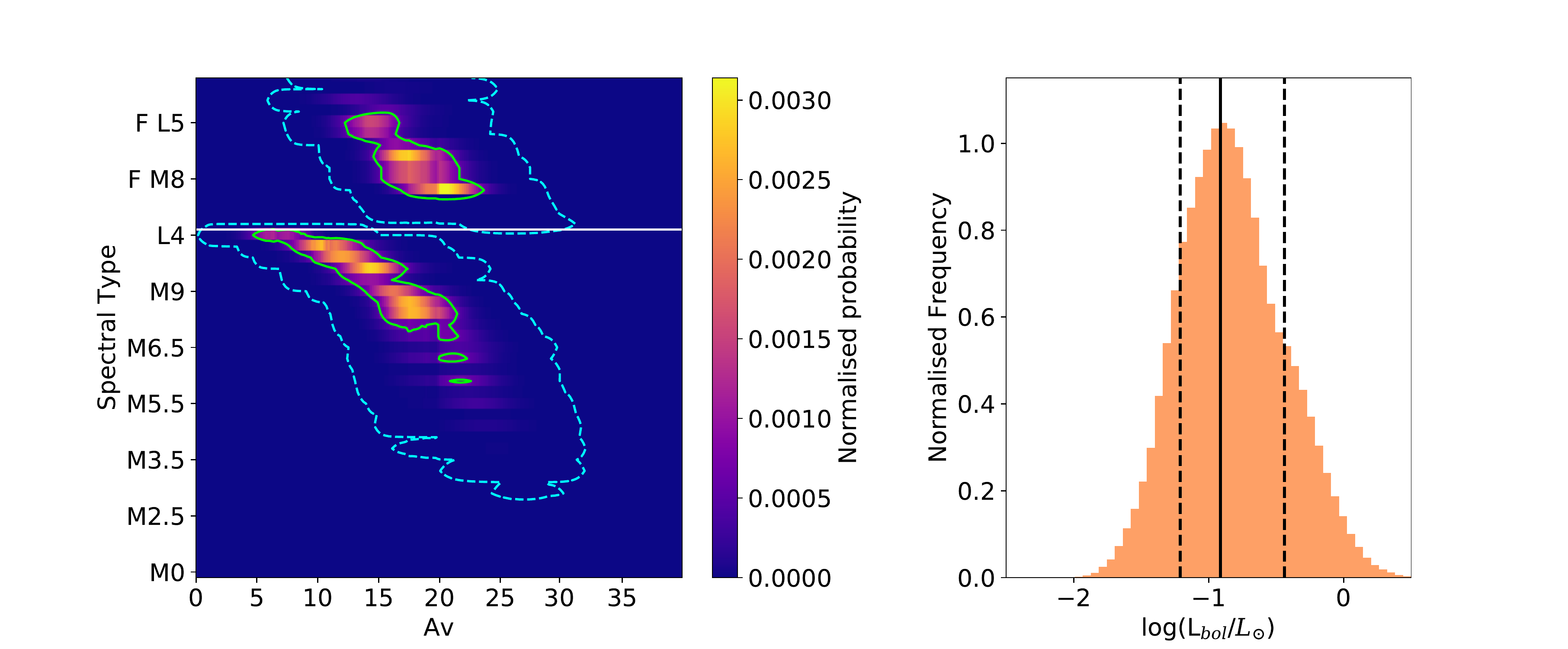}
    \caption{Left: Normalised probability map for SS182953-015639. Contours show 1$\sigma$ (blue,dashed) and 2$\sigma$ (green,solid) levels, respectively. Field and young solutions are separated by the white line (field denoted by F). Right: Histogram of log($L_{\text{bol}}/L_{\odot})$ solutions for SS182953-015639, derived from the normalised probability map. Also shown is the peak value (solid line), and the 68\% (or 1$\sigma$) Bayesian credible intervals (dashed lines), which reflect the asymmetry of the distribution.  }
    \label{fig:props_SS182953}
\end{figure*}

\begin{figure*}
    \centering
    \includegraphics[width=\textwidth]{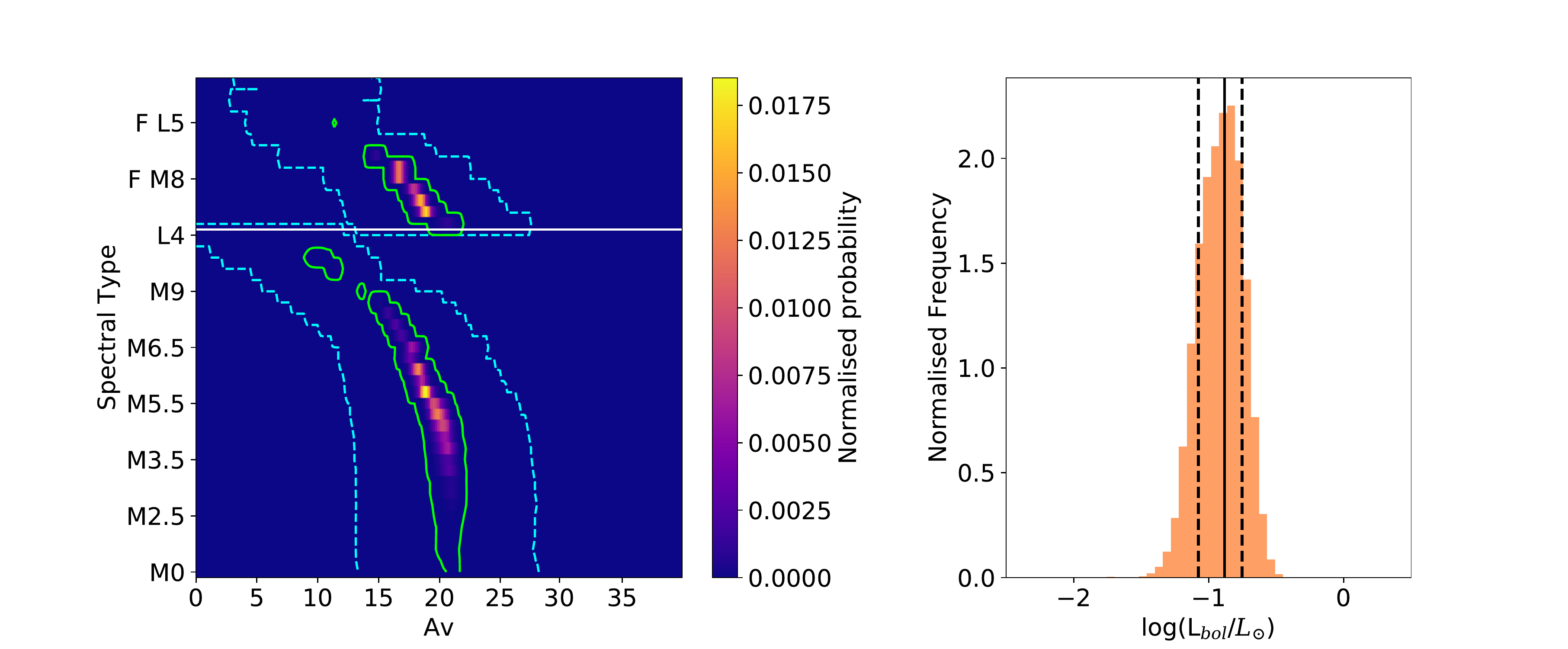}
    \caption{Left: Normalised probability map for SS182955-020416. Contours show 1$\sigma$ (blue,dashed) and 2$\sigma$ (green,solid) levels, respectively. Field and young solutions are separated by the white line (field denoted by F). Right: Histogram of log($L_{\text{bol}}/L_{\odot})$ solutions for SS182955-020416, derived from the normalised probability map. Also shown is the peak value (solid line), and the 68\% (or 1$\sigma$) Bayesian credible intervals (dashed lines), which reflect the asymmetry of the distribution. }
    \label{fig:props_SS182955}
\end{figure*}

\begin{figure*}
    \centering
    \includegraphics[width=\textwidth]{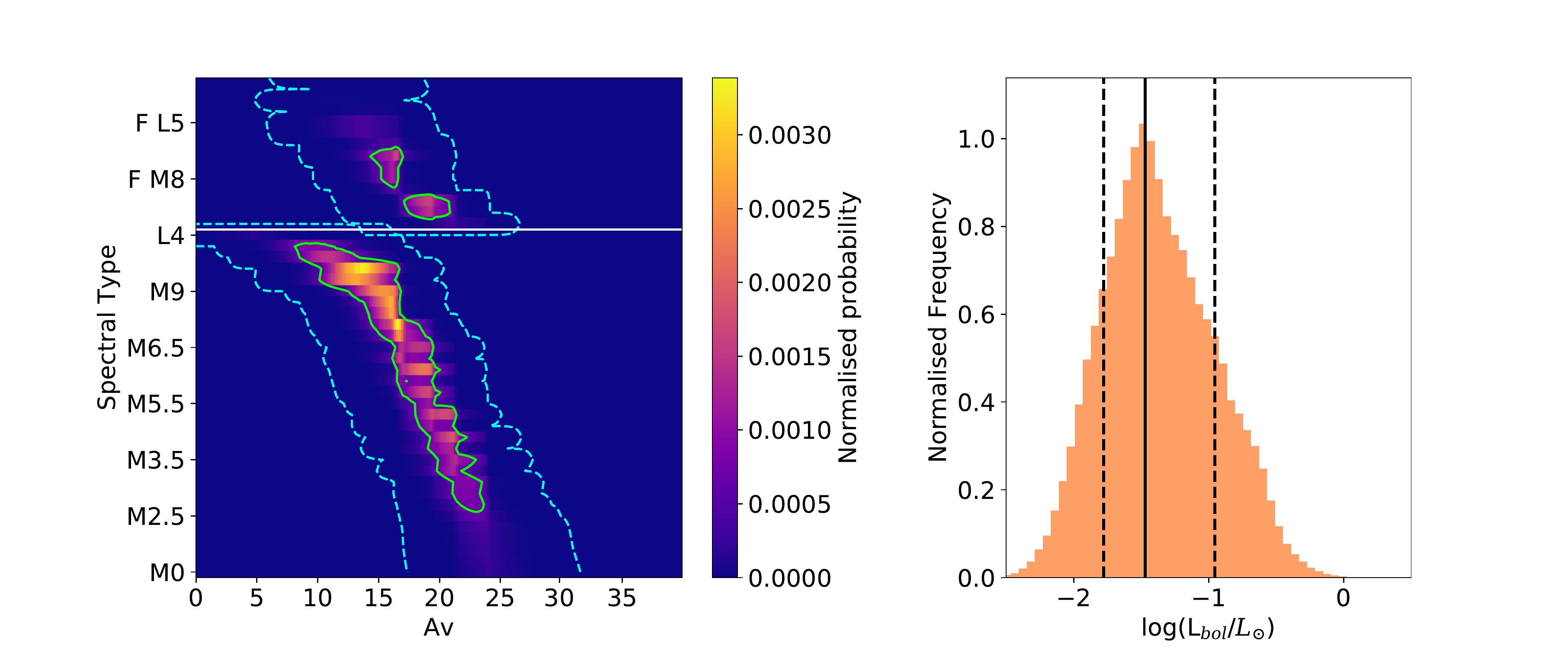}
    \caption{Left: Normalised probability map for SS182959-020335. Contours show 1$\sigma$ (blue,dashed) and 2$\sigma$ (green,solid) levels, respectively. Field and young solutions are separated by the white line (field denoted by F). Right: Histogram of log($L_{\text{bol}}/L_{\odot})$ solutions for SS182959-020335, derived from the normalised probability map. Also shown is the peak value (solid line), and the 68\% (or 1$\sigma$) Bayesian credible intervals (dashed lines), which reflect the asymmetry of the distribution. }
    \label{fig:props_SS182959}
\end{figure*}

\begin{figure*}
    \centering
    \includegraphics[width=\textwidth]{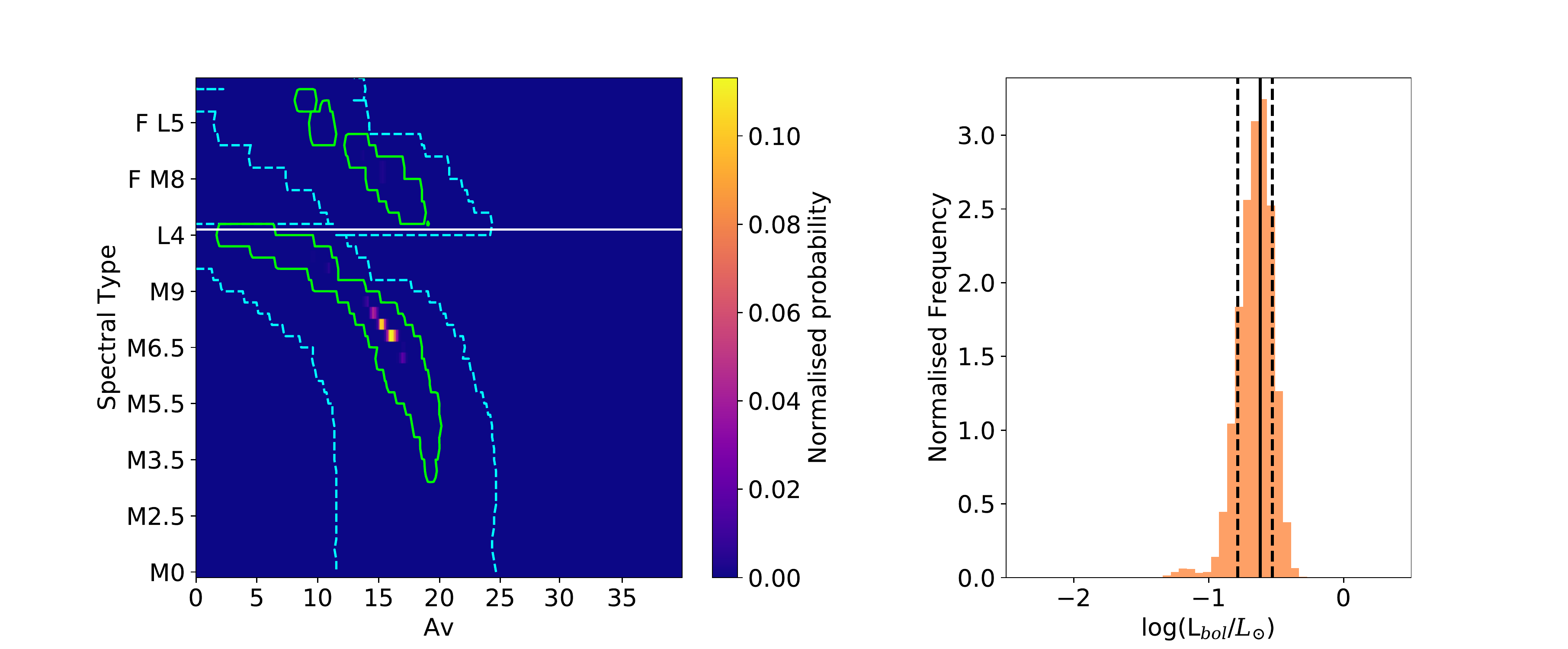}
    \caption{Left: Normalised probability map for SC182952+011618. Contours show 1$\sigma$ (blue,dashed) and 2$\sigma$ (green,solid) levels, respectively. Field and young solutions are separated by the white line (field denoted by F). Right: Histogram of log($L_{\text{bol}}/L_{\odot})$ solutions for SC182952+011618, derived from the normalised probability map. Also shown is the peak value (solid line), and the 68\% (or 1$\sigma$) Bayesian credible intervals (dashed lines), which reflect the asymmetry of the distribution.  }
    \label{fig:props_SC182952}
\end{figure*}

\section{Binary Fitting} \label{sec:app_binary}

\begin{figure*}
    \centering
    \includegraphics[width=\textwidth]{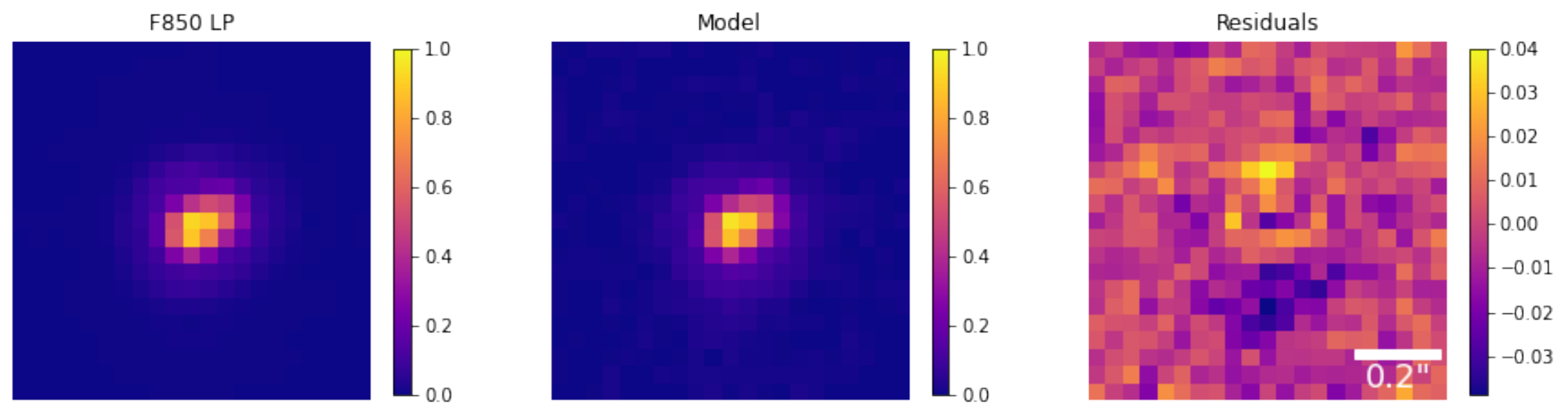}
    \caption{SS183044+020918 in the F850LP filter after PSF subtraction using five principal components using the RDI and PCA method, where the binary components are resolved. F850LP data shown in the left panel, model components shown in centre, and residuals after PSF subtraction shown in the right.}
    \label{fig:vip_binary_im}
\end{figure*}

\bsp	
\label{lastpage}
\end{document}